\documentclass[10pt,letterpaper]{article}
\usepackage[top=0.85in,left=2.75in,footskip=0.75in]{geometry}

\usepackage{amsmath,amssymb}

\usepackage{bm}

\usepackage{changepage}

\usepackage[utf8x]{inputenc}

\usepackage{textcomp,marvosym}

\usepackage{cite}

\usepackage{nameref,hyperref}

\usepackage[right]{lineno}

\usepackage{microtype}
\DisableLigatures[f]{encoding = *, family = * }

\usepackage[table]{xcolor}

\usepackage{array}

\usepackage{tikz}

\usepackage{threeparttable}

\newcolumntype{+}{!{\vrule width 2pt}}

\newlength\savedwidth

\newcommand\thickhline{\noalign{\global\savedwidth\arrayrulewidth\global\arrayrulewidth 2pt}%
\hline
\noalign{\global\arrayrulewidth\savedwidth}}

\raggedright
\usepackage{ragged2e}

\usepackage[aboveskip=1pt,labelfont=bf,labelsep=period,justification=raggedright,singlelinecheck=off, margin={-2.2in,0in}]{caption}

\makeatletter
\renewcommand{\@biblabel}[1]{\quad#1.}
\makeatother

\usepackage{lastpage,fancyhdr,graphicx}
\usepackage{epstopdf}
\fancyheadoffset[L]{2.25in}
\fancyfootoffset[L]{2.25in}
\usepackage{soul}
\usepackage{ulem}
\usepackage{cancel}

\usepackage{wrapfig}
\usepackage{rotating}
\usepackage{booktabs}
\usepackage{makecell}

\usepackage{algpseudocode}
\usepackage{algorithm}

\newcommand{\affilPasteur}{Institut Pasteur, Universit\'e Paris Cit\'e, CNRS UMR 3751, Decision and Bayesian Computation, Paris, France.}

\newcommand{\affilINRIA}{Epim\'eth\'ee, Inria, Paris, France}

\DeclareMathOperator{\argmin}{argmin}

\newcommand{\E}[1]{\left\langle #1 \right\rangle}
\newcommand{\pare}[1]{\left(#1 \right)}

\newcommand{\vdiag}{v_D}

\newcommand{\Vset}{\mathcal{N}}
\newcommand{\Eset}{\mathcal{E}}
\newcommand{\Vnumber}{N}
\newcommand{\Enumber}{E}
\newcommand{\A}{\mathbf{A}}

\newcommand{\sym}{\mathrm{sym}}
\newcommand{\asym}{\mathrm{asym}}

\newcommand{\Ednumber}{E_d}
\newcommand{\Emnumber}{E_m}

\newcommand{\vset}{{\nu}}
\newcommand{\eset}{{\epsilon}}
\newcommand{\vnumber}{n}
\newcommand{\enumber}{e}

\newcommand{\graphlet}{\alpha}

\newcommand{\graphletInstance}{g_\graphlet}

\newcommand{\graphletSize}[1]{n_{#1}}

\newcommand{\subgraphNeighborhood}{{\partial\subgraph}}

\newcommand{\model}[1]{P_{#1}}
\newcommand{\modelSet}{\mathcal{M}}
\newcommand{\params}{\theta}
\newcommand{\baseParams}{\phi}
\newcommand{\paramsSet}{\Theta}
\newcommand{\partitionFunction}[1]{\Omega_{#1}}
\newcommand{\codelength}[1]{L_{#1}}
\newcommand{\compressibility}[1]{\Delta\codelength{#1}}
\newcommand{\entropy}[1]{S_{#1}}
\newcommand{\graphletSet}{\Gamma}
\newcommand{\graphletSubset}{\mathcal{A}}
\newcommand{\subgraph}{s}
\newcommand{\subgraphSet}{\mathcal{S}}
\newcommand{\supernodes}{\mathcal{V}}
\newcommand{\nmax}{m_{\max}}
\newcommand{\ngraphlet}{m_\graphlet}
\newcommand{\integerCodelength}[1]{\codelength{\mathbb{N}}(#1)}
\newcommand{\seqCodelength}[1]{\codelength{\mathrm{seq}}(#1)}
\newcommand{\setSize}[1]{|#1|}
\newcommand{\Aut}{\mathrm{Aut}}
\newcommand{\freq}{r}

\newcommand{\minibatch}[1]{\mathcal{B}_{#1}}
\newcommand{\subgraphOccurrences}{\mathcal{C}}

\newcommand{\baseCost}{\codelength{}(H, \baseParams)}
\newcommand{\graphletCost}{\codelength{}(\graphletSet, \subgraphSet)}
\newcommand{\supernodeCost}{\codelength{}(\supernodes|H,\subgraphSet)}
\newcommand{\reconstructionCost}{\codelength{}(G|H, \supernodes, \subgraphSet, \graphletSet)}
\newcommand{\rewiringCost}{\ell_{\textrm{rew}}(i_s, H)}

\newcommand{\degree}{k}
\newcommand{\indegree}{\degree^-}
\newcommand{\outdegree}{\degree^+}
\newcommand{\degrees}{\mathbf{\degree}}
\newcommand{\indegrees}{\degrees^-}
\newcommand{\outdegrees}{\degrees^+}
\newcommand{\mindegree}{\delta}
\newcommand{\maxdegree}{\Delta}

\newcommand{\degreeR}{\kappa}
\newcommand{\indegreeR}{\degreeR^-}
\newcommand{\outdegreeR}{\degreeR^+}
\newcommand{\mutualdegreeR}{\degreeR^m}
\newcommand{\degreesR}{\bm{\degreeR}}
\newcommand{\indegreesR}{\degreesR^-}
\newcommand{\outdegreesR}{\degreesR^+}
\newcommand{\mutualdegreesR}{\degreesR^m}

\newcommand{\thetaER}{{(\Vnumber,\Enumber)}}
\newcommand{\thetaRER}{{(\Vnumber,\Emnumber,\Ednumber)}}
\newcommand{\thetaConf}{{(\outdegrees,\indegrees)}}
\newcommand{\thetaRConf}{{(\mutualdegreesR,\outdegreesR,\indegreesR)}}

\newcommand{\SubgraphCensus}{{S}{\footnotesize UBGRAPH}{C}{\footnotesize ENSUS}}
\newcommand{\SubgraphBatches}{{S}{\footnotesize UBGRAPH}{B}{\footnotesize ATHCHES}}
\newcommand{\BestCompressingSubgraph}{{M}{\footnotesize OST}{C}{\footnotesize OMPRESSING}{S}{\footnotesize UBGRAPH}}
\newcommand{\SubgraphContraction}{{S}{\footnotesize UBGRAPH}{C}{\footnotesize ONTRACTION}}

\makeatletter
\newcommand{\labelname}[1]{
  \def\@currentlabelname{#1}}%
\makeatother

\begin{document}
\vspace*{0.2in}

\begin{flushleft}
{\Large\textbf\newline{
Compression-based inference of network motif sets} 
}
\newline
\\
Alexis B\'enichou\textsuperscript{1,2*},
Jean-Baptiste Masson\textsuperscript{1,2},
Christian L.\ Vestergaard\textsuperscript{1,2*}
\\
\bigskip
\textbf{1} \affilPasteur
\\
\textbf{2} \affilINRIA
\bigskip

%
%





* alexis.benichou@pasteur.fr; christian.vestergaard@cnrs.fr

\end{flushleft}

\section*{Abstract}

Physical and functional constraints on biological networks lead to complex topological patterns across multiple scales in their organization.
A particular type of higher-order network feature that has received considerable interest is network motifs, defined as statistically regular subgraphs. 
These may implement fundamental logical and computational circuits and are referred to as “building blocks of complex networks”. 
Their well-defined structures and small sizes also enable the testing of their functions in synthetic and natural biological experiments.
Here, we develop a framework for motif mining based on lossless network compression using subgraph contractions. 
This provides an alternative definition of motif significance which allows us to compare different motifs and select the collectively most significant set of motifs as well as other prominent network features in terms of their combined compression of the network.
Our approach inherently accounts for multiple testing and correlations between subgraphs and does not rely on \textit{a priori} specification of an appropriate null model.
It thus overcomes common problems in hypothesis testing-based motif analysis and guarantees robust statistical inference.
We validate our methodology on numerical data and then apply it on synaptic-resolution biological neural networks, as a medium for comparative connectomics, by evaluating their respective compressibility and characterize their inferred circuit motifs.

\section*{Author summary}

Networks provide a useful abstraction to study complex systems by focusing on the interplay of the units composing a system rather than on their individual function.
Network theory has proven particularly powerful for unraveling how the structure of connections in biological networks influence the way they may process and relay information in a variety of systems ranging from the microscopic scale of biochemical processes in cells to the macroscopic scales of social and ecological networks. 
Of particular interest are small stereotyped circuits in such networks, termed \textit{motifs}, which may correspond to building blocks implementing fundamental operations, e.g., logic gates or filters.
We here present a new tool that finds sets of motifs in networks based on an information-theoretic measure of how much they allow to compress the network.
This approach allows us to evaluate the collective significance of sets of motifs, as opposed to only individual motifs.
We apply our methodology to compare the neural wiring diagrams, termed ``connectomes", of the tadpole larva \textit{Ciona intestinalis}, the ragworm \textit{Platynereis dumerelii}, and the nematode \textit{Caenorhabditis elegans} and the fruitfly \textit{Drosophila melanogaster} at different developmental stages.


\section*{Introduction}

Network theory has highlighted remarkable topological features of many biological and social networks~\cite{newman_structure_2003, fornito_fundamentals_2016, alon_introduction_2019}. 
Some of the main ones are 
the \textit{small world} property~\cite{watts_Collective_1998, newman2018networks, sporns_small_2004, bassett2017small}, which refers to a simultaneous high local clustering of connections and short global distances between nodes; 
scale-free features, most notably witnessed by a broad distribution of node degrees~\cite{barabasi_emergence_1999, cohen2000resilience,pastor2001epidemic,seyed2006scale};
mesoscopic, and in particular modular, structuring~\cite{newman2006modularity, ravasz2009detecting,cimini2019statistical};
and higher-order topological features~\cite{battiston_networks_2020}, such as a statistical over-representation of certain types of subgraphs, termed \textit{network motifs}~\cite{milo_network_2002, sporns_motifs_2004, tran_current_2015}. 

We here focus on network motifs.
They were first introduced to study local structures in social networks~\cite{holland_method_1977, holland_local_1976, stone_network_2019}. 
In biological networks, they are hypothesized to capture functional subunits (e.g., logic gates or filters) and have been extensively studied in systems ranging from transcription and protein networks to brain and ecological networks~\cite{milo_network_2002, milo_superfamilies_2004, sporns_motifs_2004, yeger-lotem_network_2004, bascompte_simple_2005, tran_current_2015, fornito_fundamentals_2016}.
In contrast to most other remarkable features of biological networks, the well-defined structure and small size of network motifs mean that their function may be probed experimentally, both in natural~\cite{alon_network_2007, jovanic_competitive_2016} and in synthetic experiments~\cite{alon_network_2007}.

The prevailing approach to network motif inference involves counting or estimating the frequency of each subgraph type, termed a \textit{graphlet}, and comparing it to its frequency in random networks generated by a predefined null model.
Subgraphs that appear significantly more frequently in the empirical network than in the random networks are deemed motifs.
While this procedure has offered valuable insights, 
it also  suffers from several fundamental limitations  which can make it statistically unreliable~\cite{artzy-randrup_comment_2004, ginoza_network_2010, beber_artefacts_2012, orsini_quantifying_2015, fodor_intrinsic_2020, stivala_testing_2021} (see~\nameref{SIsec:hypothesis-testing} for an overview).
Additionally, a flaw of testing-based approaches is that they cannot compare the significance of different motifs. Candidate motifs are usually treated independently. With increasingly richer and larger datasets, such methods thus risk detecting an exceedingly large amount of motifs  (see, e.g.,~\nameref{SIfig:analysis-classic-testing} Fig), which defies the original intent behind motif analysis as a means to capture essential, low-dimensional, mesoscopic properties of a network.

Information theory tells us that the presence of statistical regularities in a network makes it compressible~\cite{cover_elements_2012}.
Inspired by this fact, we here propose a methodology, based on lossless compression~\cite{bloem_large-scale_2020} as a measure of significance, that implicitly defines a generative model through the correspondence between universal codes and probability distributions~\cite{cover_elements_2012, grunwald_minimum_2007}. 
Through the minimum description length (MDL) principle~\cite{grunwald_minimum_2007, grunwald_minimum_2020}, our method infers the set of most significant motifs, as well as other node- and edge-level features, such as node degrees and edge reciprocity, by measuring how much they collectively allow to compress the network. 
We demonstrate how this approach allows to address the shortcomings of hypothesis testing-based motif inference. 
First, it naturally lets us account for multiple testing and correlations between different motifs. 
Furthermore, we can evaluate and compare even highly significant collections of motifs. 
Finally, our method selects not only the most significant motif configuration, but also node- and edge-level features, without needing to select the null model beforehand.

We first validate our approach on numerically generated networks with known absence or presence of motifs.
We then apply our methodology to discover microcircuit motifs in synapse-resolution neuron wiring diagrams, the \textit{connectomes}, of small animals which have recently become available thanks to advances in electron microscopy techniques and image segmentation~\cite{saalfeld2009catmaid,ohyama_multilevel_2015,witvliet_connectomes_2021, winding_connectome_2023}. 
We compare the compressibility induced by motif sets and other network features found in different brain regions of different animals. 
We namely analyze the connectome of \textit{Caenorhabditis elegans} at different developmental stages, and the connectomes of different brain regions of both larval and adult \textit{Drosophila melanogaster}, in addition to the complete connectomes of \textit{Platynereis dumerelii} and larval \textit{Ciona intestinalis}.
We stress the exhaustive aspect of this diverse dataset: these constitute \textit{all} the animals for which the complete anatomical, microscale wiring diagrams have presently been mapped.
We find that all the connectomes are compressible, implying significant non-random structure. 
We find that the compressibility varies between connectomes, with larger connectomes generally being more compressible.
We infer motif sets in most connectomes, but we do not find significant evidence for motifs in several of the smaller connectomes.
The typical motifs tend to be dense subgraphs. 
We compare several topological measures of the motif sets, which show high similarity between connectomes, although with some significant differences.

\section*{Materials and methods}

In this section, we develop our methodology for compression-based inference of network motif sets.
In ``\nameref{methods:graphlets_motifs}'', we first brush up on graph theory basics.
In ``\nameref{methods:subgraph-census}'', we describe the subgraph census procedure deployed to list subgraph occurrences. 
In ``\nameref{methods:model_selection}'', we briefly review the MDL principle for model selection based on lossless compression. 
In ``\nameref{methods:algorithm}'' we develop our code, corresponding to a probabilistic model, for network motif inference using subgraph contractions. 
In particular, we model a network with a prescribed motif set as an \textit{expanded latent graph}, where expansion points designate the subset of latent nodes that embody motifs. 
In ``\nameref{methods:base-codes}'' we list the codes supporting the latent graph description, as well as codes providing purely dyadic representations. The latter serve as references that allow to quantify the significance of motif sets as compared to their respective best-fitting motif-free null model. 
In ``\nameref{methods:optimization-algorithm}'' we describe our stochastic greedy optimization algorithm for selecting motif sets.
Finally, in ``\nameref{methods:datasets}'' we present the artificial networks used for numerical validation and the neural connectomes that serve as real-world applications of our motif-based inference framework.
All code and scripts are publicly available at \href{https://gitlab.pasteur.fr/sincobe/brain-motifs}{gitlab.pasteur.fr/sincobe/brain-motifs}.

\subsection*{Graphlets and motifs}\label{methods:graphlets_motifs}

Network motif analysis is concerned with the discovery of statistically significant classes of subgraphs in empirically recorded graphs.
We here restrict ourselves to directed unweighted graphs, but the concepts apply similarly to undirected graphs and may be extended to weighted graphs~\cite{onnela_intensity_2005, picciolo_weighted_2022}, time-evolving and multilayer graphs~\cite{kovanen_temporal_2011, paranjape_motifs_2017, battiston_multilayer_2017, sallmen_graphlets_2022}, and hypergraphs~\cite{lee_hypergraph_2020, lotito_higher_2022}. 
As is usual in motif analysis, we consider weakly connected subgraphs~\cite{milo_network_2002, alon_network_2007}. 
This ensures that the subgraph may represent a functional subunit where all nodes can participate in information processing.

Let $G=(\Vset,\Eset)$ denote the directed graph we want to analyze.
For simplicity in comparing different representations of $G$, we consider $G$ to be node-labeled.
Thus, the nodes $\Vset=(1, 2, \ldots, N)$ constitute an ordered set. 
The set of edges, $\Eset \subseteq \Vset\times\Vset$ indicates how the nodes are connected. By convention, a link $(i,j)\in\Eset$ indicates that $i$ connects to $j$. 
Note that, since $G$ is directed, the presence of $(i,j)\in\Eset$ does not imply the existence of $(j,i)\in\Eset$.
We denote by $E=|\Eset|$ the number of edges.
We only consider network data that form \textit{simple} directed graphs, where $\mathcal{\Eset}$ does not contain repeated elements: this is the definition of a set.
The model we propose, however, makes use of \textit{multigraphs} where $\mathcal{E}$ is a multi-set, which may contain repetitions.

A standard representation of a graph's connectivity is its \textit{adjacency matrix}---denoted $\mathbf{A}$---with entries given by $A_{ij} = |\{(i',j') \in  \Eset : (i',j') = (i,j)\}|$.
If the graph is simple, then the adjacency matrix is boolean, i.e., $A_{ij} = 1$ if $(i,j)\in\Eset$, otherwise $A_{ij} = 0$.
When dealing with a multigraph, entries of the adjacency matrix take non-negative integer values, i.e., $A_{ij} \geq 1$ if $(i,j)\in\Eset$.

An \textit{induced subgraph} $g=(\vset,\eset)$ of $G$ is the graph formed by a given subset $\vset\in \Vset$ of the nodes of $G$ and all the edges $\eset=\{(i,j): i,j\in \vset \land (i,j)\in \Eset\}$ connecting these nodes in $G$.

An undirected graph $G_{\rm un}$ is called \textit{connected} if there exists a path between all pairs of nodes in $G_{\rm un}$.
A directed graph $G$ is \textit{weakly connected} if the undirected graph obtained by replacing all the directed edges in $G$ with undirected ones is connected.

Two graphs $g=(\vset,\eset)$ and $g'=(\vset',\eset')$ are isomorphic if there exists a permutation $\sigma$ of the node indices of $g'$, such that the edges in the graphs perfectly overlap, i.e., $(i,j)\in\eset$ if and only if $(\sigma(i),\sigma(j))\in\eset'$.
A \textit{graphlet}, denoted by $\graphlet$, is an isomorphism class of weakly connected, induced subgraphs~\cite{przulj_biological_2007}, i.e., the set $\graphlet = \{ g : g \cong \graphletInstance \}$ of all graphs that are isomorphic to a given graph, $\graphletInstance$.

Finally, a \textit{motif} is a graphlet that is statistically significant.  
Traditionally, a significant graphlet is defined as one whose number of occurrences in $G$ is significantly higher than in random graphs generated by a null model~\cite{milo_network_2002}.
Instead, we propose a method that selects a set of graphlets based on how well they allow to compress $G$. 
This lets us treat motif mining as a model selection problem through the MDL principle as we detail below.

\subsection*{Subgraph census}\label{methods:subgraph-census}

The first step of a motif inference procedure is to perform a \textit{subgraph census}, consisting in counting the graphlet occurrences.
Subgraph census is computationally hard and many methods have been developed to tackle it~\cite{ribeiro_survey_2019}.

For graphs with a small number of nodes, e.g., hundreds of nodes, we implemented the parallelized  FaSe algorithm~\cite{paredes_towards_2013}, while for larger graphs, i.e., comprising a thousand nodes or more, we rely on its stochastic version, Rand-FaSe~\cite{paredes_rand-fase_2015}. 
The algorithms use Wernicke's ESU method (or Rand-ESU for large graphs)~\cite{wernicke_faster_2005} for counting graphlet occurrences. It employs a trie data structure, termed \textit{g-trie}~\cite{ribeiro_g-tries_2010}, to store the graphlet labels in order to minimize the number of computationally costly subgraph isomorphism checks.

Since our algorithm relies on contracting individual subgraphs, we also need to store the location of each subgraph in $G$. 
Due to the large number of subgraphs, the space required to store this information may exceed working memory for larger graphs or graphlets (see discussion in~\nameref{SISec:memo-time-costs}).  
Our most computationally challenging application---inference of motifs amongst all 3- to 5-node graphlets in the right mushroom body of the adult \textit{Drosophila} connectome---requires storing 1.3 TB of data. 
In such cases, we write heavy textfiles of subgraph lists, one per graphlet, on the computer static memory, which are then retrieved individually from disk, at inference time (see~\nameref{SIsec:subgraph-census}).

All scripts were run on the HPC cluster of the Institut Pasteur, but the less computationally challenging problem of inferring 3- to 4-node motifs can be run on a local workstation (see~\nameref{SISec:memo-time-costs}).

\subsection*{Compression, model selection, and hypothesis testing}\label{methods:model_selection}

The massive number of possible graphlet combinations and the correlations between graphlet counts within a network make classic hypothesis testing-based approaches for motif mining ill-suited for discovering motif sets.
For example, there are approximately 10\,000 different five-node graphlets and exponentially more possible combinations of such graphlets, making multiplicity a critical problem for hypothesis testing.
Additionally, these approaches define motif significance by comparison with a random graph null model, and the results may depend on the choice of null model~\cite{artzy-randrup_comment_2004, beber_artefacts_2012} (see ``\nameref{sec:numerical-validation}'' in the results below). 
In the context of motif mining, this choice can lead to ambiguities~\cite{artzy-randrup_comment_2004, beber_artefacts_2012, orsini_quantifying_2015}, thus rendering the analysis unreliable.

To address these problems, we cast motif mining as a model selection problem. We wish to select as motifs the multiset of graphlets, $\subgraphSet^* = [\graphlet^*]$ that, together with a tractable dyadic graph model, provides the most adequate model for $G$.
The minimum description length principle~\cite{grunwald_minimum_2007}  states that, within an inductive inference framework with finite data, the most faithful representation of the observed system is given by the model that leads to the highest compression of the data---that is, of \textit{minimum codelength}.
It relies on an equivalence between codelengths and probabilities~\cite{cover_elements_2012} and formalizes the well-known Occam's razor, or principle of parsimony. 
It is similar to Bayesian model selection and can be seen as a generalization of it~\cite{grunwald_minimum_2020}. 

To  each dataset,  model and parameter values, we associate a unique code, {i.e.}, a label that identifies one representation. The code should be lossless, which means full reconstruction  of the data from the compressed representation is possible~\cite{grunwald_minimum_2007, cover_elements_2012}. 
In practice, we are not interested in finding an actual code, but only in calculating the codelength of an optimal code~\cite{cover_elements_2012}, corresponding to our model. 

Suppose we know the generative probability distribution of  $G$, $\model{\params}$, parameterized by $\params$. 
Then, we can encode $G$ using a code whose length is equal to the negative log-likelihood~\cite{grunwald_minimum_2007}, 
\begin{equation}\label{eq:codelength-loglikelihood}
  \codelength{\params}(G) = - \log \model{\params}(G) ,
\end{equation}
where $\log$ denotes the base-2 logarithm. 
(Note that an actual code would be between 1 to 2 bits longer than Eq.~(\ref{eq:codelength-loglikelihood}) since real codewords are integer-valued and not continuous~\cite{grunwald_minimum_2007}).
When the correct model and its parameters are unknown beforehand, we must encode both the model and the graph. 
To do this, we consider two-part codes, and, more generally, multi-part codes (see below). 
In a two-part code, we first encode the model and its parameters, using $\codelength{}(\params)$ bits, and then encode the data, $G$, conditioned on this model, using  $- \log \model{\params}(G)$ bits. 
This results in a total codelength of 
\begin{equation}\label{eq:two-part}
  \codelength{}(G,\params) = -\log \model{\params}(G) + \codelength{}(\params) .
\end{equation}
With multi-part codes, we encode a hierarchical model following the same schema, where we first encode the model, then encode latent variables conditioned on the model, and then encode the data conditioned on the latent variables and the model.

When performing model selection, we consider a predefined set of models, $\modelSet = \{ \model{\params} : \params \in \paramsSet \}$, and we look for the one that, in an information-theoretic sense, best describes $G$. 
Following the MDL principle we select the parametrization $\params^*\in\paramsSet$ that minimizes the description length, 
\begin{equation}\label{eq:MLD-principle}
\params^* = \argmin_{\params \in \paramsSet} \codelength{}(G,\params) .
\end{equation}
Note that the second term in Eq.~\eqref{eq:two-part}, $\codelength{}(\params)$, quantifies the \textit{model complexity}, which measures, in bits, the volume for storing the model parameters---this is a lossless encoding. 
Thus, one must strike a balance between model likelihood and model complexity to minimize the description length, inherently penalizing overfitting.

While we focus on model selection, we also provide the absolute compression of the optimal model as an indicator of statistical significance. 
The link between compression and statistical significance is based on the \textit{no-hypercompression inequality}~\cite{grunwald_minimum_2007}. 
It states that the probability that a given model, different from the true generating model, compresses the data more than the true model is exponentially small in the codelength difference.
Formally, given a dataset $G$ (e.g., a graph) drawn from the distribution $\model{0}$ and another  description $\model{\params}$, then
\begin{equation}
    \model{0}\left[ - \log\model{0}(G) + \log\model{\params}(G) \geq K \right] \leq 2^{-K} .
\end{equation}
By identifying $\model{0}$ with a null model and $\model{\params}$ with an alternative model, the no-hypercompression inequality thus provides an upper bound on the $p$-value, i.e., $p \leq 2^{-K}$.
Note, however, that the above relation is not guaranteed to be conservative for composite null models (such as the configuration models that we consider below)~\cite{grunwald_minimum_2020, grunwald_safe_2021}.

\subsection*{Graph compression based on subgraph contractions}\label{methods:algorithm}

\begin{figure}[!ht]
    \begin{adjustwidth}{-2.in}{0.in}
        \begin{tikzpicture}
            \node at (0,0){\includegraphics[height=0.45\textheight]{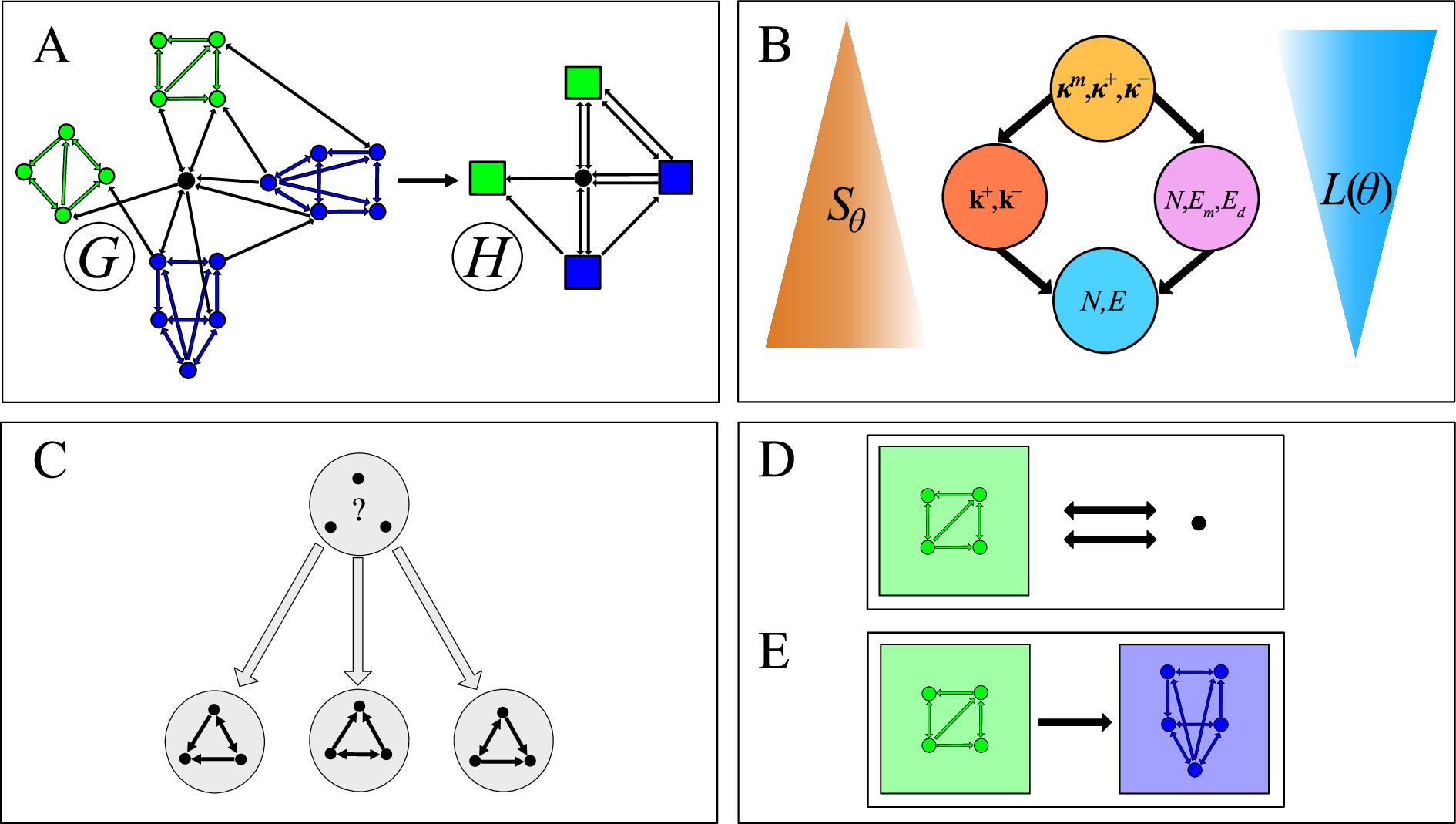}};
        \end{tikzpicture}
    \end{adjustwidth}
    
    \caption{\small
    {\bf Graphlet-based graph compression.}
    (A) Reduced representation of a graph $G$ obtained by contracting subgraphs into colored \textit{supernodes} representing the subgraphs. 
    (In this example, two different graphlets, colored in blue and green, are selected) 
    The cost for encoding the reduced representation can be split into two parts: (i) encoding the multigraph $H$  obtained by contracting subgraphs in $G$, $\baseCost$ (See ``\nameref{methods:base-codes}" section), and (ii) encoding which nodes in $H$ are supernodes and their color, designating which graphlet they represent, $\supernodeCost$ [Eq.~\eqref{eq:supernodeCost}].
    (B) Hierarchy of the four different dyadic graph models~\cite{gauvin_randomized_2022} used as base codes.
    Each node in the diagram represents a model. An edge between two nodes indicates that the upper model is less random than the lower.  The models are:
    {the Erd\H{o}s-Rényi model} $\model{\thetaER}$ (cyan); 
    {the directed configuration model} $\model{\thetaConf}$  (orange); 
 	the reciprocal Erd\H{o}s-Rényi model $\model{\thetaRER}$ (pink); 
 	and {the reciprocal configuration model} $\model{\thetaRConf}$ (yellow).
    (C-E) Encoding the additional information necessary for lossless reconstruction of $G$ from $H$, incurs a cost $\reconstructionCost$ (Eq.~\eqref{eq:reconstructionCost}) that is equal to the sum of three terms for each supernode, corresponding to encoding the labels of the nodes inside the graphlet, i.e., the graphlet's orientation (C), and how the graphlet's nodes are wired to other nodes in $H$ (D,E).  
    (C) Encoding the orientation of a graphlet is equivalent to specifying its automorphism class. For the graphlet shown in the example there are 3 possible distinguishable orientations, leading to a codelength of $\log 3$.
    (D) Encoding the connections between a simple node and a supernode involves designating to which nodes in the graphlet the in- and out-going edges to the supernode are connected.
    In this example, there are $\binom{4}{2}$ possible wiring configurations for both the in- and out-going edges, leading to a wiring cost of $\log 36$ (see Eq.~\eqref{eq:rewiringCost}).
    (E) Encoding the wiring configuration of the edges from a supernode $i$ to another supernode $j$ involves designating the edges from the group of nodes of supernode $i$ to the group of nodes in $j$ in the bipartite graph composed of the two groups (the edges from $j$ to $i$ are accounted for in the encoding of $j$). Here, there are $\binom{20}{1}$ such configurations, leading to a rewiring cost of $\log 20$ bits.}\label{fig:model}
\end{figure}

In practice, we compress the input graph by iteratively performing subgraph contractions each chosen from a set of possible graphlets, extending the approach of Bloem and de Rooij~\cite{bloem_large-scale_2020} which focused on a single graphlet.  
The model describes $G$ by a reduced representation, a multigraph $H$, with $N(H) < N(G)$ and $E(H) < E(G)$, in which a subset $\supernodes\subseteq\Vset(H)$ of nodes are marked as \textit{supernodes}, each formed by contracting  a subgraph of $G$ into a single node (Fig~\ref{fig:model}A).

We let $\graphletSet$ designate a predefined set of graphlets, which is the set of all graphlets we are interested in.
In the following, we will generally consider all graphlets from three to five nodes---in which case $\setSize{\graphletSet}=9579$---but any predefined set of graphlets, or even a single graphlet, may be used.
We define $\subgraphSet=[\graphlet]$ as a multiset of graphlets, corresponding to the subgraphs in $G$ that we contracted to obtain $H$.
We define $\graphletSubset = \{\graphlet\}$ as the set containing the unique elements of $\subgraphSet$ and let $\ngraphlet = \setSize{[\beta\in\subgraphSet: \beta=\alpha]}$ be the number of repetitions of $\graphlet$ in $\subgraphSet$.
We finally let $\model{\baseParams}$ designate a dyadic random graph model, which is used to encode $H$. 
We consider four possible such \textit{base} models (see Fig~\ref{fig:model}B and ``\nameref{methods:base-codes}" below).

The full set of parameters and latent variables of our model is $\params = \{H,\baseParams, \subgraphSet, \supernodes,\graphletSet\}$, and its codelength can be decomposed into four terms,
\begin{equation}\label{eq:total-codelength}
    \codelength{}(G, \params) 
    = \graphletCost
    + \baseCost 
    + \supernodeCost
    + \reconstructionCost
\end{equation}
where 
(i) $\graphletCost$ is the codelength for encoding the motif set;
(ii) $\baseCost$ is the codelength needed to encode the reduced multigraph $H$ using a base code corresponding to $\model{\baseParams}$; 
(iii) $\supernodeCost$ accounts for encoding which nodes of $H$ are supernodes and to which graphlets they correspond (i.e., their colors, Fig~\ref{fig:model}A);
(iv) $\reconstructionCost$ corresponds to the information needed to reconstruct $G$ from $H$ (node labels, the orientations of the contracted subgraphs, and how the subgraph's nodes are wired to their respective external neighborhoods, see Fig~\ref{fig:model}C-\ref{fig:model}D). 
We detail each of the four terms in turn. 

The first term in Eq.~\eqref{eq:total-codelength}, $\graphletCost$ is given by
\begin{equation}\label{eq:graphlets-codelength}
    \graphletCost
    = \sum_{\graphlet\in\graphletSubset} \log \setSize{\graphletSet}
    + \integerCodelength{|\graphletSet|} 
    + \sum_{\graphlet\in\graphletSubset} \log \nmax  
    + \integerCodelength{\nmax},
\end{equation}
where $\nmax = \max_{\graphlet\in\graphletSubset}\ngraphlet$ is the maximal number of repetitions of any of the graphlets in $\graphletSubset$, and $\integerCodelength{n} = \log[n(n+1)]$ is the codelength needed to encode an integer~\cite{grunwald_minimum_2007}. 
The first term in Eq.~\eqref{eq:graphlets-codelength} is the codelength needed to encode the identity of each inferred motif. Since there are $\setSize{\graphletSet}$ possible graphlets, this requires $\log\setSize{\graphletSet}$ bits per motif. 
The second term is the cost of encoding the number $\setSize{\graphletSet}$. 
The third term is the cost of encoding the number of times each of the motifs appears, requiring $\log\nmax$ bits per motif. 
The fourth term is the cost of encoding $\nmax$.

The second term in Eq.~\eqref{eq:total-codelength}, $\baseCost$, depends on the base model used to encode $H$. We consider several models and detail their codelengths in the ``\nameref{methods:base-codes}'' section below.

The third term of Eq.~\eqref{eq:total-codelength} is equal to
\begin{equation}\label{eq:supernodeCost}
  \supernodeCost 
  =  \log \binom{N(H)}{\setSize{\subgraphSet}} + \log\frac{\setSize{\subgraphSet}!}{\prod_\graphlet \ngraphlet!},
\end{equation}
where the first part corresponds to the cost of labeling $\setSize{\subgraphSet}$ nodes of $H$ as supernodes---equal to the logarithm of the number of ways to distribute the labels---and the second part corresponds to the labeling of the supernodes to show which graphlet they each correspond to---equal to the logarithm of the number of distinguishable ways to order $\subgraphSet$. 

The fourth and last term in Eq.~\eqref{eq:total-codelength} is given by
\begin{equation}\label{eq:reconstructionCost}
    \reconstructionCost 
    = \log\frac{N(G)!}{N(H)!}
    + \sum_\graphlet \ngraphlet \log\frac{\graphletSize{\graphlet}!}{|\text{Aut}(\graphlet)|}
    + \sum_{i_s\in\supernodes} \rewiringCost .
\end{equation}
Here, the first term is the cost of recovering the  original node labeling of $G$ from $H$. 
The second term encodes the orientation of each graphlet to recover the subgraphs found in $G$ (Fig~\ref{fig:model}C)---for a given graphlet $\graphlet$ (consisting of $\graphletSize{\graphlet}$ nodes) there are $\graphletSize{\graphlet}!/\setSize{\Aut(\graphlet)}$ distinguishable orientations, where $\setSize{\Aut(\graphlet)}$ denotes the size of the automorphism group of $\graphlet$.
The third term is the \textit{rewiring cost} which accounts for encoding how edges in $H$ involving a supernode are connected to the nodes of the corresponding graphlet.
Denoting by $n_{s}$ the number of nodes in the subgraph $s$ that the supernode $i_s$ replaces, the rewiring cost for one supernode is given by
\begin{equation}\label{eq:rewiringCost}
    \rewiringCost
    =  \sum_{j\in\Vset(H)\backslash\supernodes } \log \binom{\graphletSize{s}}{A_{i_sj}} \binom{\graphletSize{s}}{A_{ji_s}}
    + \sum_{j_{s'}\in\supernodes} \log \binom{\graphletSize{s}\graphletSize{{s'}}}{A_{i_sj_{s'}}} ,
\end{equation}
where the first term is the cost for designating which of the possible wiring configurations involving the nodes inside a supernode and adjacent regular nodes corresponds to the configuration found in $G$ (Fig~\ref{fig:model}D), and the second term is the cost of encoding the wiring configurations for edges from the nodes of the given supernode to the nodes of its adjacent supernodes (Fig~\ref{fig:model}E).   

\subsection*{Base codes and null models}\label{methods:base-codes}
\subsubsection*{The latent graph code}

To encode the latent reduced graph $H$, we use two-part codes of the form  $\codelength{}(H,\baseParams) = -\log \model{\baseParams}(H) + \codelength{}(\baseParams)$ (Eq.~\eqref{eq:two-part}), where $\codelength{}(\baseParams)$ encodes the parameters of the chosen dyadic random graph model---the model's \textit{parametric codelength}---and $\model{\baseParams}(H)$ is a uniform probability distribution over a multigraph ensemble conditioned on the value of $\baseParams$. 
Note that, while $G$ is a simple graph, the subgraph contractions may generate multiple edges between the same nodes in $H$, which consequently is a multigraph.
The models $\model{\baseParams}$ correspond to maximum entropy microcanonical graph ensembles~\cite{gauvin_randomized_2022, jaynes_information_1957, presse_principles_2013}, i.e., uniform distributions over graphs with certain structural properties  $\baseParams(H)$, e.g., the node degrees, set to match exactly a given value, $\baseParams(H) = \baseParams^*$.
The microcanonical distribution is given by
\begin{equation}\label{eq:microcanonical_likelihood}
  \model{\baseParams}(H) = 
    \begin{cases}
      \frac{1}{\partitionFunction{\baseParams}} & \mathrm{for}\ \baseParams(H) = \baseParams^*, \\
      0 & \mathrm{elsewise,} 
    \end{cases}
\end{equation} 
where the normalizing constant $\partitionFunction{\baseParams} = |\{H : \baseParams(H) = \baseParams^*  \}|$ is known as the microcanonical partition function.
The codelength for encoding $H$ using the model $\model{\baseParams}$ can be identified with the microcanonical entropy,
\begin{equation}
    -\log\model{\baseParams}(H) = \log\partitionFunction{\baseParams} \equiv \entropy{\baseParams} ,
\end{equation}
leading to a total codelength for encoding the model and the reduced graph of
\begin{equation}\label{eq:baseCodelength}
\codelength{}(H,\baseParams) = \entropy{\baseParams(H)} + \codelength{}(\baseParams(H)) .
\end{equation}

As base codes  we consider four different paradigmatic random graph models, namely the Erd\H{o}s-R\'{e}nyi (ER) model, the configuration model (CM), and their reciprocal versions (RER and RCM, respectively).
For the ER model, the parameters are the number of nodes and edges, while the configuration model constrains the nodes' in- and out-degrees and their reciprocal versions additionally constrain the number of reciprocated edges. 
Both the degree distributions and the edge reciprocity have been found  to be significantly non-random in biological networks, and they have been shown to influence the networks' topology and function~\cite{barabasi_emergence_1999, cohen2000resilience, pastor2001epidemic, seyed2006scale, fosdick_configuring_2018, jovanic_competitive_2016, winding_connectome_2023, gilbert2013top, bahl2020neural, jarrell2012connectome}.
Thus, it is natural to include these features in the base models, and the corresponding models have been widely employed as null models for hypothesis testing-based motif inference~\cite{milo_network_2002, milo_superfamilies_2004, yeger-lotem_network_2004, sporns_motifs_2004, alon_network_2007, fornito_fundamentals_2016, stone_network_2019, alon_introduction_2019}.

Microcanonical models are defined by the features of a graph that they keep fixed~\cite{gauvin_randomized_2022} (see Eq.~(\ref{eq:microcanonical_likelihood})).
We list them for each of the four models below and we give in Table~\ref{tab:codelengths} expressions for their entropy $\entropy{\baseParams}$ and their parametric codelength $\codelength{}(\baseParams)$ (see Section~A in~\nameref{SIsec:codelengths} for details).
\begin{itemize}
    \item \textbf{The Erd\H{o}s-R\'{e}nyi model (ER)} fixes the number of nodes and edges, $\baseParams = \thetaER$. 
    
    \item \textbf{The configuration model (CM)} fixes the nodes' in- and out-degrees (the number of incoming and outgoing edges), $\baseParams = \thetaConf$.
    
    \item \textbf{The reciprocal Erd\H{o}s-R\'{e}nyi model (RER)} fixes the number of nodes, the number of reciprocal edges, and the number of non-reciprocated edges, $\baseParams = \thetaRER$.
    Formally, for a simple graph, a (non-)reciprocated edge is conveyed by (a)symmetric entries of the adjacency matrix.
    That is, an edge $(i,j)$ is reciprocal if $A_{ij} = 1$ and $A_{ji} = 1$ and non-reciprocated if $A_{ij} = 1$ and $A_{ji} = 0$.
    The definition for multigraphs extends this idea to integer counts by defining the reciprocal part of a multiedge as the minimum of $A_{ij}$ and $A_{ji}$ and the non-reciprocated part as the rest~\cite{squartini2013reciprocity} (details can be found in Section~A in~\nameref{SIsec:codelengths}).
    
    \item \textbf{The reciprocal configuration model (RCM)} fixes the nodes' reciprocal degrees---the number of reciprocal edges each node partakes in---as well as the non-reciprocated in- and out-degrees, $\baseParams = \thetaRConf$.
\end{itemize}
The different base models respect a partial order in terms of how random they are, i.e., how large their entropy is (Fig~\ref{fig:model}B)~\cite{gauvin_randomized_2022}.
We stress that the model with the smallest entropy does not necessarily provide the shortest description of a graph $H$ due to its higher model complexity (see Section~A in~\nameref{SIsec:codelengths}).

\begin{table}
    \centering
    \caption{\small {\bf Base- and null-model codelenghts.} 
    The codelength of a model is equal to $L(H,\baseParams) = \entropy{\baseParams} + L(\baseParams)$ (Eq.~(\ref{eq:baseCodelength})), with the entropy $\entropy{\baseParams}$ and the model complexity $\codelength{}(\baseParams)$ given by the appropriate expressions in the table.
    The entropy of multigraph models are given in the first four lines and the entropy of the simple graph models are given in the next four. 
    The parametric complexity of the models is the same for multi- and simple graphs and are listed in the following four lines. 
    Finally, expressions for common parametric codelengths are given in the last four lines.
    For multigraph codes, the asymmetric and symmetric parts of the adjency matrix are denoted by $A_{ij}^\asym = \max(A_{ij}-A_{ji},0)$ and $A_{ij}^\sym  = \min(A_{ij},A_{ji})$, respectively.
    For reciprocal models (RER and RCM), $\Ednumber = \sum_{i,j} A_{ij}^\asym$ is the number of non-reciprocated edges and $\Emnumber = \sum_{i<j} A_{ij}^\sym$ is the number of reciprocated edges.
    For the configuration model (CM), $\outdegree_i = \sum_j A_{ij}$ denotes the out-degrees and $\indegree_i = \sum_j A_{ji}$ the in-degrees. 
    For the reciprocal CM (RCM), $\outdegreeR_i = \sum_j A_{ij}^\asym$ and $\indegreeR_i = \sum_j A_{ji}^\asym$ are the non-reciprocated out- and in-degrees, and  $\mutualdegreeR_i = \sum_j A_{ij}^\sym$ are the reciprocal degrees. 
    (Details can be found in~\nameref{SIsec:codelengths}.)}
    \label{tab:codelengths}

    \begin{adjustwidth}{-1.8in}{0.in}
    \renewcommand\arraystretch{2}
    \begin{tabular}{|l|l|}
        \thickhline
        \multicolumn{1}{|l|}{\bf Model} & \multicolumn{1}{|l|}{\bf Multigraph entropy $\entropy{\baseParams}$}  \\\hline
        \thickhline
        ER  & $E\log[N(N-1)] - \log\frac{E!}{\prod_{i \neq j}A_{ij}!}$ 
        \\\hline
         RER & $(\Emnumber + \Ednumber)\log[N(N-1)]  
        -\log\frac{(2\Emnumber)!!\Ednumber!}{\prod_{i<j} A_{ij}^\sym! A_{ij}^\asym! A_{ji}^\asym!}$ 
        \\\hline
        CM  & $\log \frac{E!}{\prod_i\outdegree_i!\indegree_i!}  - \sum_{i \neq j} \log A_{ij}!$  
        \\\hline
         RCM & $\log\frac{ (2E-1)!!}{\prod_i\mutualdegreeR_i!\outdegreeR_i!\indegreeR_i!} 
                                   - \sum_{i < j} \log A_{ij}^\sym!A_{ij}^\asym!A_{ji}^\asym!$
        \\        
        \thickhline
          &  \multicolumn{1}{|l|}{\bf Simple graph entropy $\entropy{\baseParams}$}   \\\thickhline
        \hline 
        ER      & $\log \binom{N(N-1)}{E}$ 
        \\\hline 
        RER     & $\log \frac{[N(N-1)/2]!}{[N(N-1)/2 - \Emnumber - \Ednumber]!  \Emnumber! \Ednumber!} + \Ednumber$ 
        \\\hline
        CM      & $\log \frac{E!}{\prod_i k_i^+!k_i^-!} - 
        \frac{1}{2\ln 2}\frac{\langle {k^+_i}^2 \rangle \langle {k^-_i}^2 \rangle}{\langle k^+_i \rangle \langle k^-_i \rangle}$ 
        \\\hline
        RCM & $\log \frac{(2\Emnumber)!!}{\prod_i \mutualdegreeR_i!} + \log \frac{\Ednumber!}{\prod_i\outdegreeR_i!\indegreeR_i!} - \frac{1}{2\ln 2} \left( 
        \frac{1}{2}\frac{\E {\pare{\mutualdegreeR_i}^2}^2}{\E{\mutualdegreeR_i}^2} 
        + \frac{\E{({\outdegreeR_i})^2} \E{({\indegreeR_i})^2}}{\E{\outdegreeR_i} \E{\indegreeR_i}} 
        + \frac{\E{\outdegreeR_i\indegreeR_i}^2}{\E{\outdegreeR_i}\E{\indegreeR_i}} 
        + \frac{\E{\mutualdegreeR_i\outdegreeR_i}\E{\mutualdegreeR_i\indegreeR_i}}{\E{\mutualdegreeR_i}\E{ \kappa_i^+}} \right) $\rule{0pt}{0.cm} 
        \\
        \thickhline
       &  \multicolumn{1}{|l|}{\bf Model complexity $\codelength{}(\baseParams)$}   \\\thickhline
        ER  & $\integerCodelength{N} + \integerCodelength{E}$ \\\hline
        RER & $\integerCodelength{N} + \integerCodelength{\Emnumber} + \integerCodelength{\Ednumber}$ \\\hline
        CM  & $\seqCodelength{\outdegrees} + \seqCodelength{\indegrees}$ \\\hline
        RCM & $\seqCodelength{\mutualdegreesR} + \seqCodelength{\outdegreesR} + \seqCodelength{\indegreesR}$ \\
        \thickhline
        & \multicolumn{1}{|l|}{\bf Parametric codelengths} \\\thickhline
        Integer  & $\integerCodelength{n} = n(n+1)$ \\\hline
        Sequence  & $\seqCodelength{\mathbf{x}} = \min\{ L_U(\mathbf{x}),L_{\lambda=1}(\mathbf{x}),L_{\lambda=1/2}(\mathbf{x}) \} + \log 3 + \integerCodelength{n}$, with $n = |\mathbf{x}|$ \\
        \hline 
        Uniform   & $L_U(\mathbf{x}) = n\log(\Delta-\delta+1) + \integerCodelength{\Delta} + \integerCodelength{\delta}$, with $n = |\mathbf{x}|$, $\Delta = \max(\mathbf{x})$ and $\delta = \min(\mathbf{x})$ \\
        \hline
        Dirichlet-&  $L_\lambda(\mathbf{x})=  -\log\frac{\Gamma(\Lambda)}{\Gamma(n+\Lambda)} + \frac{\Lambda}{\lambda}\log\Gamma(\lambda) -  \sum_{\substack{\delta\leq \mu \leq \Delta \\ \mu\in \mathbb{N}}} \log\Gamma\left[\lambda + \sum_{i=1}^n \delta(x_i,\mu)\right]$, \vspace{-6pt}\\
        multinomial &  with $n = |\mathbf{x}|$, $\Delta = \max(\mathbf{x})$, $\delta = \min(\mathbf{x})$, and $\Lambda = (\Delta - \delta + 1)\lambda$ \\
        \hline
    \end{tabular}
    \end{adjustwidth}
\end{table}

\subsubsection*{Motif-free reference codes}\label{methods:null_models}

To assess the significance of inferred motif sets, we compare the motif-based graph codes to their purely dyadic counterparts.
In Table~\ref{tab:codelengths}, we also list expressions for the entropy of dyadic simple graph codes for the ER, CM, RER, and RCM models (see Section~A in~\nameref{SIsec:codelengths} for details and Section~B in~\nameref{SIsec:codelengths} for a derivation of the entropy of the simple graph RCM).
The parametric complexity of the simple graph models are identical to the ones of the multigraph base models.
Including these purely dyadic codes in the set of possible models $\modelSet$ ensures that our motif inference is conservative and does not find spurious motifs in random networks (see ``\nameref{sec:numerical-validation}" in Results below).

\subsection*{Optimization algorithm}\label{methods:optimization-algorithm}

\begin{figure}[ht!]
    \begin{adjustwidth}{-2.25in}{0.in}
    \begin{tikzpicture}
        \node at (0,0){\includegraphics[height=0.5\textheight]{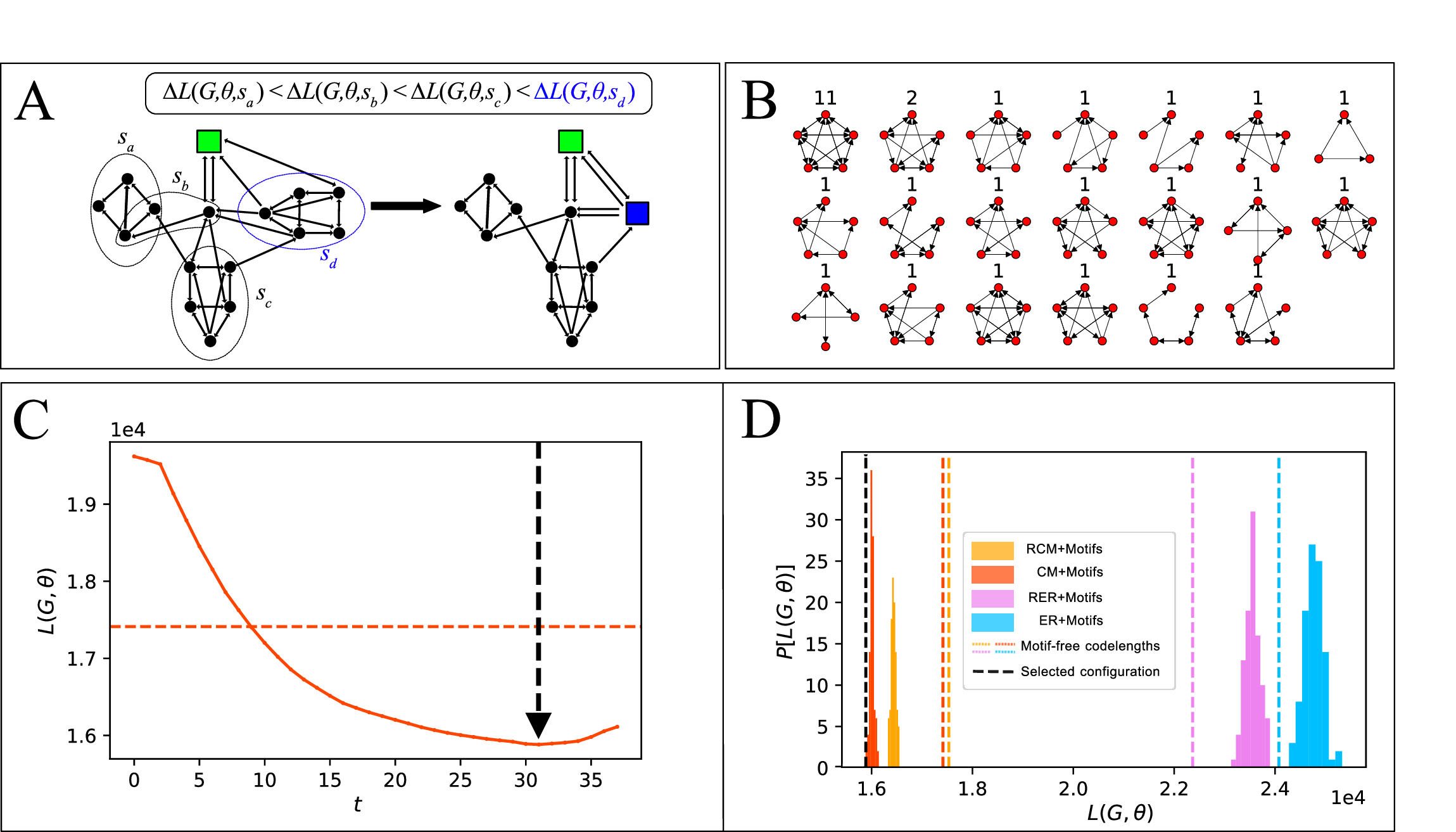}};
    \end{tikzpicture}
    \end{adjustwidth}
    \caption{\small
    {\bf Greedy optimization algorithm.}
    (A) Illustration of a single step of the greedy stochastic algorithm.
    The putative compression $\compressibility{}(G,\params,\subgraph)$ that would be obtained by contracting each of the subgraphs in the minibatch is calculated, and the subgraph contraction resulting in the highest compression is selected (highlighted in blue).
    (B) Example of motif set inferred in the connectome of the right hemisphere of the mushroom bodies (MB right) of the \textit{Drosophila} larva.
    (C) Evolution of the codelength during a single algorithm run. 
    The algorithm is continued until no more subgraphs can be contracted. 
    The representation $\params^* = \params_t$ with the shortest codelength is selected; here, after the 31st iteration (indicated by a vertical black dashed line). The horizontal orange dashed line indicates the codelength of the corresponding simple graph model without motifs (see \nameref{methods:null_models}).
    (D) The algorithm is run a hundred times for each dyadic base model and the most compressing model $\hat{\params}$ is selected.
    Histograms represent the codelengths of models with motifs after each run of the greedy algorithm;
    colors correspond to the different base models (blue: ER model, orange: configuration model, pink: reciprocal ER model, yellow: reciprocal configuration model, see Fig~\ref{fig:model}B and Table~\ref{tab:codelengths}); 
    vertical dashed lines represent the codelengths of models without motifs, and the black dashed line indicates the codelength of the shortest-codelength model---here the configuration model with motifs.}
\label{fig:greedy_algorithm}
\end{figure}

To infer a motif set, we apply a greedy iterative algorithm that contracts the most compressing subgraph in each iteration. Since the number of $\vnumber$-node subgraphs grows super-exponentially in $\vnumber$, it is not convenient to consider all subgraphs at once. Thus, we developed a stochastic algorithm that randomly samples a mini-batch of subgraphs in each iteration and contracts the one that compresses the most among these (Fig~\ref{fig:greedy_algorithm}).
We give in Algorithms~\ref{algo:greedy_compression}--\ref{algo:subgraph_contraction} pseudocode for its implementation and describe below each of the main steps involved.

\begin{algorithm}
    \caption{Greedy motif inference}
    \label{algo:greedy_compression}
    \begin{algorithmic}[1]
    \Require Graph $G$, graphlet set $\Gamma$, base model $\model{\baseParams}$, subgraph minibatch size $B$\vspace{.2cm}

    \State $t \gets 0$
    \State $H_0\gets G$
    \State $\subgraphSet_0,\supernodes_0 \gets \emptyset,\emptyset$
    \State $\params_0\gets (H_0,\phi(H_0),S_0,\supernodes_0,\Gamma)$
    \State $\Theta \gets \{\params_0\}$
    \State $\mathcal{C} \gets \Call{SubgraphCensus}{G,\graphletSet}$
    
    \While{$\mathcal{C}$ is not $\emptyset$}

        \State $t \gets t + 1$
        \State $\subgraphOccurrences,\minibatch{} \gets \Call{SubgraphBatches}{B,\graphletSet,\subgraphOccurrences}$ 
        \State $\graphlet,\subgraph_{\graphlet} \gets \Call{MostCompressingSubgraph}{G,\minibatch{},\theta_{t-1}}$ 
        \State $H_{t},\subgraphSet_t,\supernodes_t\gets\Call{SubgraphContraction}{H_{t-1},\mathcal{V}_{t-1},\mathcal{S}_{t-1},\graphlet,s_{\graphlet}}$ 
        \State $\params_t \gets (H_t,\baseParams(H_t),\subgraphSet_t,\supernodes_t,\Gamma)$
        \State $\Theta\gets \Theta\cup \{\theta_t\}$
    
    \EndWhile
    \Ensure $\argmin_{\theta\in\Theta}\{L(G,\theta)\}$ 
    \end{algorithmic}
\end{algorithm}

\begin{algorithm}
    \caption{Sample subgraph batches}\label{algo:subgraph_batches}
    \begin{algorithmic}[1]
    \Function{SubgraphBatches}{$B,\Gamma,\mathcal{C}$}
        \State $\mathcal{B} \gets \emptyset$
        \For{$\alpha \in \Gamma$}
            \State $\mathcal{B}_{\alpha} \gets \emptyset$
            \While{$|\mathcal{B}_{\alpha}| < B$ and $|S_{\alpha}| > 0$}
                \State $s_{\alpha} \gets \Call{SampleGraphletInstance}{\mathcal{C}_{\alpha}}$ 
                \If{\Call{NonOverlappingSubgraph}{$H,s_{\alpha}$}}  \
                    \State $\mathcal{B}_{\alpha} \gets \mathcal{B}_{\alpha}\cup \{s_{\alpha}\}$
                \Else{ $\mathcal{C}_\alpha\gets \mathcal{C}_\alpha  \backslash \{s_\alpha\}$}
                \EndIf
            \EndWhile
            \State $\mathcal{B} \gets \mathcal{B}\cup\mathcal{B}_{\alpha}$
            \State $\mathcal{C}_\alpha \gets \mathcal{C}_\alpha\backslash\mathcal{B}_\alpha$
        \EndFor
        \State\Return $\mathcal{C},\mathcal{B}$ 
    \EndFunction
    \end{algorithmic}
    
    \vspace{4pt}
    
    \begin{algorithmic}[1]
        \Function{SampleGraphletInstance}{$\mathcal{C}_\alpha$}
            \State\Return $s_\alpha$, a subgraph sampled uniformly from  $\mathcal{C}_\alpha$ 
        \EndFunction
    \end{algorithmic}
    
    \vspace{4pt}
        
    \begin{algorithmic}[1]
    \Function{NonOverlappingSubgraph}{$H,s_\alpha$}
        \State $b \gets \boldsymbol{\mathrm{True}}$
        \For{$i \in s_\alpha$}
            \If{$i \notin \Vset(H)$}
                \State $b \gets \boldsymbol{\mathrm{False}}$ 
                \Comment{ Delete node labels of already contracted subgraphs }
            \EndIf
        \EndFor
        \State\Return $b$
    \EndFunction
    \end{algorithmic}
\end{algorithm}

\begin{algorithm}          
    \caption{\label{algo:most_compressing_subgraph}Find most compressing subgraph.}
    \begin{algorithmic}[1]
    \Function{MostCompressingSubgraph}{$G$,$\mathcal{B},\theta$}
        \State $ s^* \gets \mathrm{argmax}_{s \in \mathcal{B}}\{ \Delta L (G,\params,s)\}$ 
        \Comment see Section~C in~\nameref{SIsec:codelengths}.
        \State Let $\alpha\in\Gamma$ be the graphlet such that $g_\alpha \cong s^*$. 
        \State \Return $\alpha,s^*$
    \EndFunction
    \end{algorithmic} 
\end{algorithm}

\begin{algorithm}
    \caption{\label{algo:subgraph_contraction}Subgraph contraction}
    \begin{algorithmic}[1]
    \Function{SubgraphContraction}{$H,\mathcal{V},\mathcal{S},\graphlet,\subgraph_\graphlet$}
        
        \For {$(i,j) \in \mathcal{E}(\subgraph_\alpha)$}
            \State $A_{ij}(H)\gets 0$
        \EndFor
        \State $\Vset(H)\gets \Vset(H)\backslash \subgraph_\graphlet$
        \State Let $i_{\alpha}$ be the label of a new supernode
        \State $\Vset(H)\gets \Vset(H)\cup \{i_{\alpha}\}$
        \State $\supernodes \gets \supernodes \cup \{i_\alpha\}$ 
        \State $\subgraphSet \gets \subgraphSet \cup \{\alpha\}$
        \For {$l \in \partial s_\alpha$}
            \State $A_{i_{\alpha}l}(H) \gets 0$
            \For {$i \in \Vset(s_\alpha)$}
                \State $A_{i_{\alpha}l}(H) \gets A_{i_{\alpha}l}(H) + A_{il}(H)$
            \EndFor
        \EndFor
        \State \Return $H,\mathcal{V},\mathcal{S}$
    \EndFunction
    \end{algorithmic}
\end{algorithm}

\paragraph*{Subgraph census.} (\SubgraphCensus{} in Algorithm~\ref{algo:greedy_compression}).
We first perform a subgraph census to provide a set of lists of the graphlet occurrences in $G$, $\subgraphOccurrences = \{\subgraphOccurrences_\graphlet : \graphlet\in\graphletSet\}$ with $\subgraphOccurrences_\graphlet = \{g \equiv G[\nu] : \nu \subseteq \Vset \land g \simeq \alpha \}$  (see the ``\nameref{methods:subgraph-census}'' section above).
We consider in the ``Results" section below $\graphletSet$ to be \textit{all} graphlets of three, four, and five nodes, but any predefined set of graphlets may be specified in the algorithm.
\vspace{12pt} 

Once the subgraph census is completed, we perform stochastic greedy optimization by iterating the following steps. 

\paragraph*{Subgraph sampling} (\SubgraphBatches{}, Algorithm~\ref{algo:subgraph_batches}). 
In each step, the algorithm samples a minibatch of subgraphs, $\minibatch{t}$, consisting of $B$ subgraphs per graphlet selected uniformly from $\subgraphOccurrences$. 
The \SubgraphBatches{} function also discards subgraphs in $\subgraphOccurrences$ that overlap with already contracted subgraphs. Indeed, from a biological point of view, overlapping supernodes correspond to nested circuit motifs, whose significance differs from the standard circuit motifs, where each node is identified with a single unit (e.g., a neuron). 
Furthermore, this constraint guarantees a faster algorithmic convergence by progressively  excluding many subgraphs candidates. 
The number of subgraphs per graphlet, $B$, is a hyperparameter of the algorithm. 
We tested different values of $B$ and found similar results for values in the range $10$--$100$ (see~\nameref{SIfig:minibatch_size_effect} Fig). 

The check of overlap is performed by a boolean sub-function \Call{NonOverlappingSubgraph}{} (see Algorithm~\ref{algo:subgraph_batches}). It asserts whether a node of a subgraph $\subgraph$ is already part of a supernode of $H_{t-1}$. 

\paragraph*{Finding the most compressing subgraph.} (\BestCompressingSubgraph{}, Algorithm~\ref{algo:most_compressing_subgraph}).
We calculate for each subgraph $\subgraph\in\minibatch{t}$ how much it would allow to further compress $G$ compared to the representation of the previous iteration, i.e., the codelength difference $\Delta \codelength{}(G,\theta_t,s) = \codelength{}(G,\params_t) - \codelength{}(G, \tilde{\params}_t(s)) $, 
where $\tilde{\params}_t(s)$ represents the putative parameter set after contraction of $s$ (see Section~C in~\nameref{SIsec:codelengths} for expressions of codelength differences).
The subgraph $s^*$ for which $\Delta\codelength{}$ is maximal is selected for contraction. 

\paragraph{Subgraph contraction.} (\SubgraphContraction{}, Algorithm~\ref{algo:subgraph_contraction}). 
The reduced graph $H_t$ is obtained by contraction of the subgraph $s^*\equiv\subgraph_\graphlet$ (isomorphic to the graphlet $\graphlet$) in $H_{t-1}$. 
The subgraph contraction consists of deleting in $H_{t-1}$ the regular nodes and simple edges of $s_\alpha$, and replacing them with a supernode $i_\alpha$ that connects to the union of the neighborhoods of the nodes of $s_\alpha$, denoted $\partial s_\graphlet$, through multiedges.
We refer to $\partial s_\graphlet$ as the subgraph's neighborhood, which, by design, is identical to the supernode's neighborhood.
Nodes of $s_{\alpha}$ that share neighbors will result in the formation of parallel edges,  affecting the adjacency matrix according to $A_{i_{\alpha}j} = \sum_{i \in s_{\alpha}} A_{ij}$.

\paragraph*{Stopping condition and selection of most compressed representation.}
At each iteration $t$, the algorithm generates a compressed version of $G$, parametrized by $\params_t$. 
We run the algorithm until no more subgraphs can be contracted, i.e., until there are no more subgraphs that are isomorphic to a graphlet in $\graphletSet$ and do not involve a supernode in $H_t$.
We then select the representation that achieves the minimum codelength among them (Fig~\ref{fig:greedy_algorithm}C),
\begin{equation}
    \params^* = \argmin \{\codelength{}(G,\params_t)\} .
\end{equation}

\paragraph*{Repeated inferences for each base code.}
Since different base models lead to different inferred motif sets (see~\nameref{SIfig:base-model-diff} Fig), we run the optimization algorithm independently for each base model, and since the algorithm is stochastic, we run it 100 times per connectome and base model to gauge its variability and check that the inference is reasonable (Fig~\ref{fig:greedy_algorithm}D).
We select the model $\hat{\params}$ with the shortest codelength among all these, and its corresponding motif set if the best model is one with motifs,
\begin{equation}
    \hat{\params} = \argmin\{L(G,\theta^*)\} .
\end{equation}
%

\subsection*{Datasets}\label{methods:datasets}

\subsubsection*{Artificial datasets\label{datasets:artificial}} 

\paragraph*{Randomized networks.}\label{methods:randomized-networks}
To quantify the propensity of our approach and of hypothesis testing-based methods to infer spurious motifs (i.e., false positives), we apply them to random networks without motifs. 
To generate random networks corresponding to the different null models, 
we apply the same Markov-chain edge swapping procedures~\cite{fosdick_configuring_2018} used for hypothesis-testing based motif inference (see more details in~\nameref{SIsec:hypothesis-testing}).

\paragraph*{Planted motif model.}\label{methods:planted-motif-model}
To test the ability of our method to detect motifs that genuinely are present in a network (i.e., true positives), we generated random networks according to a \textit{planted motif model} which generates networks with placed motifs by inverting our compression algorithm according to the following steps: 
(i) generate a random latent multigraph $H$ according to the ER model;
(ii) designate at random a predetermined number of the nodes as supernodes;
(iii) expand the supernodes by replacing them with the motif of choice, oriented at random and with its nodes wired at random to the supernode's neighbors in $H$.

\subsubsection*{Empirical datasets}
We apply our method to infer microcircuit motifs in synapse-resolution neural connectomes of different small animals obtained from serial electron microscopy (SEM) imaging (see Table~\ref{tab:datasets} for descriptions and references of the datasets). 
All input raw and processed connectomes can be found in our GitLab project, in the \texttt{data} folder(\href{https://gitlab.pasteur.fr/sincobe/brain-motifs/-/tree/master/data}{gitlab.pasteur.fr/sincobe/brain-motifs/-/tree/master/data}).

\begin{table}[!ht]
\caption{\small
{\bf Connectome datasets.}\label{tab:datasets}
For each connectome, we list its number of non-isolated nodes, $N$, its number of directed edges, $E$, its density $\rho = E/[N(N-1)]$, the features of the most compressing model for the connectome, its compressibility $\compressibility{}^*$, the difference in codelengths between the best models with and without motifs, $\compressibility{\textrm{motifs}}$, and the reference to the original publication of the dataset. 
The absolute compressibility $\compressibility{}^*$ measures the number of bits that the shortest-codelength model compresses compared to a simple Erd\H{o}s-Rényi model (Eq.~\eqref{eq:compression}). 
The difference in compression with and without motifs, $\compressibility{\textrm{motifs}}$, quantifies the significance of the inferred motif sets as the number of bits gained by the motif-based encoding compared to the optimal motif-free, dyadic model. 
For datasets where no motifs are found, this column is marked as ``N/A''.
All datasets are available at \href{https://gitlab.pasteur.fr/sincobe/brain-motifs/-/tree/master/data}{https://gitlab.pasteur.fr/sincobe/brain-motifs/-/tree/master/data}}
\begin{adjustwidth}{-2.25in}{0in}
\centering
\begin{tabular}{|l|l|l|l|l|l|l|l|l|}
\hline
\multicolumn{1}{|l|}{Species} & \multicolumn{1}{|l|}{Connectome} & \multicolumn{1}{|l|}{ $N$} & \multicolumn{1}{|l|}{$E$} & \multicolumn{1}{|l|}{$\rho$} &
\multicolumn{1}{|l|}{ Best model  } & 
\multicolumn{1}{|l|}{$\compressibility{}^*$} & 
\multicolumn{1}{|l|}{$\compressibility{\textrm{motifs}}$} &
\multicolumn{1}{|l|}{Ref.}\\ 
\thickhline
\textit{C. elegans} & Head Ganglia---Hour 0 & 187 & 856 & 0.025 & RCM   & 354  & N/A &\cite{witvliet_connectomes_2021}\\ \hline
\textit{C. elegans} & Head Ganglia---Hour 5 & 194 & 1108 & 0.030 & RCM  & 494  & N/A &\cite{witvliet_connectomes_2021}\\ \hline
\textit{C. elegans} & Head Ganglia---Hour 8 & 198 & 1104 & 0.028 & RCM  & 626  & N/A &\cite{witvliet_connectomes_2021}\\ \hline
\textit{C. elegans} & Head Ganglia---Hour 15 & 204 & 1342 & 0.032 & RCM & 722  & N/A & \cite{witvliet_connectomes_2021}\\ \hline
\textit{C. elegans} & Head Ganglia---Hour 23 & 211 & 1801 & 0.041 & RCM& 957  & N/A &\cite{witvliet_connectomes_2021}\\ \hline
\textit{C. elegans} & Head Ganglia---Hour 27 & 216 & 1737 & 0.037 & RCM & 939  & N/A &\cite{witvliet_connectomes_2021}\\ \hline
\textit{C. elegans} & Head Ganglia---Hour 50 & 222 & 2476 & 0.050 & RCM & 1428 & N/A &\cite{witvliet_connectomes_2021}\\ \hline
\textit{C. elegans} & Head Ganglia---Hour 50 & 219 & 2488 & 0.052 & RCM& 1562 & N/A & \cite{witvliet_connectomes_2021}\\ \hline
\textit{C. elegans} & Hermaphrodite---nervous system & 309 & 2955 & 0.031 & RCM+Motifs& 2167 & \textbf{286} & \cite{white_structure_1986}\\ \hline
\textit{C. elegans} & Hermaphrodite---whole animal & 454 & 4841 & 0.024 & CM+Motifs & 7605 & \textbf{2661} & \cite{cook_whole-animal_2019}\\ \hline
\textit{C. elegans} & Male---whole animal & 575 & 5246 & 0.016 & CM+Motifs & 8979 & \textbf{2759} &\cite{cook_whole-animal_2019}\\ \hline
\textit{Drosophila} & Larva---left AL & 96 & 2142 & 0.235 & RCM  & 1550 & N/A &\cite{berck2016wiring} \\ \hline
\textit{Drosophila} & Larva---right AL & 96 & 2218 & 0.244 & RCM & 1527 & N/A &\cite{berck2016wiring} \\ \hline 
\textit{Drosophila} & Larva---left \& right ALs & 174 & 4229 & 0.140 &RCM+Motifs & 4117 & \textbf{105} & \cite{berck2016wiring} \\ \hline 
\textit{Drosophila} & Larva---left MB & 191 & 6449 & 0.167 & CM+Motifs & 8050 & \textbf{1369} & \cite{eichler2017complete}\\ \hline
\textit{Drosophila} & Larva---right MB & 198 & 6499 & 0.178 & CM+Motifs & 8191 & \textbf{1529} & \cite{eichler2017complete}\\ \hline
\textit{Drosophila} & Larva---left \& right MBs & 387 & 16956 & 0.114 & RCM+Motifs & 23764 & \textbf{5348} & \cite{eichler2017complete}\\ \hline
\textit{Drosophila} & Larva---motor neurons & 426 & 3795 & 0.021 & CM &  4762 & N/A & \cite{zarin2019drosophila} \\ \hline 
\textit{Drosophila} & Larva---whole brain & 2952 & 110140 & 0.013 & RCM+Motifs & 149521 & \textbf{28793} &\cite{winding_connectome_2023} \\ \hline
\textit{Drosophila} & Adult---right AL & 761 & 36901 & 0.064 & RCM+Motifs& 76007 & \textbf{61} &\cite{scheffer_connectome_2020} \\ \hline
\textit{Drosophila} & Adult---right LH & 3008 & 100914 & 0.011 &RCM+Motifs& 109473 & \textbf{583} &\cite{scheffer_connectome_2020} \\ \hline
\textit{Drosophila} & Adult---right MB & 4513 & 247863 & 0.012 & RCM+Motifs & 429773 & \textbf{13657} & \cite{scheffer_connectome_2020} \\ \hline
\textit{C. intestinalis} & Larva---whole brain & 222 & 3085 & 0.063 & RCM+Motifs & 3805 & \textbf{263} & \cite{ryan2016cns} \\ \hline
\textit{P. dumerelii} & Larva---whole brain & 2728 & 11433 & 0.002 & RCM+Motifs& 15733 & \textbf{325} & \cite{veraszto2020whole} \\ \hline
\end{tabular}
\label{table1}
\end{adjustwidth}
\end{table}


\section*{Results}

\subsection*{Numerical validation}\label{sec:numerical-validation}
To test the validity and performance of our motif inference procedure, we apply it to numerically generated networks with a known absence or presence of higher-order structure in the form of motifs (see ``\nameref{datasets:artificial}''in Methods). 

\subsubsection*{Null networks}

We first test the stringency of our inference method and compare it to classic, hypothesis testing-based approaches. We test whether they infer spurious motifs in random networks generated by the four dyadic random graph models (See ``\nameref{methods:randomized-networks}'' in the Methods). 
Since these random networks do not have any non-random higher-order structure, a trustworthy inference procedure should find no, or at least very few, significant motifs.

Hypothesis testing-based approaches to motif inference consist of checking whether each graphlet is significantly over-represented with respect to a predefined null model (we detail the procedure in~\nameref{SIsec:hypothesis-testing}). 
This approach is highly sensitive to the choice of null model and infers spurious motifs if the chosen null model does not correspond to the generative model  (Fig~\ref{fig:validity}A-\ref{fig:validity}D). 
Nevertheless, when the chosen null model is the generative model, almost no spurious motifs are found using the approach (Fig~\ref{fig:validity}A-\ref{fig:validity}D). 
However, since there is no general protocol for the choice of null model in the frequentist approach, this sensitivity to null model choice is a major concern in practice.

By casting motif inference as a model selection problem, our approach allows us to select the most appropriate model, including amongst a selection of null models. 
In our test, our approach consistently selects the true generative model for the networks, i.e., one of the four null models, and thus does not infer any spurious motifs (Fig~\ref{fig:validity}A-\ref{fig:validity}D).

\subsubsection*{Planted motifs}

To evaluate the efficiency of our method in finding motifs that are present in a network, we apply it to synthetic networks  with planted motifs (see ``\nameref{methods:planted-motif-model}'' in the Methods).

\begin{figure}[ht!]    
    \begin{adjustwidth}{-2.2in}{0.in}
         \begin{tikzpicture}
             \node at (0,0){\includegraphics[height=0.53\textheight]{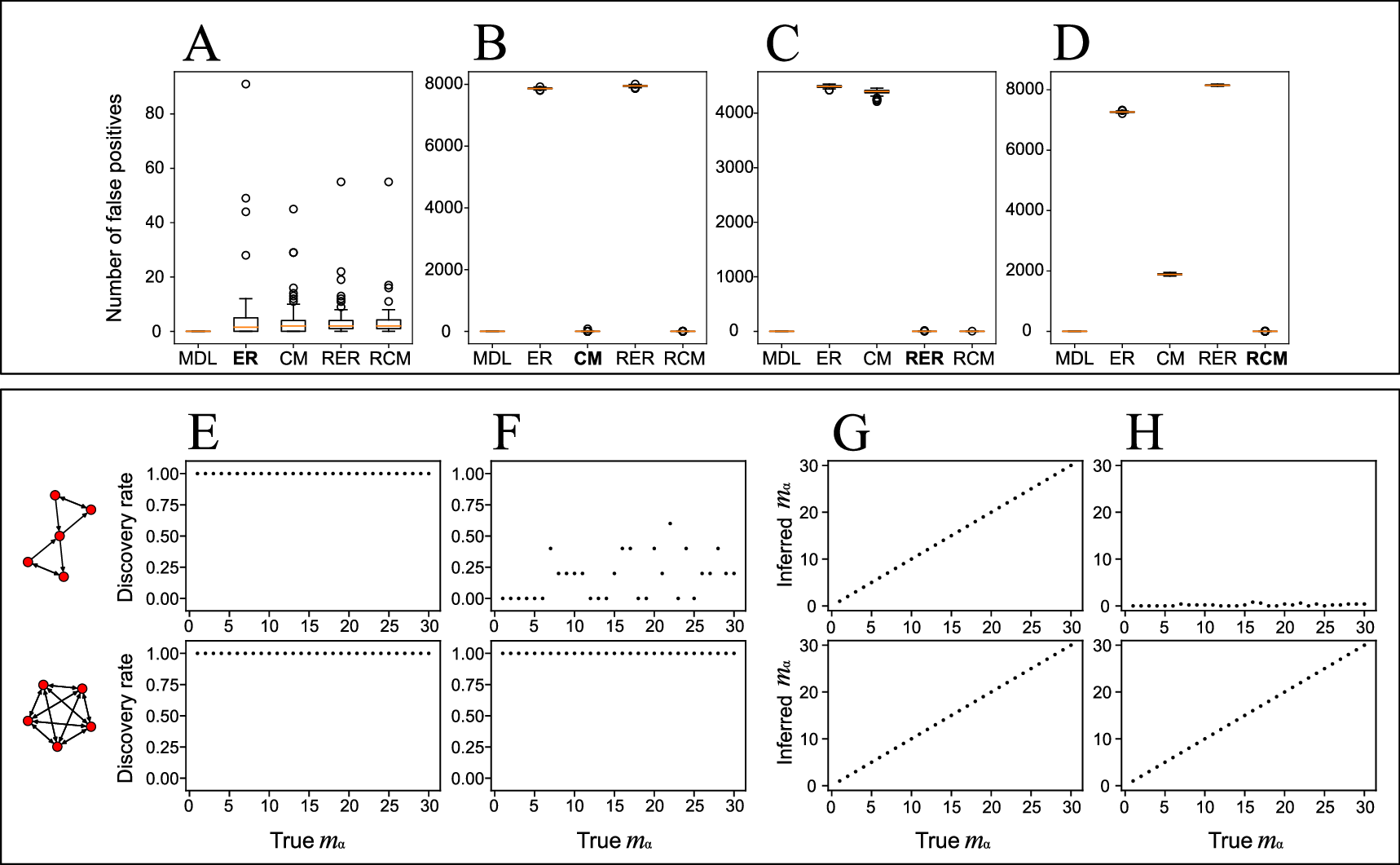}};
         \end{tikzpicture}
    \end{adjustwidth}
   
    \caption{\small
    {\bf Performance of compression-based motif inference on numerically generated networks.}\label{fig:validity}
    (A-D) Number of spurious motifs inferred using our compression-based method with MDL-based model selection and using hypothesis testing with four different null models in random networks generated from the same four null models: 
    (A) the Erd\H{o}s-R\'{e}nyi model (ER);
    (B) the configuration model (CM); 
    (C) the reciprocal ER model (RER); and
    (D) the reciprocal CM (RCM).  
    The x-axis labels indicate which method was used for motif inference: our method (MDL) or classic hypothesis testing with each of the four null models as reference. 
    The corresponding generative model is highlighted in boldface.
    To make hypothesis testing as conservative as possible, we applied a Bonferroni correction, which multiplies the raw $p$-values by $\setSize{\graphletSet} = 9\,576$ and we set the uncorrected significance threshold to $0.01$. 
    The random networks in (A-D) are all generated by fixing the values of each null model's parameters to those of the \textit{Drosophila} larva right MB connectome (e.g., $N=198$ and $E=6\,499$ for the ER model).
    (E-H) Ability of our method to correctly identify a placed graphlet as a motif as a function of the number of times it is repeated, $m_{\graphlet}$. 
    We show results for two selected 5-node graphlets: an hourglass structure (top row) and a clique (bottom row). The clique is the densest graphlet and is totally symmetric (the number of orientations, i.e., the number of non-automorphic node permutations, is equal to one). The hourglass has intermediary density, $\rho_\graphlet = 2/5$, and symmetry, with 60 non-automorphic orientations within a possible range of 1 to $5! = 120$.
    The generated networks in (E-H) contain $N = 300$ nodes and an edge density of either $\rho = E/N(N-1) = 0.025$ (E,G) or $\rho=0.1$ (F,H). 
    Each point is an average over five independently generated graphs. 
    (E,F) The discovery rate is the estimated probability that the planted motif belongs to the inferred motif set, i.e., $\langle 1-\delta(m_\graphlet,0) \rangle$. 
    (G,H) Average inferred number of repetitions of the planted motif, $\langle m_\graphlet\rangle$.}
    
\end{figure}

We show in Fig~\ref{fig:validity}E-\ref{fig:validity}H the ability of our algorithm to identify a motif (Fig~\ref{fig:validity}E and~\ref{fig:validity}G) and its occurrences (Fig~\ref{fig:validity}F and \ref{fig:validity}H) in numerically generated networks as a function of the number of times the motif is repeated in the network.
We show in~\nameref{SIfig:motif_detection_rate_300}-\nameref{SIfig:motifs_detected_100} Figs a more in-depth analysis including additional motifs, different network sizes, and an extended range of network densities. 
The performance of the algorithm is affected by both the  frequency of the planted motif (Fig~\ref{fig:validity}E-\ref{fig:validity}H) and its topology, with denser motifs generally being easier to identify (Figs~\ref{fig:validity}E-\ref{fig:validity}H, see also~\nameref{SIfig:motif_detection_rate_300} and \nameref{SIfig:motifs_detected_100}). 
The size of the network does not have a significant effect on our ability to detect motifs, but its edge density does (compare~\nameref{SIfig:motif_detection_rate_300} and~\nameref{SIfig:motifs_detected_300} Figs to~\nameref{SIfig:motif_detection_rate_100} and~\nameref{SIfig:motifs_detected_100} Figs). 
The latter is expected since motifs whose density differs significantly from the network's average density are easier to identify than motifs with a similar density.
This is similar to hypothesis testing-based approaches based on graphlet frequencies where dense motifs tend to be highly unlikely under the null model and thus easier to detect. However, we stress that our method does not rely on the same definition of significance---compression instead of over-representation---so the motifs that are easiest to infer are not necessarily the same with the different approaches (\nameref{SIfig:base-model-diff} Fig).

\subsection*{Neural connectomes}

We apply our method to infer circuit motifs in structural connectomes and characterize the  regularity of the connectivity of synapse-resolution brain networks of different species at different developmental stages (see Table~\ref{tab:datasets}).  
We consider boolean connectivity matrices that represent neural wiring as a binary, directed network where each node represents a neuron and an edge represents the presence of synaptic connections from one pre-synaptic neuron to a post-synaptic neuron. To keep in line with the usual definition of a motif, we exclude self-connections of neurons onto themselves, but they can be included if one wants to investigate such motifs. 

\begin{figure}[ht!]

    \begin{adjustwidth}{-2.5in}{0.in}

    \begin{tikzpicture}
        \node at (0,0) {\includegraphics[height=0.45\textheight]{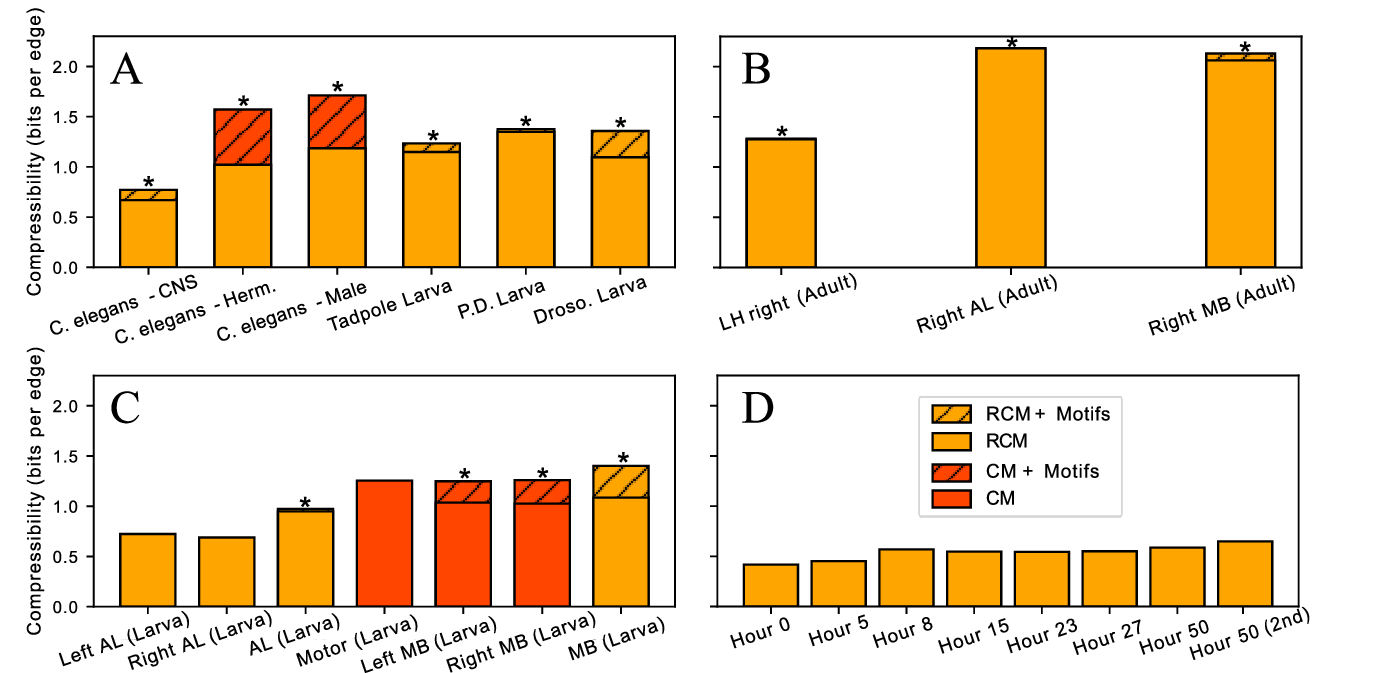}};
    \end{tikzpicture}    
    \end{adjustwidth}
    \caption{\small
    {\bf Compressibility of neural connectomes}. 
    Compressibility (measured in number of bits per edge in the network) $\Delta\codelength{}^*/E$ of different connectomes
    as compared to encoding the edges independently using the Erd\H{o}s-Rényi simple graph model (see Table~\ref{tab:codelengths}). 
    Two types of models are shown for the datasets: the best simple network encoding and
    the best motif-based encoding when this compresses more than the simple encoding. Asterisks highlight connectomes where motifs permit a higher compression than the reference models.
    (A) Whole-CNS and whole-animal connectomes. 
    (B) Connectomes of three different regions of the adult \textit{Drosophila} right hemibrain. Note that while the relative increase in compressibility of these connectomes obtained using motifs is relatively small, the motifs are highly significant due to the large size of these connectomes (Table~\ref{tab:datasets}). 
    (C) Connectomes of different brain regions of first instar \textit{Drosophila} larva. 
    (D) Connectomes of \textit{C.\ elegans} head ganglia at different developmental stages, from 0 hours to 50 (adult).
    While no higher-order motifs are found, the compressibility increases with maturation (and thus the size) of the connectome.}
    \label{fig:compressibility}
\end{figure}

We measure the compressibility of a connectome $G$ as the difference in codelength between its encoding using a simple Erd\H{o}s-R\'enyi model, i.e., encoding the edges individually, and its encoding using the most compressing model,
\begin{equation}\label{eq:compression}
    \Delta\codelength{}^* =  \codelength{}(G,\thetaER) - \codelength{}(G,\theta^*)  .
\end{equation}
As Fig~\ref{fig:compressibility} and Table~\ref{tab:datasets} show, all the empirical connectomes are compressible, confirming their non-random structure (see~\nameref{SIfig:compressibility_B} Fig for a comparison of all the models considered).
Significant higher-order structures in the form of motifs are found in all the whole-CNS and whole-nervous-system connectomes studied here (Fig~\ref{fig:compressibility}A) as well as in many connectomes of individual brain regions (Fig~\ref{fig:compressibility}B and~\ref{fig:compressibility}C).
Besides motifs, we find significant non-random degree distributions of the nodes in all connectomes (Fig~\ref{fig:compressibility}).
This is consistent with node degrees being a salient feature of many biological networks, including neuronal networks~\cite{fornito_fundamentals_2016}.
Reciprocal connections are also a significant feature of almost all connectomes studied, in alignment with empirical observations from \textit{in vivo} experiments \cite{winding_connectome_2023,cervantes2017reciprocal,bahl2020neural,singer2021recurrent,gilbert2013top} where modulation of neural activity is often implemented through recurrent patterns. 
Note that reciprocal connections are often considered a two-node motif. We chose to encode it as a dyadic feature of the base model since this is more efficient and allows for a higher compression, but it is entirely possible to encode them as graphlets by allowing also two-node graphlets as supernodes in the reduced graph (instead of restricting to 3--5 node graphlets as we did here). 

For several smaller, regional connectomes, we do not find statistical evidence for higher-order motifs (Fig~\ref{fig:compressibility}C and~\ref{fig:compressibility}D), indicating the absence of significant higher-order circuit patterns (i.e., involving more than two neurons) in these connectomes. 
Note that network size did not have a significant effect on motif detectability in our numerical experiments above (see~\ref{SIfig:motif_detection_rate_300}--\ref{SIfig:motifs_detected_100} Figs), so the absence of motifs in these connectomes are likely due to their structural particularities rather than simply their smaller size.
In particular, we do not find evidence for motifs in the \textit{C.\ elegans} head ganglia (brain) connectomes at any developmental stage (Fig~\ref{fig:compressibility}D). 
Note, however, that we do detect significant edge and node features (as encoded by the reciprocal configuration model), highlighting the non-random distribution of neuron connectivity and the importance of feedback connections in these connectomes. 
Furthermore, we do find higher-order motifs in the more complete \textit{C.\ elegans} connectomes that also include sensory and motor neurons (Fig~\ref{fig:compressibility}A), in line with what was found earlier using hypothesis-testing based motif mining~\cite{milo_network_2002, cook_whole-animal_2019}.

\begin{figure}[ht!]
    
    \begin{adjustwidth}{-2.4in}{0.in}
    \begin{tikzpicture}
        \node at (0,0){\includegraphics[height=0.38\textheight]{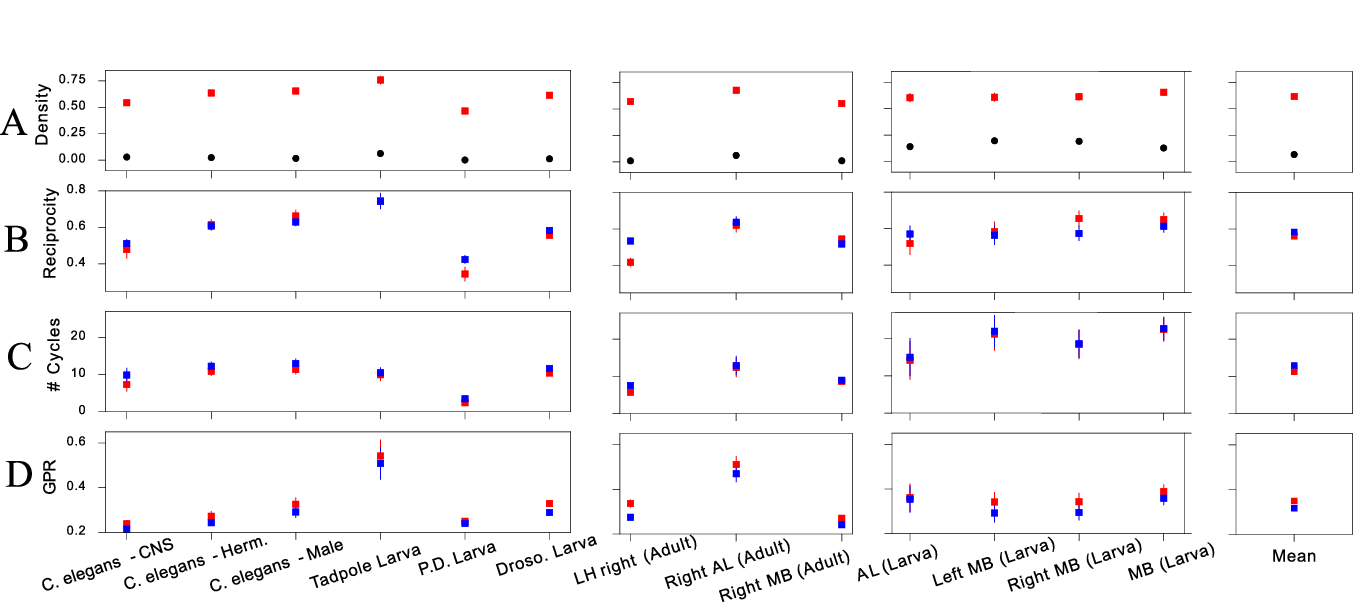}};
    \end{tikzpicture}
    
    \end{adjustwidth}
    \caption{\small {\bf Topological properties of motif sets}.
    Graph measures averaged over the inferred graphlet multiset, $\subgraphSet$, i.e., for a network measure $\varphi$, one point corresponds to the quantity $\mu_\varphi(\subgraphSet) = \sum_{\graphlet\in\subgraphSet}\varphi(\graphlet)/|\subgraphSet|$. The density (A), reciprocity (B) and number of cycles (C) and are standard properties of directed networks~\cite{hagberg2020networkx}. 
    The graph polynomial root (D) measures the structural symmetry of the motifs~\cite{dehmer2020orbit}. 
    Details can be found in~\nameref{SIsec:gpr}. 
    Red squares indicate averages over the connectomes' inferred motif sets. Blue squares are reference values, computed from  average over randomized graphlets with their density conserved. 
    To obtain the fixed-density references per motif set, we generate for each graphlet a collection of a hundred randomized configurations sharing the same density. The black dots of panel (A) show the connectomes' global density.}
    \label{fig:graphlet_measures}
\end{figure}

To study the structural properties of the inferred motif sets, we computed different average network measures of the motifs of each connectome (see definitions in~\nameref{SIsec:gpr}). 
The density of inferred motifs is much higher than the average density of the connectome (Fig~\ref{fig:graphlet_measures}A). 
While the density of motifs is high for all connectomes, it does vary significantly between them in a manner that is seemingly uncorrelated with the average connectome density. 
The motifs' high density means that half of their node pairs or more are connected on average, which would lead to high numbers of reciprocal connections even if the motifs were wired at random. 
We indeed observe a high reciprocity of connections in the inferred motifs, and that this reciprocity is in large part explained by their high average density (Fig~\ref{fig:graphlet_measures}B), although we do observe significant variability and differences from this random baseline. 
The average number of cycles in the motifs is, on the other hand, in general completely explained by the motifs' high density (Fig~\ref{fig:graphlet_measures}C).    
To probe the higher-order structure of the inferred motifs we measure their symmetry as measured by the graph polynomial root (GPR)~\cite{dehmer2020orbit}.
As Fig~\ref{fig:graphlet_measures}D shows, the motifs are on average more symmetric than random graphlets of the same density even if the individual differences are often not significant. 
Thus, of the four aggregate topological features we investigated, the elevated density is the most salient feature of the motif sets. 
This does not exclude the existence of salient (higher-order) structural particularities of the motifs beyond their high density, only that such features are not captured well by these simple aggregate measures. 

\begin{figure}[hbt!]

    \begin{adjustwidth}{-2.5in}{0.in}
        \begin{tikzpicture}
            \node at (0,0){\includegraphics[height=0.27\textheight]{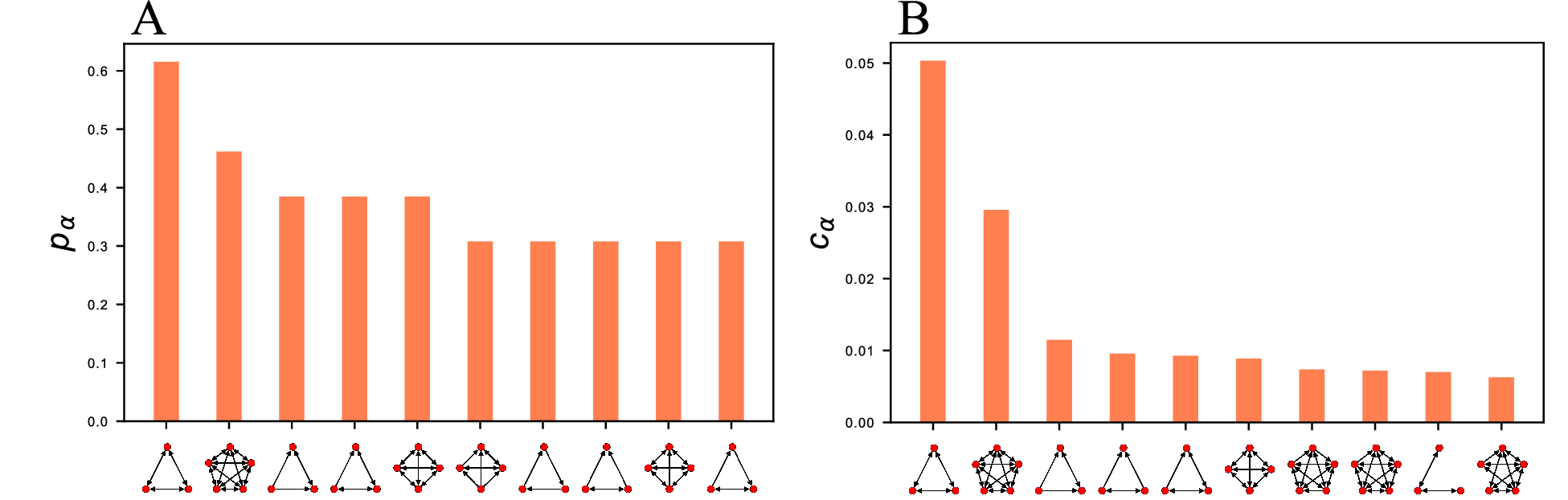}};
        \end{tikzpicture}
    \end{adjustwidth}
    
    \caption{\small {\bf Connectomes share common motifs.} 
    Most frequently appearing motifs in the motif sets inferred for all connectomes.
    (A) Most frequently found motifs: fraction of connectomes in which each motif is found, $p_\graphlet = \frac{1}{|\mathcal{G}|}\sum_{G\in\mathcal{G}}(1 - \delta_{m_\graphlet(G),0})$.
    (B) Most repeated motifs: average graphlet concentration $c_\graphlet = \frac{1}{|\mathcal{G}|}\sum_{G\in\mathcal{G}}\frac{m_\graphlet(G)}{\sum_{\graphlet}m_\graphlet(G)}$.
    }
    \label{fig:motif_sets}
\end{figure}

Even though the inferred motif sets are highly diverse, we observe that several motifs are found in a large fraction of the connectomes (Fig~\ref{fig:motif_sets}A).
The same motifs also tend to be among the most frequent motifs, i.e., the ones making up the largest fraction of the inferred motif sets on average (Fig~\ref{fig:motif_sets}B).
These tend to be highly dense graphlets, with the two most frequent motifs being the three and five node cliques, which are each found in roughly half of the connectomes and are also the most frequent motifs in the motif sets on average.
The ten most frequently found motifs (Fig~\ref{fig:motif_sets}A) and the most repeated motifs (Fig~\ref{fig:motif_sets}B) do not perfectly overlap, though six of the ten motifs are the same between the two lists. 

\section*{Discussion}

We have developed a methodology to infer sets of network motifs and evaluate their collective significance based on lossless compression. 
Our approach defines an implicit generative model and lets us cast motif inference as a model selection problem through the MDL principle. 
It overcomes several common limitations of traditional hypothesis testing-based methods, which are unable to compare the significance of different motifs and have difficulties dealing with multiple testing, correlations between motif counts, the evaluation of low $p$-values, and the often ill-defined problem of choosing the proper null model to compare against.

Our compression-based methodology accounts for multiple testing and correlations between motifs, and it does not rely on approximations of the null distribution of a test statistic. 
Note that such approximations are generally necessary for statistical hypothesis testing to be computationally feasible.
For example, there are about 10\,000 possible five-node motifs, so
to control for false positives using the Bonferroni correction, raw $p$-values must be multiplied by 10\,000.
Thus, one needs to be able to reliably estimate raw $p$ values smaller than $5\cdot10^{-6}$ to evaluate significance at a nominal level of 0.05. To obtain an exact test, we must generate of the order of a million random networks and perform a subgraph census of each, a typically unfeasible computational task.
Furthermore, constrained null models are hard to sample uniformly~\cite{orsini_quantifying_2015}, and even in models that are simple enough for the Markov chain edge swap procedure to be ergodic, correlations may persist for a long time, inducing an additional risk of spurious results~\cite{ginoza_network_2010, beber_artefacts_2012}.

Our method furthermore allows us to infer not only significant motif sets but also compare and rank the significance of different motifs and sets of motifs and other network features such as node degrees and reciprocity of edges.
It thus overcomes the need for choosing the null model {a priori}, which leads to spurious motifs if this choice is not appropriate.

Note that while our method enables statistically grounded inference of motif sets, it does not provide an estimate of their intrinsic statistical variability since it relies on a greedy optimization algorithm---in the language of Bayesian inference, inferred motif sets correspond to maximum a posteriori estimates. This variability  could in principle be estimated via Markov chain Monte Carlo (MCMC) sampling around the optimum motif set, but the development of an efficient MCMC algorithm is an open problem. Thus, for the time being, the variability can only be assessed experimentally by comparing multiple measurements. 

Our method is conceptually close to the subgraph covers proposed in~\cite{wegner_subgraph_2014} which models a graph with motifs as the projection of overlapping subgraphs onto a simple graph and relies on information theoretic principles to select an optimum cover.
That approach modeled the space of subgraph covers as a microcanonical ensemble instead of the observed graph directly. This makes it harder to fix node- and edge-level features such as degrees and reciprocity since these are functions of the cover's latent variables~\cite{wegner_atomic_2021}, although progress in inferring such features has recently been made~\cite{wegner2024nonparametric}.
We instead based our methodology on subgraph contractions as proposed in \cite{bloem_large-scale_2020}, whose approach we extended to allow for collective inference of motif sets and selection of base model features. 
In particular, we let the number of distinct graphlets be free in our method, instead of being limited to one; to deal with the problem of selecting between thousands of graphlets, we developed a stochastic greedy algorithm that selects the most compressing subgraph at each step; we simplified the model for the reduced graph by using multigraph codes, avoiding multiple prequential plug-in codes to account for parallel edges and providing exact codelengths; and we developed two new base models to account for reciprocal edges. 

We emphasize that the method we extended \cite{bloem_large-scale_2020} and ours are not the first ones to rely on the MDL principle for network pattern mining (see, e.g., the survey in~\cite{liu_graph_2018}).
The SUBDUE \cite{holder1994substucture} and VoG \cite{koutra2014vog} algorithms in particular are precursors of our work, though their focus was on graph summarization rather than motif mining.
The SUBDUE algorithm \cite{holder1994substucture} deterministically (but not optimally) extracts the graphlet that can compress a fixed encoding of the adjacency matrix and edge list when a sample of isomorphic (and quasi-isomorphic) subgraphs are contracted. The VoG algorithm \cite{koutra2014vog} uses a set of graphlet types, e.g., cliques or stars, and looks for the set of subgraphs (belonging exactly or approximately to these graphlet types) that best compresses a fixed encoding of the adjacency matrix; the latter being distinct from the one used in SUBDUE. 
These algorithms differ conceptually from ours in focusing not on motif mining but on more specific regularities for the problem of graph summarization. 
Their advantage is mainly computational as their implementations scale better with the input graph size. 
While being computationally more expensive, our approach does not impose or reduce a graphlet dictionary and the representation of the reduced graph is not  constrained by a specific functional form. 

Exponential random graph models (ERGMs) provide another generative framework for the problem of inferring important subgraphs of a network~\cite{robins_introduction_2007, lusher_exponential_2013}. 
Different from our approach, ERGMs generally rely on global graphlet counts and not on contracting specific subgraphs. 
This tends to make them unstable for general graphlets, making them hard to fit, due to issues of near-degeneracy~\cite{schweinberger_instability_2011, lusher_exponential_2013}.
This severely conditions the flexibility of ERGMs for motif inference since only a constrained set of particular combinations of motifs are known to ensure convergence of model fits \cite{stivala_testing_2021, snijders_new_2006, schweinberger_local_2015}. 

We applied our approach to uncover and characterize motifs and other structural regularities in synapse-resolution neural connectomes of several species of small animals. 
We find that the connectomes contain significant structural regularities in terms of a high number of feedback connections (high reciprocity), non-random degrees, and higher-order circuit motifs. 
In some smaller connectomes we do not find significant evidence for higher order motifs. This is in particular the case for connectomes of the head ganglia of \textit{C. elegans}, both at maturity and during its development. We still find significant reciprocity and non-random degrees in these connectomes though, confirming the fundamental importance of these measures in biological connectomes.
A high reciprocity in particular translates to a large number of feedback connections in the animals' neural networks, a feature whose biological importance has frequently been observed~\cite{jovanic_competitive_2016,winding_connectome_2023,gilbert2013top,bahl2020neural,jarrell2012connectome}.

The functional importance of higher-order motifs is less well known, but dense subgraphs are known to have an impact on information propagation in a network~\cite{pastor_epidemic_2015} and several circuit motifs have been proposed to carry out fundamental computations (e.g., feedforward and feedback regulation~\cite{milo_network_2002, alon_network_2007, alon_introduction_2019},  cortical computations~\cite{douglas_canonical_1989, harris_neocortical_2015, sterling_principles_2015}, predictive coding~\cite{bastos_canonical_2012}, and decision making~\cite{jovanic_competitive_2016}).
With the advent of synaptic resolution connectomes, the stage is now set for testing these hypotheses and comparing the structural characteristics of different networks with robust statistical tools such the method we introduced here. 
While we demonstrated our methodology's ability to detect the most significant circuit patterns in a network among all possible graphlets, it may directly be applied to test for the presence of pre-specified motifs such as the ones cited above by simply changing the graphlet set to include only those circuits. 

The mere presence of statistically regular features does not reveal their potential function, nor their origin~\cite{mazurie_evolutionary_2005}. 
These questions must be explored through computational modeling and, ultimately, biological experiments~\cite{bascompte_simple_2005, alon_network_2007, jovanic_competitive_2016, jovanic_neural_2019}. 
In this aspect our methodology offers an additional advantage over frequency-based methods since it infers not only motifs but also their localization in the network, making it possible to better inform physical models of circuit dynamics and to test their function directly \textit{in vivo} experiments. 

The compressibility of all the neural connectomes investigated here can be seen as a manifestation of the \textit{the genomic bottleneck principle}~\cite{zador_critique_2019}, which states that the information stored in an animal's genome about the wiring of its neural connectome must be compressed or the quantity of information needed to store it would exceed the genome's capacity. 
Note however that the codelengths needed to describe the connectomes we infer are necessarily lower bounds on the actual codelengths needed to encode the neural wiring blueprints. 
First, our model is a crude approximation to reality, and a more realistic (and thus more compressing) model would incorporate the physical constraints on neural wiring such as its embedding in 3D space, steric constraints, and the fact that the nervous system is the product of morphogenesis.
Second, our code is lossless, which means we perfectly encode the placement of each link in the connectome, while the wiring of neural connections may partially be the product of randomness. Thus a lossy encoding would be a more appropriate measure of a connectome's compressibility~\cite{koulakov_encoding_2021} but it introduces the difficulty of defining the appropriate distortion measure.
Third, subgraph census quickly becomes computationally unfeasible for larger motifs, which generally limits the size of motifs we can consider to less than ten nodes.
Allowing for overlapping contractions could be a way to infer larger motifs as combinations of smaller ones (similar to \cite{elhesha_identification_2016}). 

We proposed four different base models for our methodology, which allows to select and constrain the important edge- and node-level features of reciprocity and degrees in our model. 
It is straightforward to incorporate additional base models as long as their microcanonical entropy can be evaluated efficiently. 
We envisage two important extensions to the base models. 
First, block structure, which may be incorporated as a stochastic block model~\cite{peixoto2014hierarchical, peixoto2017nonparametric}, is ubiquitous in biological and other empirical networks and has been shown to have an important impact on signal propagation~\cite{hens2019spatiotemporal}. 
Second, the network's embedding in physical space, as modelled using geometric graphs or other latent space models~\cite{boguna2021network,bianconi20215}, is also meaningful. 
It should matter for neuronal networks due to considerations such as wiring cost~\cite{sterling_principles_2015}, signal latency~\cite{sterling_principles_2015}, and steric constraints~\cite{sterling_principles_2015}.

Our approach contributes both to the burgeoning field of higher-order networks~\cite{battiston_networks_2020} and to the growing push towards principled statistical inference of network data~\cite{peel2022statistical} by providing a robust generative framework for motif inference.
The field of statistical network analysis is still in its infancy and much work is still needed to make inference methods more robust. 
Here, we have for example not considered the problems of noisy data and incomplete sampling~\cite{crane2018probabilistic} which can influence the apparent structure and dynamic of network data in complex ways~\cite{granovetter1976network, achlioptas2009bias, genois2015compensating, crane2018probabilistic}.
It should be interesting to extend statistical inference to non-local higher order structures, such as symmetry-group based structures  \cite{morone2019symmetry} or, e.g., hierarchically nested motifs which might be incorporated in a similar manner to the recent hierarchical extensions of stochastic block models~\cite{peixoto2014hierarchical, lyzinski2016community}.
A common barrier to the development of principled statistical inference of many network models is that they do not admit easily tractable likelihoods. This is in particular the case for many higher-order models, such as the one of \cite{morone2019symmetry}, and, more famously, for the small world model of Watts and Strogatz~\cite{watts_Collective_1998} and the preferential attachment model of Barab\'asi and Albert~\cite{barabasi_emergence_1999}.
Simulation-based inference \cite{cranmer2020frontier} provides a promising framework for bridging the gap between such models and statistical inference \cite{goyal2023framework}.

\section*{Acknowledgments}
We acknowledge the help of the HPC Core Facility of the Institut Pasteur for this work. 



\nolinenumbers

%
%
%

\bibliography{motifs-compression}

\begin{thebibliography}{100}

\bibitem{newman_structure_2003}
Newman ME.
\newblock The structure and function of complex networks.
\newblock SIAM review. 2003;45(2):167--256.

\bibitem{fornito_fundamentals_2016}
Fornito A, Zalesky A, Bullmore E.
\newblock Fundamentals of brain network analysis.
\newblock Academic Press; 2016.

\bibitem{alon_introduction_2019}
Alon U.
\newblock An introduction to systems biology: design principles of biological
  circuits.
\newblock CRC press; 2019.

\bibitem{watts_Collective_1998}
Watts DJ, Strogatz SH.
\newblock Collective dynamics of ‘small-world’networks.
\newblock nature. 1998;393(6684):440--442.

\bibitem{newman2018networks}
Newman M.
\newblock Networks.
\newblock Oxford university press; 2018.

\bibitem{sporns_small_2004}
Sporns O, Zwi JD.
\newblock The small world of the cerebral cortex.
\newblock Neuroinformatics. 2004;2:145--162.

\bibitem{bassett2017small}
Bassett DS, Bullmore ET.
\newblock Small-world brain networks revisited.
\newblock The Neuroscientist. 2017;23(5):499--516.

\bibitem{barabasi_emergence_1999}
Barab{\'a}si AL, Albert R.
\newblock Emergence of scaling in random networks.
\newblock science. 1999;286(5439):509--512.

\bibitem{cohen2000resilience}
Cohen R, Erez K, Ben-Avraham D, Havlin S.
\newblock Resilience of the internet to random breakdowns.
\newblock Physical review letters. 2000;85(21):4626.

\bibitem{pastor2001epidemic}
Pastor-Satorras R, Vespignani A.
\newblock Epidemic spreading in scale-free networks.
\newblock Physical review letters. 2001;86(14):3200.

\bibitem{seyed2006scale}
Seyed-Allaei H, Bianconi G, Marsili M.
\newblock Scale-free networks with an exponent less than two.
\newblock Physical Review E. 2006;73(4):046113.

\bibitem{newman2006modularity}
Newman ME.
\newblock Modularity and community structure in networks.
\newblock Proceedings of the national academy of sciences.
  2006;103(23):8577--8582.

\bibitem{ravasz2009detecting}
Ravasz E.
\newblock Detecting hierarchical modularity in biological networks.
\newblock Computational Systems Biology. 2009; p. 145--160.

\bibitem{cimini2019statistical}
Cimini G, Squartini T, Saracco F, Garlaschelli D, Gabrielli A, Caldarelli G.
\newblock The statistical physics of real-world networks.
\newblock Nature Reviews Physics. 2019;1(1):58--71.

\bibitem{battiston_networks_2020}
Battiston F, Cencetti G, Iacopini I, Latora V, Lucas M, Patania A, et~al.
\newblock Networks beyond pairwise interactions: {Structure} and dynamics.
\newblock Physics Reports. 2020;874:1--92.
\newblock doi:{10.1016/j.physrep.2020.05.004}.

\bibitem{milo_network_2002}
Milo R, Shen-Orr S, Itzkovitz S, Kashtan N, Chklovskii D, Alon U.
\newblock Network {Motifs}: {Simple} {Building} {Blocks} of {Complex}
  {Networks}.
\newblock Science. 2002;298(5594):824--827.
\newblock doi:{10.1126/science.298.5594.824}.

\bibitem{sporns_motifs_2004}
Sporns O, K{\"o}tter R.
\newblock Motifs in {Brain} {Networks}.
\newblock PLOS Biology. 2004;2(11):e369.
\newblock doi:{10.1371/journal.pbio.0020369}.

\bibitem{tran_current_2015}
Tran NTL, Mohan S, Xu Z, Huang CH.
\newblock Current innovations and future challenges of network motif detection.
\newblock Briefings in Bioinformatics. 2015;16(3):497--525.
\newblock doi:{10.1093/bib/bbu021}.

\bibitem{holland_method_1977}
Holland PW, Leinhardt S.
\newblock A {Method} for {Detecting} {Structure} in {Sociometric} {Data}.
\newblock In: Leinhardt S, editor. Social {Networks}. Academic Press; 1977. p.
  411--432.

\bibitem{holland_local_1976}
Holland PW, Leinhardt S.
\newblock Local {Structure} in {Social} {Networks}.
\newblock Sociological Methodology. 1976;7:1--45.
\newblock doi:{10.2307/270703}.

\bibitem{stone_network_2019}
Stone L, Simberloff D, Artzy-Randrup Y.
\newblock Network motifs and their origins.
\newblock PLOS Computational Biology. 2019;15(4):e1006749.
\newblock doi:{10.1371/journal.pcbi.1006749}.

\bibitem{milo_superfamilies_2004}
Milo R, Itzkovitz S, Kashtan N, Levitt R, Shen-Orr S, Ayzenshtat I, et~al.
\newblock Superfamilies of {Evolved} and {Designed} {Networks}.
\newblock Science. 2004;303(5663):1538--1542.
\newblock doi:{10.1126/science.1089167}.

\bibitem{yeger-lotem_network_2004}
Yeger-Lotem E, Sattath S, Kashtan N, Itzkovitz S, Milo R, Pinter RY, et~al.
\newblock Network motifs in integrated cellular networks of
  transcription{\textendash}regulation and protein{\textendash}protein
  interaction.
\newblock Proceedings of the National Academy of Sciences.
  2004;101(16):5934--5939.
\newblock doi:{10.1073/pnas.0306752101}.

\bibitem{bascompte_simple_2005}
Bascompte J, Meli{\'a}n CJ.
\newblock Simple {Trophic} {Modules} for {Complex} {Food} {Webs}.
\newblock Ecology. 2005;86(11):2868--2873.

\bibitem{alon_network_2007}
Alon U.
\newblock Network motifs: theory and experimental approaches.
\newblock Nature Reviews Genetics. 2007;8(6):450--461.

\bibitem{jovanic_competitive_2016}
Jovanic T, Schneider-Mizell CM, Shao M, Masson JB, Denisov G, Fetter RD, et~al.
\newblock Competitive {Disinhibition} {Mediates} {Behavioral} {Choice} and
  {Sequences} in {Drosophila}.
\newblock Cell. 2016;167(3):858--870.e19.
\newblock doi:{10.1016/j.cell.2016.09.009}.

\bibitem{artzy-randrup_comment_2004}
Artzy-Randrup Y, Fleishman SJ, Ben-Tal N, Stone L.
\newblock Comment on "{Network} {Motifs}: {Simple} {Building} {Blocks} of
  {Complex} {Networks}" and "{Superfamilies} of {Evolved} and {Designed}
  {Networks}".
\newblock Science. 2004;305(5687):1107--1107.
\newblock doi:{10.1126/science.1099334}.

\bibitem{ginoza_network_2010}
Ginoza R, Mugler A.
\newblock Network motifs come in sets: {Correlations} in the randomization
  process.
\newblock Phys Rev E. 2010;82(1):011921.
\newblock doi:{10.1103/PhysRevE.82.011921}.

\bibitem{beber_artefacts_2012}
Beber ME, Fretter C, Jain S, Sonnenschein N, M{\"u}ller-Hannemann M, H{\"u}tt
  MT.
\newblock Artefacts in statistical analyses of network motifs: general
  framework and application to metabolic networks.
\newblock Journal of The Royal Society Interface. 2012;9(77):3426--3435.
\newblock doi:{10.1098/rsif.2012.0490}.

\bibitem{orsini_quantifying_2015}
Orsini C, Dankulov MM, Colomer-de Sim{\'o}n P, Jamakovic A, Mahadevan P, Vahdat
  A, et~al.
\newblock Quantifying randomness in real networks.
\newblock Nat Commun. 2015;6(1):1--10.
\newblock doi:{10.1038/ncomms9627}.

\bibitem{fodor_intrinsic_2020}
Fodor J, Brand M, Stones RJ, Buckle AM.
\newblock Intrinsic limitations in mainstream methods of identifying network
  motifs in biology.
\newblock BMC Bioinformatics. 2020;21(1):165.
\newblock doi:{10.1186/s12859-020-3441-x}.

\bibitem{stivala_testing_2021}
Stivala A, Lomi A.
\newblock Testing biological network motif significance with exponential random
  graph models.
\newblock Appl Netw Sci. 2021;6(1):1--27.
\newblock doi:{10.1007/s41109-021-00434-y}.

\bibitem{cover_elements_2012}
Cover TM, Thomas JA.
\newblock Elements of {Information} {Theory}.
\newblock John Wiley \& Sons; 2012.

\bibitem{bloem_large-scale_2020}
Bloem P, de~Rooij S.
\newblock Large-scale network motif analysis using compression.
\newblock Data Min Knowl Disc. 2020;34(5):1421--1453.
\newblock doi:{10.1007/s10618-020-00691-y}.

\bibitem{grunwald_minimum_2007}
Gr\"{u}nwald PD.
\newblock The {Minimum} {Description} {Length} {Principle}.
\newblock Penguin Book; 2007.

\bibitem{grunwald_minimum_2020}
Gr{\"u}nwald P, Roos T.
\newblock Minimum description length revisited.
\newblock International Journal of Mathematics for Industry.
  2020;doi:{10.1142/S2661335219300018}.

\bibitem{saalfeld2009catmaid}
Saalfeld S, Cardona A, Hartenstein V, Toman{\v{c}}{\'a}k P.
\newblock CATMAID: collaborative annotation toolkit for massive amounts of
  image data.
\newblock Bioinformatics. 2009;25(15):1984--1986.

\bibitem{ohyama_multilevel_2015}
Ohyama T, Schneider-Mizell CM, Fetter RD, Aleman JV, Franconville R,
  Rivera-Alba M, et~al.
\newblock A multilevel multimodal circuit enhances action selection in
  {Drosophila}.
\newblock Nature. 2015;520(7549):633--639.
\newblock doi:{10.1038/nature14297}.

\bibitem{witvliet_connectomes_2021}
Witvliet D, Mulcahy B, Mitchell JK, Meirovitch Y, Berger DR, Wu Y, et~al.
\newblock Connectomes across development reveal principles of brain maturation.
\newblock Nature. 2021;596(7871):257--261.
\newblock doi:{10.1038/s41586-021-03778-8}.

\bibitem{winding_connectome_2023}
Winding M, Pedigo BD, Barnes CL, Patsolic HG, Park Y, Kazimiers T, et~al.
\newblock The connectome of an insect brain.
\newblock Science. 2023;379(6636):eadd9330.

\bibitem{onnela_intensity_2005}
Onnela JP, Saram{\"a}ki J, Kert{\'e}sz J, Kaski K.
\newblock Intensity and coherence of motifs in weighted complex networks.
\newblock Phys Rev E. 2005;71(6):065103.
\newblock doi:{10.1103/PhysRevE.71.065103}.

\bibitem{picciolo_weighted_2022}
Picciolo F, Ruzzenenti F, Holme P, Mastrandrea R.
\newblock Weighted network motifs as random walk patterns.
\newblock New J Phys. 2022;24(5):053056.
\newblock doi:{10.1088/1367-2630/ac6f75}.

\bibitem{kovanen_temporal_2011}
Kovanen L, Karsai M, Kaski K, Kert{\'e}sz J, Saram{\"a}ki J.
\newblock Temporal motifs in time-dependent networks.
\newblock J Stat Mech. 2011;2011(11):P11005.
\newblock doi:{10.1088/1742-5468/2011/11/P11005}.

\bibitem{paranjape_motifs_2017}
Paranjape A, Benson AR, Leskovec J.
\newblock Motifs in {Temporal} {Networks}.
\newblock In: Proceedings of the {Tenth} {ACM} {International} {Conference} on
  {Web} {Search} and {Data} {Mining}. {WSDM} '17. New York, NY, USA:
  Association for Computing Machinery; 2017. p. 601--610.

\bibitem{battiston_multilayer_2017}
Battiston F, Nicosia V, Chavez M, Latora V.
\newblock Multilayer motif analysis of brain networks.
\newblock Chaos: An Interdisciplinary Journal of Nonlinear Science. 2017;27(4).

\bibitem{sallmen_graphlets_2022}
Sallmen S, Nurmi T, Kivel{\"a} M.
\newblock Graphlets in multilayer networks.
\newblock Journal of Complex Networks. 2022;10(2):cnac005.

\bibitem{lee_hypergraph_2020}
Lee G, Ko J, Shin K.
\newblock Hypergraph motifs: Concepts, algorithms, and discoveries.
\newblock arXiv preprint arXiv:200301853. 2020;.

\bibitem{lotito_higher_2022}
Lotito QF, Musciotto F, Montresor A, Battiston F.
\newblock Higher-order motif analysis in hypergraphs.
\newblock Communications Physics. 2022;5(1):79.

\bibitem{przulj_biological_2007}
Pr{\v z}ulj N.
\newblock Biological network comparison using graphlet degree distribution.
\newblock Bioinformatics. 2007;23(2):e177--e183.
\newblock doi:{10.1093/bioinformatics/btl301}.

\bibitem{ribeiro_survey_2019}
Ribeiro P, Paredes P, Silva MEP, Aparicio D, Silva F.
\newblock A {Survey} on {Subgraph} {Counting}: {Concepts}, {Algorithms} and
  {Applications} to {Network} {Motifs} and {Graphlets}.
\newblock arXiv:191013011 [cs]. 2019;.

\bibitem{paredes_towards_2013}
Paredes P, Ribeiro P.
\newblock Towards a faster network-centric subgraph census.
\newblock In: Proceedings of the 2013 {IEEE}/{ACM} {International} {Conference}
  on {Advances} in {Social} {Networks} {Analysis} and {Mining}. {ASONAM} '13.
  Niagara, Ontario, Canada: Association for Computing Machinery; 2013. p.
  264--271.

\bibitem{paredes_rand-fase_2015}
Paredes P, Ribeiro P.
\newblock Rand-{FaSE}: fast approximate subgraph census.
\newblock Soc Netw Anal Min. 2015;5(1):17.
\newblock doi:{10.1007/s13278-015-0256-2}.

\bibitem{wernicke_faster_2005}
Wernicke S.
\newblock A {Faster} {Algorithm} for {Detecting} {Network} {Motifs}.
\newblock In: Casadio R, Myers G, editors. Algorithms in {Bioinformatics}.
  Lecture {Notes} in {Computer} {Science}. Berlin, Heidelberg: Springer; 2005.
  p. 165--177.

\bibitem{ribeiro_g-tries_2010}
Ribeiro P, Silva F.
\newblock g-tries: an efficient data structure for discovering network motifs.
\newblock In: Proceedings of the 2010 {ACM} {Symposium} on {Applied}
  {Computing}. {SAC} '10. Sierre, Switzerland: Association for Computing
  Machinery; 2010. p. 1559--1566.

\bibitem{grunwald_safe_2021}
Gr{\"u}nwald P, de~Heide R, Koolen W. Safe {Testing}; 2021.
\newblock Available from: \url{http://arxiv.org/abs/1906.07801}.

\bibitem{gauvin_randomized_2022}
Gauvin L, G{\'e}nois M, Karsai M, Kivel{\"a} M, Takaguchi T, Valdano E, et~al.
\newblock Randomized {Reference} {Models} for {Temporal} {Networks}.
\newblock SIAM Rev. 2022;64(4):763--830.
\newblock doi:{10.1137/19M1242252}.

\bibitem{jaynes_information_1957}
Jaynes ET.
\newblock Information {Theory} and {Statistical} {Mechanics}.
\newblock Phys Rev. 1957;106(4):620--630.
\newblock doi:{10.1103/PhysRev.106.620}.

\bibitem{presse_principles_2013}
Press{\'e} S, Ghosh K, Lee J, Dill KA.
\newblock Principles of maximum entropy and maximum caliber in statistical
  physics.
\newblock Rev Mod Phys. 2013;85(3):1115--1141.
\newblock doi:{10.1103/RevModPhys.85.1115}.

\bibitem{fosdick_configuring_2018}
Fosdick BK, Larremore DB, Nishimura J, Ugander J.
\newblock Configuring {Random} {Graph} {Models} with {Fixed} {Degree}
  {Sequences}.
\newblock SIAM Rev. 2018;60(2):315--355.
\newblock doi:{10.1137/16M1087175}.

\bibitem{gilbert2013top}
Gilbert CD, Li W.
\newblock Top-down influences on visual processing.
\newblock Nature Reviews Neuroscience. 2013;14(5):350--363.

\bibitem{bahl2020neural}
Bahl A, Engert F.
\newblock Neural circuits for evidence accumulation and decision making in
  larval zebrafish.
\newblock Nature neuroscience. 2020;23(1):94--102.

\bibitem{jarrell2012connectome}
Jarrell TA, Wang Y, Bloniarz AE, Brittin CA, Xu M, Thomson JN, et~al.
\newblock The connectome of a decision-making neural network.
\newblock science. 2012;337(6093):437--444.

\bibitem{squartini2013reciprocity}
Squartini T, Picciolo F, Ruzzenenti F, Garlaschelli D.
\newblock Reciprocity of weighted networks.
\newblock Scientific reports. 2013;3(1):2729.

\bibitem{white_structure_1986}
White JG, Southgate E, Thomson JN, Brenner S.
\newblock The structure of the nervous system of the nematode {Caenorhabditis}
  elegans.
\newblock Philosophical Transactions of the Royal Society of London B,
  Biological Sciences. 1986;314(1165):1--340.
\newblock doi:{10.1098/rstb.1986.0056}.

\bibitem{cook_whole-animal_2019}
Cook SJ, Jarrell TA, Brittin CA, Wang Y, Bloniarz AE, Yakovlev MA, et~al.
\newblock Whole-animal connectomes of both {Caenorhabditis} elegans sexes.
\newblock Nature. 2019;571(7763):63--71.
\newblock doi:{10.1038/s41586-019-1352-7}.

\bibitem{berck2016wiring}
Berck ME, Khandelwal A, Claus L, Hernandez-Nunez L, Si G, Tabone CJ, et~al.
\newblock The wiring diagram of a glomerular olfactory system.
\newblock Elife. 2016;5:e14859.

\bibitem{eichler2017complete}
Eichler K, Li F, Litwin-Kumar A, Park Y, Andrade I, Schneider-Mizell CM, et~al.
\newblock The complete connectome of a learning and memory centre in an insect
  brain.
\newblock Nature. 2017;548(7666):175--182.

\bibitem{zarin2019drosophila}
Zarin AA, Mark B, Cardona A, Litwin-Kumar A, Doe CQ.
\newblock A Drosophila larval premotor/motor neuron connectome generating two
  behaviors via distinct spatio-temporal muscle activity.
\newblock BioRxiv. 2019; p. 617977.

\bibitem{scheffer_connectome_2020}
Scheffer LK, Xu CS, Januszewski M, Lu Z, Takemura Sy, Hayworth KJ, et~al.
\newblock A connectome and analysis of the adult Drosophila central brain.
\newblock Elife. 2020;9:e57443.

\bibitem{ryan2016cns}
Ryan K, Lu Z, Meinertzhagen IA.
\newblock The CNS connectome of a tadpole larva of Ciona intestinalis (L.)
  highlights sidedness in the brain of a chordate sibling.
\newblock Elife. 2016;5:e16962.

\bibitem{veraszto2020whole}
Veraszt{\'o} C, Jasek S, G{\"u}hmann M, Shahidi R, Ueda N, Beard JD, et~al.
\newblock Whole-animal connectome and cell-type complement of the
  three-segmented Platynereis dumerilii larva.
\newblock BioRxiv. 2020; p. 2020--08.

\bibitem{cervantes2017reciprocal}
Cervantes-Sandoval I, Phan A, Chakraborty M, Davis RL.
\newblock Reciprocal synapses between mushroom body and dopamine neurons form a
  positive feedback loop required for learning.
\newblock Elife. 2017;6:e23789.

\bibitem{singer2021recurrent}
Singer W.
\newblock Recurrent dynamics in the cerebral cortex: Integration of sensory
  evidence with stored knowledge.
\newblock Proceedings of the National Academy of Sciences.
  2021;118(33):e2101043118.

\bibitem{dehmer2020orbit}
Dehmer M, Chen Z, Emmert-Streib F, Mowshowitz A, Varmuza K, Feng L, et~al.
\newblock The orbit-polynomial: a novel measure of symmetry in networks.
\newblock IEEE access. 2020;8:36100--36112.

\bibitem{hagberg2020networkx}
Hagberg A, Conway D.
\newblock Networkx: Network analysis with python.
\newblock URL: https://networkx github io. 2020;.

\bibitem{wegner_subgraph_2014}
Wegner AE.
\newblock Subgraph covers: an information-theoretic approach to motif analysis
  in networks.
\newblock Physical Review X. 2014;4(4):041026.

\bibitem{wegner_atomic_2021}
Wegner AE, Olhede S.
\newblock Atomic subgraphs and the statistical mechanics of networks.
\newblock Physical Review E. 2021;103(4):042311.

\bibitem{wegner2024nonparametric}
Wegner AE, Olhede SC.
\newblock Nonparametric inference of higher order interaction patterns in
  networks.
\newblock arXiv preprint arXiv:240315635. 2024;.

\bibitem{liu_graph_2018}
Liu Y, Safavi T, Dighe A, Koutra D.
\newblock Graph summarization methods and applications: A survey.
\newblock ACM computing surveys (CSUR). 2018;51(3):1--34.

\bibitem{holder1994substucture}
Holder LB, Cook DJ, Djoko S, et~al.
\newblock Substucture Discovery in the SUBDUE System.
\newblock In: KDD workshop. Citeseer; 1994. p. 169--180.

\bibitem{koutra2014vog}
Koutra D, Kang U, Vreeken J, Faloutsos C.
\newblock Vog: Summarizing and understanding large graphs.
\newblock In: Proceedings of the 2014 SIAM international conference on data
  mining. SIAM; 2014. p. 91--99.

\bibitem{robins_introduction_2007}
Robins G, Pattison P, Kalish Y, Lusher D.
\newblock An introduction to exponential random graph (p*) models for social
  networks.
\newblock Social Networks. 2007;29(2):173--191.
\newblock doi:{10.1016/j.socnet.2006.08.002}.

\bibitem{lusher_exponential_2013}
Lusher D, Koskinen J, Robins G.
\newblock Exponential random graph models for social networks: Theory, methods,
  and applications.
\newblock Cambridge University Press; 2013.

\bibitem{schweinberger_instability_2011}
Schweinberger M.
\newblock Instability, {Sensitivity}, and {Degeneracy} of {Discrete}
  {Exponential} {Families}.
\newblock Journal of the American Statistical Association.
  2011;106(496):1361--1370.
\newblock doi:{10.1198/jasa.2011.tm10747}.

\bibitem{snijders_new_2006}
Snijders TA, Pattison PE, Robins GL, Handcock MS.
\newblock New specifications for exponential random graph models.
\newblock Sociological methodology. 2006;36(1):99--153.

\bibitem{schweinberger_local_2015}
Schweinberger M, Handcock MS.
\newblock Local dependence in random graph models: characterization, properties
  and statistical inference.
\newblock Journal of the Royal Statistical Society: Series B (Statistical
  Methodology). 2015;77(3):647--676.
\newblock doi:{10.1111/rssb.12081}.

\bibitem{pastor_epidemic_2015}
Pastor-Satorras R, Castellano C, Van~Mieghem P, Vespignani A.
\newblock Epidemic processes in complex networks.
\newblock Reviews of modern physics. 2015;87(3):925.

\bibitem{douglas_canonical_1989}
Douglas RJ, Martin KAC, Whitteridge D.
\newblock A {Canonical} {Microcircuit} for {Neocortex}.
\newblock Neural Computation. 1989;1(4):480--488.
\newblock doi:{10.1162/neco.1989.1.4.480}.

\bibitem{harris_neocortical_2015}
Harris KD, Shepherd GMG.
\newblock The neocortical circuit: themes and variations.
\newblock Nat Neurosci. 2015;18(2):170--181.
\newblock doi:{10.1038/nn.3917}.

\bibitem{sterling_principles_2015}
Sterling P, Laughlin S.
\newblock Principles of neural design.
\newblock MIT press; 2015.

\bibitem{bastos_canonical_2012}
Bastos AM, Usrey WM, Adams RA, Mangun GR, Fries P, Friston KJ.
\newblock Canonical {Microcircuits} for {Predictive} {Coding}.
\newblock Neuron. 2012;76(4):695--711.
\newblock doi:{10.1016/j.neuron.2012.10.038}.

\bibitem{mazurie_evolutionary_2005}
Mazurie A, Bottani S, Vergassola M.
\newblock An evolutionary and functional assessment of regulatory network
  motifs.
\newblock Genome Biology. 2005;6(4):R35.
\newblock doi:{10.1186/gb-2005-6-4-r35}.

\bibitem{jovanic_neural_2019}
Jovanic T, Winding M, Cardona A, Truman JW, Gershow M, Zlatic M.
\newblock Neural {Substrates} of {Drosophila} {Larval} {Anemotaxis}.
\newblock Current Biology. 2019;29(4):554--566.

\bibitem{zador_critique_2019}
Zador AM.
\newblock A critique of pure learning and what artificial neural networks can
  learn from animal brains.
\newblock Nat Commun. 2019;10(1):1--7.
\newblock doi:{10.1038/s41467-019-11786-6}.

\bibitem{koulakov_encoding_2021}
Koulakov A, Shuvaev S, Lachi D, Zador A.
\newblock Encoding innate ability through a genomic bottleneck.
\newblock BiorXiv. 2021; p. 2021--03.

\bibitem{elhesha_identification_2016}
Elhesha R, Kahveci T.
\newblock Identification of large disjoint motifs in biological networks.
\newblock BMC Bioinformatics. 2016;17(1):408.
\newblock doi:{10.1186/s12859-016-1271-7}.

\bibitem{peixoto2014hierarchical}
Peixoto TP.
\newblock Hierarchical block structures and high-resolution model selection in
  large networks.
\newblock Physical Review X. 2014;4(1):011047.

\bibitem{peixoto2017nonparametric}
Peixoto TP.
\newblock Nonparametric Bayesian inference of the microcanonical stochastic
  block model.
\newblock Physical Review E. 2017;95(1):012317.

\bibitem{hens2019spatiotemporal}
Hens C, Harush U, Haber S, Cohen R, Barzel B.
\newblock Spatiotemporal signal propagation in complex networks.
\newblock Nature Physics. 2019;15(4):403--412.

\bibitem{boguna2021network}
Boguna M, Bonamassa I, De~Domenico M, Havlin S, Krioukov D, Serrano M{\'A}.
\newblock Network geometry.
\newblock Nature Reviews Physics. 2021;3(2):114--135.

\bibitem{bianconi20215}
Bianconi G.
\newblock 5. Information theory of spatial network ensembles.
\newblock Handbook on Entropy, Complexity and Spatial Dynamics: A Rebirth of
  Theory? 2021; p.~61.

\bibitem{peel2022statistical}
Peel L, Peixoto TP, De~Domenico M.
\newblock Statistical inference links data and theory in network science.
\newblock Nature Communications. 2022;13(1):6794.

\bibitem{crane2018probabilistic}
Crane H.
\newblock Probabilistic foundations of statistical network analysis.
\newblock Chapman and Hall/CRC; 2018.

\bibitem{granovetter1976network}
Granovetter M.
\newblock Network sampling: Some first steps.
\newblock American journal of sociology. 1976;81(6):1287--1303.

\bibitem{achlioptas2009bias}
Achlioptas D, Clauset A, Kempe D, Moore C.
\newblock On the bias of traceroute sampling: or, power-law degree
  distributions in regular graphs.
\newblock Journal of the ACM (JACM). 2009;56(4):1--28.

\bibitem{genois2015compensating}
G{\'e}nois M, Vestergaard CL, Cattuto C, Barrat A.
\newblock Compensating for population sampling in simulations of epidemic
  spread on temporal contact networks.
\newblock Nature communications. 2015;6(1):8860.

\bibitem{morone2019symmetry}
Morone F, Makse HA.
\newblock Symmetry group factorization reveals the structure-function relation
  in the neural connectome of Caenorhabditis elegans.
\newblock Nature communications. 2019;10(1):4961.

\bibitem{lyzinski2016community}
Lyzinski V, Tang M, Athreya A, Park Y, Priebe CE.
\newblock Community detection and classification in hierarchical stochastic
  blockmodels.
\newblock IEEE Transactions on Network Science and Engineering.
  2016;4(1):13--26.

\bibitem{cranmer2020frontier}
Cranmer K, Brehmer J, Louppe G.
\newblock The frontier of simulation-based inference.
\newblock Proceedings of the National Academy of Sciences.
  2020;117(48):30055--30062.

\bibitem{goyal2023framework}
Goyal R, De~Gruttola V, Onnela JP.
\newblock Framework for converting mechanistic network models to probabilistic
  models.
\newblock Journal of Complex Networks. 2023;11(5):cnad034.

\bibitem{wernicke_fanmod_2006}
Wernicke S, Rasche F.
\newblock {FANMOD}: a tool for fast network motif detection.
\newblock Bioinformatics. 2006;22(9):1152--1153.
\newblock doi:{10.1093/bioinformatics/btl038}.

\bibitem{peixoto_nonparametric_2017}
Peixoto TP.
\newblock Nonparametric {Bayesian} inference of the microcanonical stochastic
  block model.
\newblock Phys Rev E. 2017;95(1):012317.
\newblock doi:{10.1103/PhysRevE.95.012317}.

\bibitem{bianconi_entropy_2009}
Bianconi G.
\newblock Entropy of network ensembles.
\newblock Phys Rev E. 2009;79(3):036114.
\newblock doi:{10.1103/PhysRevE.79.036114}.

\bibitem{johnson1975finding}
Johnson DB.
\newblock Finding all the elementary circuits of a directed graph.
\newblock SIAM Journal on Computing. 1975;4(1):77--84.

\bibitem{mckay2014practical}
McKay BD, Piperno A.
\newblock Practical graph isomorphism, II.
\newblock Journal of symbolic computation. 2014;60:94--112.

\end{thebibliography}

\clearpage
\appendix 
\renewcommand{\thesection}{}

\newgeometry{top=1in, bottom=1in, left=1.in, right=1.in}
\captionsetup{aboveskip=1pt,labelfont=bf,labelsep=period,justification=raggedright,singlelinecheck=off, margin={-0.5in,0in}}

{\huge \bf Supporting information}

\section*{S1 Text: Classic motif mining based on hypothesis testing}
\labelname{S1 Text}
\label{SIsec:hypothesis-testing}

The prevailing approach to network motif mining involves counting or estimating the frequency of each graphlet~\cite{przulj_biological_2007}, and comparing it to its frequency in random networks generated by a null model~\cite{milo_network_2002, fosdick_configuring_2018}.
Subgraphs that appear significantly more frequently in the empirical network than in the random networks are deemed motifs.
We here briefly describe the approach and discuss several of its main limitations.

\subsection*{Limitations}

Hypothesis-testing-based motif mining  suffers from several fundamental statistical limitations. Each of these problems can make such inferences statistically unreliable. 

\paragraph*{Gaussian assumption.}
First, motifs are inferred based either on a $Z$-test or on direct estimation of $p$ values from sampling of random networks. 
The former approach assumes Gaussian statistics under the null, which is often not a good approximation~\cite{fodor_intrinsic_2020}. 
In the latter approach, it is only possible to evaluate $p$-values that are larger than $1/M$ where $M$ is the number of random networks analyzed.
This is computationally expensive and precludes the evaluation of low $p$ values, which in turn makes it practically impossible to correct for multiple testing using standard approaches, such as the Bonferroni correction, which effectively decreases the significance threshold by a factor of the order of the number of tests. 

\paragraph*{Dependence on the choice of null model.}
Second, the appropriate null model is often not known~\cite{artzy-randrup_comment_2004, beber_artefacts_2012, orsini_quantifying_2015} or it may be computationally unfeasible to sample it~\cite{ginoza_network_2010, beber_artefacts_2012, orsini_quantifying_2015}. 
However, results may crucially depend on the choice of null model~\cite{artzy-randrup_comment_2004, beber_artefacts_2012}, potentially leading to the inference of spurious motifs (see Figs~3A--3D in the main text).

\paragraph*{Correlated motif counts.}
Third, the frequencies of different graphlets are not guaranteed to be independent, so one should account for these correlations when performing statistical testing~\cite{fodor_intrinsic_2020}.
Moreover, one should also account for these correlations in the null model to avoid inferring spurious motifs~\cite{milo_network_2002, stivala_testing_2021}.
Given a  graphlet $\alpha$, hypothesis-based motif inference qualifies $\alpha$ as a network motif in an empirical network $G$  if its frequency $f_\alpha(G)$ in $G$ is significantly greater than in an ensemble of random networks $\mathcal{G}_{\theta}$ sampled from a null model $P_\theta$. 

\subsection*{Method} 

For uniformly sampling simple random networks, we use the shuffling algorithms described in S5 Text. 
When the edge swapping procedures are ergodic and unbiased, they are guaranteed to uniformly sample the corresponding ensembles of random networks after a large enough number of swaps~\cite{fosdick_configuring_2018}.
However, the mixing time, i.e., the number of swaps needed for the generated networks to be practically independent, is not known in general~\cite{fosdick_configuring_2018}. 
To ensure that correlations between randomized networks are not likely to influence results (i.e., we try to favor hypothesis-testing based methods as much as possible), we perform $100 E$ successful edge swaps to generate each random network. 
This does not guarantee an absence of correlations, but we note that the number of swaps is larger than what is typically prescribed in the literature (for reference 0.2E edge-swaps were used to generate each random network in  \cite{milo_network_2002}, 3E in \cite{wernicke_fanmod_2006} and 6E in \cite{ribeiro_g-tries_2010}).

We utilize the typical normality assumption of the graphlet frequencies under the null and employ as test statistic the $Z$-score given by
\begin{align}
    Z_{\alpha,\theta}(G)   & = \frac{f_\alpha(G) - \mu_{\alpha,\theta}}{{\sigma}_{\alpha,\theta}} , 
\end{align}
where
\begin{align}
    \mu_{\alpha,\theta} & = \frac{1}{|\mathcal{G}_{\theta}|}  \sum_{G'\in\mathcal{G}_\theta} f_\alpha(G') 
\end{align}
and
\begin{align}
    {\sigma}^2_{\alpha,\theta} & = \frac{1}{|\mathcal{G}_{\theta}|-1}\sum_{G'\in\mathcal{G}_\theta} \left[ f_\alpha^2(G') - {\mu}^2_{\alpha,\theta} \right].
\end{align}
In all experiments, the size of $\mathcal{G}_\theta$ is set to 100 and the significance threshold (nominal alpha-level) is fixed at 0.01. 
To correct for multiple testing (one test for each graphlet), we employ a Bonferroni correction, which multiplies the raw $p$-values obtained directly from the $Z$-scores by $|\Gamma| \approx  10^4$.  
As displayed in Figs~3A--D, depending on the choice of the null, a considerable number of motifs can be falsely detected. 
A similar effect can also be seen in empirical data, where the number of motifs found varies enormously with the choice of null model (see S1 Fig), even though we corrected for multiple testing with the maximally conservative Bonferroni correction.
S1 Fig also demonstrates that the motifs found vary significantly depending on the null, and that the smallest number of motifs is not necessarily found under the most restricted null hypothesis. 

\clearpage
\section*{S2 Text: Computational time \& memory of the motif-based inference}
\labelname{S2 Text}
\label{SISec:memo-time-costs}
\setcounter{table}{0}

Any motif mining algorithm, depending on its parameters and on the size and density of the input graph, is likely to suffer from a high computational cost. The main practical limit for our method is the consumption of hard memory and RAM, while for hypothesis testing-based method the limiting factor is typically the time complexity. 
As developed in S3 Text and Algorithm 2 in the ``Methods'' section, induced subgraphs, which ranges from 3 to 5 nodes, are listed and saved on the hard memory. 
If the subgraph maximal size is decreased to 4, the memory cost is considerably reduced. 
To give a sense of the applicability of our method in practice, we review the time and memory costs for the motif-based inference in two settings: (i) The saved induced subgraphs' sizes ranges from 3 to 4 nodes (Table~\ref{SItab:memo-time-costs-4nodes}); the subgraph census and inference are performed on a cluster, using {4 CPUs}, with single-thread parallelization; this condition mimics what can be done locally, on a common laptop. {(ii)} The saved induced subgraphs' sizes ranges from {3 to 5} nodes (Table~\ref{SItab:memo-time-costs-5nodes}); the subgraph census and inference are performed on a cluster, using {10 CPUs}, with single-thread parallelization.

\begin{table}[ht!]
    \centering
    \begin{adjustwidth}{-.48in}{0.in}
    \begin{tabular}{|l|l|l|l|l|}
    \hline 
    \multicolumn{1}{|l|}{Species} & \multicolumn{1}{|l|}{Connectome} & 
    \multicolumn{1}{|l|}{Census---Memory (MB)} & \multicolumn{1}{|l|}{Census---Time (s)} & 
    \multicolumn{1}{|l|}{Inference---Time (s)} \\
    \thickhline
        \textit{C. elegans} & Hermaphrodite---nervous system & 25 & 1.049 & 8.918\\\hline
        \textit{C. elegans} & Hermaphrodite---whole animal & 44 & 1.028 & 14.971 \\\hline
        \textit{C. elegans} & Male---whole animal & 37 & 0.968 & 14.435  \\\hline
        \textit{Drosophila} &  Larva---Left MB & 98 & 1.843 & 9.94 \\\hline
        \textit{Drosophila} &  Larva---Right MB & 95 & 1.725 & 8.817 \\\hline
        \textit{Drosophila} &  Larva---Left \& right MBs & 713 & 8.288 & 54.774\\\hline
        \textit{C. intestinalis} & Larva---whole animal & 20 & 2.089 & 5.349  \\\hline
    \end{tabular}
    \end{adjustwidth}
    \caption{\small{\bf Time and space costs for the inference of motif sets, with motif sizes ranging from 3 to 4 nodes.} The memory required and the temporal duration of the subgraph census are measured for connectomes that are compressible using our motif-based scheme. Contrary to the hypothesis testing-based approaches, the subgraph census is performed only once, instead of hundreds  of times. However, a single subgraph census takes longer as the node labels of each induced subgraph are saved. In the case of small connectomes which have up to 400 neurons, the census can be handled by common laptops. Inference was repeated a hundred times, and we indicate in the rightmost column the duration corresponding to inference of the optimal model.}
    \label{SItab:memo-time-costs-4nodes}
\end{table}

\begin{table}[ht!]
    \centering
    \begin{adjustwidth}{-.48in}{0.in}
    \begin{tabular}{|l|l|l|l|l|}
    \hline
    \multicolumn{1}{|l|}{Species} & \multicolumn{1}{|l|}{Connectome} & 
    \multicolumn{1}{|l|}{Census---Memory (GB)} & \multicolumn{1}{|l|}{Census---Time (s)} & 
    \multicolumn{1}{|l|}{Inference---Time (s)} \\
    \thickhline
        \textit{C. elegans} & Hermaphrodite---nervous system & 0.93 & 120.982 & 403.47\\\hline 
        \textit{C. elegans} & Hermaphrodite---whole animal & 1.8 & 70.739 & 339.03 \\\hline 
        \textit{C. elegans} & Male---whole animal & 1.3 & 120.982 & 271.97  \\\hline 
        \textit{Drosophila} &  Larva---Left MB & 3.7 & 138.685 & 590.75 \\\hline 
        \textit{Drosophila} &  Larva---Right MB & 3.8 & 129.007 & 673.64 \\\hline 
        \textit{Drosophila} &  Larva---Left \& right MBs & 50 & 1424.428 & 3 468.60\\\hline 
        \textit{C. intestinalis} & Larva---whole animal & 0.044 & 17.764 & 117.61  \\\hline
        \multicolumn{1}{|l}{} &
        \multicolumn{1}{l}{} &
        \multicolumn{1}{l}{} & 
        \multicolumn{1}{l}{} &
        \multicolumn{1}{l|}{}
         \\\hline
        \textit{Drosophila} & Larva---whole brain & 21 (1\%) & 435.504 & 4 604.81 \\\hline     
        \textit{Drosophila} & Adult---right MB & 1300 (1\%) & 3 501.565 & 15 039.00\\\hline 
    \end{tabular}
    \end{adjustwidth}
    \caption{\small{\bf Time and space costs for the inference of motif sets, with motif sizes ranging from 3 to 5 nodes.} The memory required and the duration of the subgraph census are measured for connectomes that are compressible using our motif-based scheme. With such large volumes of subgraph files, common laptops cannot handle the computational task. In this setting, we recommend storing subgraphs and running the inference algorithm on an HPC cluster. The last two lines confirm even more the amplitude of the demanding resources. For the two largest connectomes, we performed a stochastic subgraph census, where only 1\% of the 5-node subgraphs are saved and only 10\% of the 4-node subgraphs---overall, approximately 1\% of the total number of subgraphs. Even if we remove the 1\% of the 5-node subgraphs, the inference time complexity is still too great for common laptops.}   
    \label{SItab:memo-time-costs-5nodes}
\end{table}

\section*{S3 Text: Subgraph census: dealing with lists of induced subgraphs}
\labelname{S3 Text}
\label{SIsec:subgraph-census}

\subsection*{Writing graphlet occurrences}

Subgraph-census is a computationally hard task since it involves repeatedly solving the subgraph isomorphism problem. 
Since our algorithm uses not only the number of occurrences of each graphlet but also their placement in the original graph, we need not only to determine their number of occurrences but to list all weakly connected subgraphs from three to five nodes. 
There are about 10\,000 distinct five-node graphlets, thus their frequencies can easily be stored on any modern laptop. 
However, the exhaustive lists of all graphlet occurrences can be very large, depending on the size and density of the network at hand. 
In the case of the brain regions of the adult \textit{Drosophila melanogaster}, the magnitude of such connectomes is of the order of a thousand neurons, which, for their specific density, leads to at least several billion five-node subgraphs. 
In this case it is not possible to dynamically store all subgraph occurrences. 
Instead, we progressively write them to disk directly using textfile pointers thanks to the \texttt{ifstream} object of the C++ standard library (\href{https://cplusplus.com/reference/fstream/ifstream/}{cplusplus.com/reference/fstream/ifstream}). 
Each text file corresponds to a graphlet, containing isomorphic subgraphs divided on each line, with grouped node labels stored in CSV format. 

\subsection*{Reading graphlet occurrences}

Uniformly sampling subgraphs is part of our stochastic greedy algorithm. Randomizing the collected subgraph lists followed by sequential reading of their elements would be the most direct and simple approach to do this. 
Since it is not possible to store every induced subgraph in memory, we store the pointer positions of every line (i.e., every subgraph memory address) in a vector. 
These vector elements are shuffled and then read sequentially to perform a uniform sampling of the subgraphs. 
The memory gain, per graphlet textfile, is of the order of the textfile size times the graphlet size. 
In large connectomes, this memory gain is however not enough and it is not possible to store all subgraph pointers in memory. 
For this case, we implemented a procedure that divides the reading and shuffling of a graphlet textfile in chunks of a fixed size. 
If the number of subgraphs in a graphlet textfile (i.e., its number of lines) is lower than the chunk size, then the sampling is performed as described above. 
When the graphlet textfile is larger than the chunk size, the subgraph sampling is not uniform. Indeed, the initial node labeling of the input graph, together with the order of the subgraph mining imposed by Wernicke's algorithm, requires having access to all the listed subgraphs for a sampling to be exactly uniform. 
The larger the chunk reading size is, the less biased the sampling will be. 
For all the adult \textit{Drosophila melanogaster} connectomes, we fixed the chunk size to a million subgraphs, so that the maximum number of stored subgraph pointers in RAM per graphlet is a million. 
When all the subgraphs of the current chunk of the full graphlet textfile are forbidden by the non-overlapping supernode constraint (see Algorithm 2 in the ``Methods'' section), another chunk is read. 

\clearpage
\section*{S4 Text: Graph codelengths and subgraph contraction costs}
\labelname{S4 Text}
\label{SIsec:codelengths}
\setcounter{table}{0}

\section*{Graph codelengths}

In this supplementary note, we review the dyadic graph models used as base and null models, namely the Erd\H{o}s-Rényi model (ER), the configuration model (CM), the reciprocal Erd\H{o}s-Rényi model (RER), and the reciprocal configuration model (RCM).
Model parameters, as well as and their relationships, are detailed throughout the note, and summarized in Table~\ref{SItab:parameters}.

\begin{table}[h!]
    \centering
    \begin{adjustwidth}{.19in}{0.in}
    \caption{\small \bf Dyadic graph model parameters and simple identities.}
    \label{SItab:parameters}
    \end{adjustwidth}
    \begin{adjustwidth}{-.5in}{0.in}
        \begin{tabular}{|l|l|l|}
        \thickhline
        \multicolumn{1}{|l|}{\bf Network feature} & \multicolumn{1}{|l|}{\bf Notation} & \multicolumn{1}{|l|}{\bf Expression} \\\thickhline
            \rule{0pt}{0.4cm}Number of directed edges & $\Enumber$ & $\sum_{ij} A_{ij} = \sum_i \outdegree_i = \sum_i\indegree_i = \Ednumber + 2\Emnumber$ \\\hline
            \rule{0pt}{0.4cm}Out-degree of node $i$ &  $\outdegree_i$ & $\sum_j A_{ij}$ \\\hline
            \rule{0pt}{0.4cm}In-degree of node $i$ &  $\indegree_i$ & $\sum_j A_{ji}$ \\\hline
            \rule{0pt}{0.4cm}Number of reciprocal ({m}utual) edges & $\Emnumber$ & $1/2\sum_{i<j}(A_{ij} - A_{ji} + |A_{ij} -A_{ji}|) = 1/2 \sum_i\mutualdegreeR_i$ \\\hline
            \rule{0pt}{0.4cm}Number of non-reciprocated ({d}irected) edges & $\Ednumber$&  $1/2\sum_{ij}(A_{ij} + A_{ji} - |A_{ij} -A_{ji}|) = \sum_i \outdegreeR_i = \sum_i \indegreeR_i$ \\\hline
            \rule{0pt}{0.4cm}Reciprocal ({m}utual) degree of node $i$ &  $\mutualdegreeR_i$ & $1/2\sum_{j}(A_{ij} + A_{ji} - |A_{ij} - A_{ji}|)$ \\\hline
            \rule{0pt}{0.4cm}Non-reciprocated out-degree of node $i$ & $\outdegreeR_i$ & $1/2\sum_j(A_{ij} - A_{ji} + |A_{ij} - A_{ji}|) = \outdegree_i - \mutualdegreeR_i$ \\\hline 
            \rule{0pt}{0.4cm}Non-reciprocated in-degree of node $i$ & $\indegreeR_i$ & $1/2\sum_j(A_{ji} - A_{ij} + |A_{ji} - A_{ij}|) = \indegree_i - \mutualdegreeR_i$\\\hline
        \end{tabular}
    \end{adjustwidth}
\end{table}

\subsection*{Erd\H{o}s-R\'enyi model (ER)}

\paragraph{Multigraphs.} The microcanonical Erd\H{o}s-R\'enyi (ER) model encodes a multigraph $G$ with a fixed number of nodes, $N$, and edges, $E$. 
The microcanonical probability distribution over the space of directed loop-free multigraphs is given by~\cite{peixoto_nonparametric_2017} 
\begin{equation}\label{eq:P_ER}
	P_{(N,E)}(G) = \frac{E!}{\prod_i\prod_{j\neq i} A_{ij}!} \, [N(N-1)]^{-E} . 
\end{equation}
The second factor in Eq.~\eqref{eq:P_ER} is the number of ways to place each edge between the $N(N-1)$ pairs of nodes, and the first factor accounts for the indistinguishability of the ordering of the multiedges. 
This leads to an entropy of
\begin{equation}\label{eq:S_ER}
    \entropy{\thetaER}(H) = E\log[N(N-1)]  - \log E! + \sum_i \sum_{j\neq i} \log A_{ij}! .
\end{equation}
\paragraph{Simple graphs.} The entropy of the simple, directed Erd\H{o}s-R\'{e}nyi model is found by counting the number of ways to place $E$ edges amongst $N(N-1)$ pairs of nodes without overlap. 
This leads to
\begin{equation}
    \label{eq:S_ER_simple_dir}
	\entropy{\thetaER}(G) = \log\binom{N(N-1)}{E} = \log \frac{[N(N-1)]!}{[N(N-1)-E]! E!}.
\end{equation}	   
\paragraph{Model complexity.} The parametric complexity of the ER model requires encoding the two positive integers that are its parameters. 
We make use of the standard codelength function for encoding natural numbers~\cite{grunwald_minimum_2020}.
To a positive integer $n\in\mathbb{N}$, we attribute the cost 
\begin{equation}
    \integerCodelength{n} = \log n(n+1) . 
\end{equation}
This leads to a codelength for describing $\thetaER$ of
\begin{equation}\label{eq:model_complexity_ER}
    \codelength{}\thetaER = \integerCodelength{N} + \integerCodelength{E}.
\end{equation}

\subsection*{Configuration model (CM)}

\paragraph{Multigraphs.} The configuration model (CM) generates random networks with fixed in- and out-degrees of each node, i.e., the sequences $\outdegrees =(\outdegree_i)$ and $\indegrees = (\indegree_i)$. 
The in-degree corresponds to the number of edges pointing towards the node, $k_i^- = \sum_j A_{ji}$, whereas the out-degree is the number of edges originating at the node, $k_i^+=\sum_j A_{ij}$.
The entropy of the configuration model is given by~\cite{fosdick_configuring_2018}
\begin{equation}\label{eq:S_Conf}
    \entropy{\thetaConf}(G) = \log E! - \sum_i \left( \log\outdegree_i! + \log\indegree_i! - \sum_{j\neq i} A_{ij}! \right) .
\end{equation}
\paragraph{Simple graphs.} There are no exact closed-form expressions for the microcanonical entropy of the configuration model for simple graphs. We thus use the approximation developed in~\cite{bianconi_entropy_2009}, which provides a good approximation for sparse graphs,
\begin{equation}\label{eq:S_conf_simple_dir}
    \entropy{\thetaConf}(G) \approx \log \frac{E!}{\prod_i k_i^+!k_i^-!} - \frac{1}{2\ln 2}\frac{\langle {k^+_i}^2 \rangle \langle {k^-_i}^2 \rangle}{\langle k^+_i \rangle \langle k^-_i \rangle} .
\end{equation}

\paragraph{Model complexity.} Contrary to the Erd\H{o}s-R\'enyi model, the configuration model is a \textit{microscopic} description in the sense that it introduces two parameters per node (in addition to the number of nodes $\Vnumber$) and thus a total of $2N+1$ parameters (as compared to 2 parameters for the ER model).
Thus, while its entropy is always smaller than that of the ER model, its parametric complexity is higher. 
We consider two possible ways to encode the degree sequences $\degrees^+$ and $\degrees^-$.
The simplest and most direct approach to encode a sequence $\degrees$ is to consider each element individually as a priori uniformly distributed in the interval of integers between $\mindegree = \min\{\degree_i\in\degrees\}$ and $\maxdegree = \max\{\degree_i\in\degrees\}$. 
This leads to a codelength of
\begin{equation}
    \codelength{U}(\degrees) = N\log(\maxdegree - \mindegree + 1) + \integerCodelength{\mindegree} + \integerCodelength{\maxdegree} .
\end{equation}
Assuming that the degrees are generated according to the same unknown probability distribution, it is typically more efficient to use a so called \textit{plug-in} code \cite{grunwald_minimum_2007, bloem_large-scale_2020}, which describes them as sampled from a Dirichlet-multinomial distribution over the integers between $\mindegree$ and  $\maxdegree$. 
To each possible value  $\mindegree \leq \mu \leq \maxdegree$ that a degree may take, we calculate the frequency $\freq_\mu$ of the value $\mu$ in $\degrees$. We then have
\begin{equation}\label{eq:plug-in}
    P_{\boldsymbol{\lambda}}(\degrees) = \frac{\Gamma(\Lambda)}{\Gamma(N+\Lambda)}\prod_{\mindegree\leq\mu\leq\maxdegree} \frac{\Gamma(\freq_\mu +\lambda_\mu)}{\Gamma(\lambda_\mu)} ,
\end{equation}
where $\lambda_\mu$ are prior hyperparameters and 
$\Lambda = \sum_\mu \lambda_\mu$.
When all $\lambda_\mu  = \lambda = 1$, the priors are uniform probability distributions, while the case $\lambda_\mu = \frac{1}{2}$ corresponds to the Jeffreys prior~\cite{grunwald_minimum_2007}. 
The plug-in codelength is thus given by
\begin{equation}
    \codelength{\lambda}(\degrees) = -\log P_\lambda(\degrees) + \integerCodelength{\mindegree} + \integerCodelength{\maxdegree} .
\end{equation} 
In the implementation of our algorithm, we select the encoding of the degree sequences among $\codelength{U}(\degrees)$, $\codelength{\lambda=1}(\degrees)$ and $\codelength{\lambda=1/2}(\degrees)$ that results in the minimal codelength.
Encoding this choice takes $\log3$ bits. 
Including also the encoding of the number of nodes, $N$, the cost of encoding a degree sequence is
\begin{equation}\label{eq:sequence-codelength}
    L_{\mathrm{seq}}(\degrees) = \min\{L_U(\degrees),L_{\lambda=1}(\degrees),L_{\lambda=1/2}(\degrees)\} + \log 3 + \integerCodelength{N} .
\end{equation}
Finally, the total parametric codelength of the configuration model is the sum of two sequence codelengths, leading to
\begin{equation}
    \codelength{\mathrm{seq}}\thetaConf = \codelength{\mathrm{seq}}(\outdegrees) + \codelength{\mathrm{seq}}(\indegrees) .
\end{equation}

\subsection*{Reciprocal models}

Reciprocated (or \textit{mutual}) edges are an important feature of many biological networks~\cite{jovanic_competitive_2016,winding_connectome_2023,gilbert2013top,bahl2020neural,jarrell2012connectome}.
Reciprocal edges confer to a network a partially symmetric structure. 
If they represent an important fraction of the total number of edges, this regularity can be used to significantly compress the network. 
To account for reciprocal edges in a simple manner, we consider them as a different edge type that are placed independently of directed edges. 
Thus, we model a multigraph $G$ as the overlay of independent symmetric and asymmetric multigraphs, $G^\sym$ and $G^\asym$, respectively, where $G^\sym$ is an undirected multigraph and $G^\asym$ is a directed multigraph.
The adjacency matrix of $G$ is given by  $\A(G) = \A(G^\sym) + \A(G^\asym)$, and a reciprocal model's likelihood is equal to the product of the likelihoods of the symmetric and asymmetric parts, leading to a codelength of
\begin{equation}\label{eq:reciprocal-codelength}
    \codelength{}(G,\phi) = \codelength{}(G^\sym,\phi^\sym) + \codelength{}(G^\asym,\phi^\asym) ,
\end{equation}
where $\phi = (\phi^\sym,\phi^\asym)$ and $\phi^\sym$ and $\phi^\asym$ are the parameters of the models used to encode the symmetric and asymmetric edges of $G$, respectively.
In practice, we set for each pair $(i,j)\in\Vset(G)\times\Vset(G)$ the entries of the symmetric and asymmetric adjacency matrices to be
\begin{align}
    A_{ij}^\asym &= \max(A_{ij}-A_{ji},0)  = \frac{1}{2}(A_{ij}-A_{ji} + |A_{ij}-A_{ji}|) ,\\
    A_{ij}^\sym  &= \min(A_{ij},A_{ji}) \ \ \ \ \ \ = \frac{1}{2}(A_{ij}+A_{ji} - |A_{ij}-A_{ji}|) .
\end{align}
This maximizes the number of edges in the symmetric representation, which minimizes the codelength since the entropy of an undirected model is lower than its directed counterpart and since each reciprocal edge encoded in $G^\sym$ corresponds to two directed edges. 

\subsection*{Reciprocal Erd\H{o}s-R\'enyi model (RER)}

\paragraph{Multigraphs.} The reciprocal version of the Erd\H{o}s-R\'{e}nyi model (RER) has 3 parameters, $\thetaRER$, where $\Emnumber$ is the number of reciprocal (mutual) edges and $\Ednumber$ is the number of directed edges, and we have $\Enumber = 2\Emnumber + \Ednumber$.
The model's codelength is
\begin{equation} \label{eq:L_RER}
	\codelength{}(G,\thetaRER) = \entropy{(\Vnumber,\Ednumber)}(G^\asym) + \entropy{(\Vnumber,\Emnumber)}(G^\sym) + \codelength{}\thetaRER .
\end{equation}
The entropy of the directed graph model, $\entropy{(\Vnumber,\Ednumber)}(G^\asym)$, is given by Eq.~\eqref{eq:S_ER} with $\Enumber$ replaced by $\Ednumber$. 
The entropy of the symmetric part is given by~\cite{peixoto_nonparametric_2017} 
\begin{equation}
    \label{eq:S_ER_undir}
	\entropy{(\Vnumber,\Emnumber)}(G^\sym) = \Emnumber\log[\Vnumber(\Vnumber-1)/2] - \log \Emnumber! + \sum_i \sum_{i<j} \log A_{ij}^\sym! ,
\end{equation}	

\paragraph{Simple graphs.} Contrary to multigraphs, the placement of directed and reciprocal edges is not entirely independent for simple graphs since we do not allow the edges to overlap. 
However, we can model the placement of one type of edges (say reciprocal edges) as being entirely random and the second type (e.g., directed) as being placed randomly between the pairs of nodes not already covered by the first type. 
This leads to a number of possible configurations of 
\begin{equation}
    \label{eq:No_ER_simple_reciprocal}
    \partitionFunction{\thetaRER} = \binom{\Vnumber(\Vnumber-1)/2}{\Emnumber} 
    \binom{\Vnumber(\Vnumber-1)/2 - \Emnumber}{\Ednumber} ,
    2^{\Ednumber} ,
\end{equation}	
where the first factor is the number of ways to place the reciprocal edges, the second factor is the number of ways to place the directed edges amongst the remaining node pairs without accounting for their direction, and the third factor is the number of ways to orient the directed edges. 
Simplifying and taking the logarithm yields the following expression for the entropy of the reciprocal ER model,
\begin{equation}
    \label{eq:S_ER_simple_reciprocal}
    \entropy{\thetaRER}(G) = 
    \log \frac{[\Vnumber(\Vnumber-1)/2]!}{[\Vnumber(\Vnumber-1)/2 - \Emnumber - \Ednumber]! \Emnumber! \Ednumber!} + \Ednumber .
\end{equation}

\paragraph{Model complexity.} The RER model's parametric complexity is equal to
\begin{equation}
   \codelength{}\thetaRER 
   = \integerCodelength{N} + \integerCodelength{\Ednumber} + \integerCodelength{\Emnumber}.
\end{equation}

\subsection*{Reciprocal configuration model (RCM)} 

\paragraph{Multigraphs.} 
Similarly to the ER model, we extend the configuration model to a reciprocal version (RCM) by introducing a third degree sequence, describing each node's \textit{mutual} degree, defined as the number of reciprocal edges it partakes in. 
The model is thus defined by the set of parameters $\thetaRConf$ where 
$\mutualdegreeR_i = \sum_j A_{ij}(G^\sym)$ is the mutual degree of node $i$,
$\outdegreeR_i = \sum_j A_{ij}(G^\asym)$ is the non-reciprocated out-degree, and 
$\indegreeR_i = \sum_j A_{ji}(G^\asym)$
is the non-reciprocated in-degree. 
The codelength of the reciprocal configuration model is equal to 
\begin{equation}\label{eq:codelength_Rconf} 
    \codelength{}(G,\thetaRConf) = 
    \entropy{(\outdegreesR,\indegreesR)}(G^\asym) + \entropy{\mutualdegreesR}(G^\sym) +
    \codelength{}\thetaRConf .
\end{equation}
The entropy of the asymmetric graph is given by Eq.~\eqref{eq:S_Conf} with $\thetaConf$ replaced by $(\outdegreesR,\indegreesR)$. 
The entropy of the symmetric graph is given by~\cite{fosdick_configuring_2018}
\begin{equation}\label{eq:S_conf_undirected}
    \entropy{\mutualdegreesR}(G^\sym) = \log(2\Emnumber)! - \log(2\Emnumber)!! 
    - \sum_i \left( \log\mutualdegreeR_i! - \sum_{j\neq i} A_{ij}^\sym! \right) .
\end{equation}

\paragraph{Simple graphs.} To derive an approximation for the entropy of the reciprocal configuration model for simple graphs, we follow the same approach as in~\cite{bianconi_entropy_2009} but with the three-degree sequences $\thetaRConf$ constrained instead of only two
(see the subsequent Section for a detailed derivation).
This leads to a microcanonical entropy of
\begin{align}
    \entropy{\thetaRConf}(G) &\approx \log \frac{(2\Emnumber)!!}{\prod_i \mutualdegreeR_i!} + \log \frac{\Ednumber!}{\prod_i\outdegreeR_i!\indegreeR_i!} \\
    & - \frac{1}{2\ln 2} \left( 
    \frac{1}{2}\frac{\E {\pare{\mutualdegreeR_i}^2}^2}{\E{\mutualdegreeR_i}^2} 
    + \frac{\E{{\outdegreeR_i}^2} \E{{\indegreeR_i}^2}}{\E{\outdegreeR_i} \E{\indegreeR_i}} 
    + \frac{\E{\outdegreeR_i\indegreeR_i}^2}{\E{\outdegreeR_i}\E{\indegreeR_i}} 
    + \frac{\E{\mutualdegreeR_i\outdegreeR_i}\E{\mutualdegreeR_i\indegreeR_i}}{\E{\mutualdegreeR_i}\E{ \kappa_i^+}} \right)  . \nonumber
\end{align}

\paragraph{Model complexity.} The parametric part of the codelength is equal to
\begin{equation}
    \codelength{}\thetaRConf = \codelength{\mathrm{seq}}(\outdegreesR) + \codelength{\mathrm{seq}}(\indegreesR) + \codelength{\mathrm{seq}}(\mutualdegreesR) ,
\end{equation}
with $\codelength{mathrm{seq}}$ given by Eq.~(\ref{eq:sequence-codelength}).

\section*{Entropy of the simple reciprocal configuration model}
\label{SIsec:simple-reciprocal-configuration-model}

To derive the entropy of the reciprocal configuration model for simple graphs, we follow the approach developed in \cite{bianconi_entropy_2009}. Compared to the directed configuration model, the reciprocal version has the mutual degree sequence, i.e., the number of mutual stubs (corresponding to reciprocal edges) per node, as an additional set of parameters. 
We let $\mathbf{u}$ denote a vector of size $N$ filled with ones, and we define the microcanonical partition function as a sum over  a product of Dirac delta functions which define the constrained parameter values of the model, 
\begin{equation}
        \Omega\thetaRConf = \sum_\mathbf{A}\delta\left(\mutualdegreesR - \vdiag\left(\mathbf{AA}^T\right)\right)
        \delta\left( \outdegreesR - \mathbf{Au} + \vdiag\left(\mathbf{AA}^T\right) \right)
        \delta\left( \indegreesR - \mathbf{A}^T\mathbf{u} + \vdiag\left(\mathbf{AA}^T\right) \right) ,
\end{equation}
where $\vdiag$ is a function that maps the diagonal elements of a $N\times N$ matrix to a $N$-dimensional vector. 
The Dirac delta functions can be expanded in terms of Fourier integrals to obtain 
\begin{equation}
    \Omega\thetaRConf
        = \int \frac{\text{d}\boldsymbol{\lambda}^+\text{d}\boldsymbol{\lambda}^-\text{d}\boldsymbol{\mu}}{(2\pi)^{3N}} \ e^{{\boldsymbol{\lambda}^+}^T\outdegreesR + {\boldsymbol{\lambda}^-}^T\indegreesR + \boldsymbol{\mu}^T\mutualdegreesR} \sum_{\mathbf{A}} e^{-{\boldsymbol{\lambda}^+}^T \mathbf{Au}-{\boldsymbol{\lambda}^-}^T \mathbf{A}^T\mathbf{u}-{\boldsymbol{\mu}}^T \mathrm{diag}\left(\mathbf{AA}^T\right)} ,
\end{equation}
where $\boldsymbol{\lambda}^+,\boldsymbol{\lambda}^-,\boldsymbol{\mu}$ are identified as vectors of Lagrange multipliers. 
Compared to the classical configuration model, symmetric pairs of elements of the adjacency matrix are not independent and need to be considered simultaneously, such that
\begin{align}
    \nonumber\sum_{\mathbf{A}} e^{-{\boldsymbol{\lambda}^+}^T \mathbf{Au}-{\boldsymbol{\lambda}^-}^T \mathbf{A}^T\mathbf{u}-{\boldsymbol{\mu}}^T \mathrm{diag}\left(\mathbf{AA}^T\right)} &= \sum_{\{A_{ij},A_{ji}\} = \{\{0,0\},\{1,1\},\{0,1\},\{1,0\}\}} e^{-\sum_{ij}(\mu_i - \lambda_i^+ - \lambda_i^-) A_{ij}A_{ji}}  e^{-\sum_{ij}(\lambda_i^++\lambda_j^-)A_{ij}}\\
        &=\prod_{i<j}\left(1 + e^{-\mu_i -\mu_j} + e^{-\lambda_i^+-\lambda_j^-} + e^{-\lambda_i^--\lambda_j^+}\right) .
\end{align}
For the sake of readability, we set $s_{ij} \equiv e^{-\mu_i -\mu_j} + e^{-\lambda_i^+-\lambda_j^-} + e^{-\lambda_j^+-\lambda_i^-} $.
To estimate $\Omega\thetaRConf$, we apply a Laplace approximation to its Fourier integral form, and we thus we seek to maximize the following quantity
\begin{equation}
    NQ(\boldsymbol{\mu},\boldsymbol{\lambda}^+,\boldsymbol{\lambda}^-|  \mutualdegreesR,\outdegreesR,\indegreesR) = {\boldsymbol{\mu}}^T\mutualdegreesR +{\boldsymbol{\lambda}^+}^T\outdegreesR + {\boldsymbol{\lambda}^-}^T\indegreesR  + \sum_{i<j}\ln (1+s_{ij}) .
\end{equation}
The saddle point equations to be solved are thus
\begin{align}
    \frac{\partial Q}{\partial \mu_i} = 0 & \Leftrightarrow \mutualdegreeR_i = e^{-\mu_i}\sum_{j\neq i}\frac{e^{-\mu_j}}{1+s_{ij}} ,\\ 
    \frac{\partial Q}{\partial \lambda_i^+} = 0 & \Leftrightarrow \outdegreeR_i = e^{-\lambda_i^+}\sum_{j\neq i}\frac{e^{-\lambda_j^-}}{1+s_{ij}} ,\\ 
    \frac{\partial Q}{\partial \lambda_i^-} = 0 & \Leftrightarrow \indegreeR_i = e^{-\lambda_i^-}\sum_{j\neq i}\frac{e^{-\lambda_j^+}}{1+s_{ij}} .
\end{align}
Here, we are only interested in the sparse graph approximation. The computation of the Hessian would be compulsory in a rigorous calculation. However, when actually evaluated, it only leads to sums of logarithmic terms in the degree sequences elements, which are negligible compared to the sums of log factorial terms in the degree sequence elements (when the graph is of finite size). 
This is the same observation that is found in \cite{bianconi_entropy_2009}, while never explicitly justified there. 
For our results to be consistent with the standard form of other simple network entropy expressions, we also choose to neglect the contribution of the Hessian to the sparse graph approximation of the microcanonical partition function. 
At large $N$, it is reasonable to assume $s_{ij} = o(1)$ such that the right hand terms of the saddle point equations are finite. This leads to the simplified equations
\begin{align}
    \mutualdegreeR_i &= K_m e^{-\mu_i}, & K_m = \sum_j e^{-\mu_j} , \\
    \outdegreeR_i &= K_- e^{-\lambda^+_i}, & K_- = \sum_j e^{-\lambda_j^-} ,\\
    \indegreeR_i &= K_+ e^{-\lambda^-_i}, & K_+ = \sum_j e^{-\lambda_j^+} .
\end{align}
The constants $K_m,K_+,K_-$ are determined by global structural constraints, which are the number of asymmetrically connected (or directed) pairs of nodes and the number of mutual edges:
\begin{align}
    2\Emnumber &= \sum_i \mutualdegreeR_i = K_m^2 \Leftrightarrow K_m = \sqrt{2\Emnumber} ,\\ 
    \Ednumber  &= \sum_i \outdegreeR_i = K_+K_- \Leftrightarrow K_+ = K_- = \sqrt{\Ednumber} .
\end{align}
All is now set for a second-order estimation of the microcanonical partition function in $s_{ij}$. 
We have
\begin{align}
    \boldsymbol{\mu}^T\mutualdegreesR 
    &= -\sum_i \mutualdegreeR_i \ln \mutualdegreeR_i + \Emnumber\ln(2\Emnumber) \approx \ln\frac{(2\Emnumber)!!}{\prod_i \mutualdegreeR_i!} - \Emnumber , 
\end{align}
\begin{align}
    {\boldsymbol{\lambda}^+}^T\outdegreesR + {\boldsymbol{\lambda}^-}^T\indegreesR 
    &=  -\sum_i \outdegreeR_i\ln\outdegreeR_i - \sum_i \indegreeR_i\ln\indegreeR_i + \Ednumber\ln\Ednumber \approx \ln\frac{\Ednumber!}{\prod_i\outdegreeR_i!\indegreeR_i!} - \Ednumber ,
\end{align}
\begin{align}
    s_{ij} &= \frac{\mutualdegreeR_i\mutualdegreeR_j}{2\Emnumber} + \frac{\outdegreeR_i\indegreeR_j + \indegreeR_i\outdegreeR_j }{\Ednumber} ,
\end{align}
and 
\begin{align}
    \frac{1}{2}\sum_{ij} \left( s_{ij} - \frac{s_{ij}^2}{2} \right) 
    & = \Emnumber + \Ednumber - \frac{1}{2}\underbrace{\left( \frac{1}{2}\frac{\langle {\mutualdegreeR_i}^2\rangle^2}{\langle \mutualdegreeR_i \rangle^2} + \frac{\langle {\outdegreeR_i}^2 \rangle\langle{\indegreeR_i}^2\rangle}{\langle\outdegreeR_i\rangle\langle\indegreeR_i\rangle} + \frac{\langle\outdegreeR_i\indegreeR_i\rangle^2}{\langle\outdegreeR_i\rangle\langle\indegreeR_i\rangle} + \frac{\langle \mutualdegreeR_i\outdegreeR_i\rangle \langle\mutualdegreeR_i\indegreeR_i\rangle} {\langle\mutualdegreeR_i\rangle \langle\outdegreeR_i\rangle}\right)}_{\Psi\thetaRConf} . 
\end{align}
Putting it all together we obtain 
\begin{equation}
    L_\thetaRConf(G) = \log \frac{(2\Emnumber)!!}{\prod_i \mutualdegreeR_i!} + \log \frac{\Ednumber!}{\prod_i\outdegreeR_i!\indegreeR_i!} - \frac{1}{2\ln 2}\Psi\thetaRConf .
\end{equation}
%
Similar to the sparse approximation of the entropy of the simple graph configuration model~\cite{bianconi_entropy_2009}, we see that the entropy for the simple graph reciprocal configuration model amounts to the multigraph codelength from which is subtracted a functional cut-off that depends on the statistics of the degree sequences.

\section*{Subgraph contraction costs}\label{SIsec:contraction-cost}

We give in this note closed-form expressions for the putative difference in codelength which would be obtained by contracting a given subgraph in the reduced graph $H_t = (\Vset_t,\Eset_t)$.
These are the expressions we use in practice in our greedy algorithm to select the most compressing subgraph at each iteration.
The codelength difference for a subgraph contraction depends on the base model used to encode $H_t$, so we give below expressions for each of the four base models.
We will for notational convenience drop the subscripts pertaining to the iteration $t$ and thus simply refer to $H=(\Vset,\Eset)$ in the remainder of this section.

\subsection*{Common quantities}

In all steps, base models share a cost difference related to the transformation of a group of entries in the adjacency matrix $\mathbf{A} = \mathbf{A}(H)$ from $[A_{ij}]$ to $[A_{ij}']$ caused by the contraction of a given subgraph $\subgraph = (\vset,\eset) \in H$. 
This difference is given by
\begin{align} \label{eq:dL-common}
    \ell_\mathbf{A}(\subgraph) &\ = \sum_{i\in \vset}\log\left[\frac{ k^+_i(\subgraph)!}{\prod_{j\in\subgraphNeighborhood}A_{ij}!} \times \frac{k^-_i(\subgraph)!}{\prod_{j\in\subgraphNeighborhood}A_{ji}!}\right] ,
\end{align}
where the subgraph's neighborhood $\subgraphNeighborhood$ is the set of nodes in $H$ that are connected to a node of $\subgraph$, and
\begin{align}
    k^+_i(\subgraph) &\ = \sum_{j\in\subgraphNeighborhood} A_{ij} ,\\ 
    k^-_i(\subgraph) &\ = \sum_{j\in\subgraphNeighborhood} A_{ji}.\\
\end{align}
%
In the following, we will denote by $n=|\nu|$ the subgraph size and by $e=|\epsilon|$ the number of edges $s$ holds. We furthermore denote by $e_d=\frac{1}{2}\sum_{i,j\in\nu}|A_{ij}-A_{ji}|$ the number of directed (asymmetric) edges and by $e_m = \frac{1}{2}(e - e_d)$ the number of mutual edges of $s$. 
The number of nodes in $H$ is denoted $N$, the number of edges $\Enumber$, the number of mutual edges $\Emnumber$, and the number of (asymmetric) directed edges $\Ednumber$.

\subsection*{Base model complexity cost} 
In the inference of our model, the entropy of the base model, $\entropy{\phi}$, is not the only term that is affected when contracting a subgraph. 
The base model parameters $\phi(H)$ may also change, which, in turn, may change the codelength needed to encode them, and must thus also be taken into account in the putative codelength difference. 

\subsubsection*{Positive integer}
We let $a$ denote a positive integer. 
It could represent a number of edges or nodes, the maximum degree, etc. 
As was described in the section ``Compression, model selection, and hypothesis testing", $a$ can be encoded using $\integerCodelength{a}$ bits. 
Thus, if the contraction of a subgraph $s$ induces the variation $a\longrightarrow a + \Delta a(s)$, then the associated compression (codelength difference) is 
\begin{equation}\label{eq:subgraph-contraction-cost-integer}
    \compressibility{\mathbb{N}}(a,s) = \integerCodelength{a} - \integerCodelength{a + \Delta a(s)} = \log \frac{a(a+1)}{[a+\Delta a(s)][a+\Delta a(s)+1]}  .
\end{equation}
For instance, for the ER model, a subgraph contraction changes the base model's parameters as $\Delta E(s) = -e$ and $\Delta N(s) = 1-n$,  leading to a change in codelength of 
\begin{equation}
    \compressibility{\mathbb{N}}(E,s) 
    = \log \frac{E(E+1)}{[E-e][E-e+1]} 
\end{equation}
for describing $E$, and of
\begin{equation}\label{eq:graph_size_update}
    \compressibility{\mathbb{N}}(N,s) 
    = \log \frac{N(N+1)}{[N-n+1][N-n+2]} 
\end{equation}
for describing $N$.

\subsubsection*{Sequences of positive integers}
Let $\textbf{a}=(a_i)$ be a sequence of $N$ positive integers. 
For our purpose, \textbf{a} represents a sequence of node degrees (i.e., the in,- out-, or mutual degrees of $H$). 

The sequence $\mathbf{a}$ is described either by a uniform code or a plug-in code (Eq.~\eqref{eq:sequence-codelength}) that both depend on the range of values distributed in $\mathbf{a}$. 
We let the maximum value of the sequence be denoted $Q\equiv\max\mathbf{a}$ ($\Delta$ in the main text), and the minimum value $q\equiv\min\mathbf{a}$ ($\delta$ in the main text). 
For each distinct value $\mu$ in $\mathbf{a}$, we let $r_\mu \equiv \sum_{i=1}^{N}\delta(a_i,\mu)$ denote its frequency. 
Let $a(s) \equiv a_{i_s}$ be the new sequence element added to $\mathbf{a}$, after $s$ is contracted into a supernode, which is labeled $i_s$. It is given by
\begin{equation}
    a(s) = \sum_{i\in \nu} a_i - a_0(s) ,
\end{equation}
where $a_0(s)$ is related to the deleted internal edges of $s$ or the concatenation of subgraph neighborhoods into multiedges. 
Let us review how supernode degrees are computed in the configuration model and its reciprocal conterpart to show that the above expression holds for the present study. 

In the case of the configuration model, the evaluation of the out-degree of a new supernode is
\begin{align}
    k^+(s) &= \sum_{i\in \nu}\sum_{j\in\partial s} A_{ij} \\ 
    &= \sum_{i\in \nu} \left(\sum_{j} A_{ij} - \sum_{j\in s} A_{ij}\right)\\
    &= \sum_{i\in \nu} k_i^+ - e,
\end{align}
such that one identifies $k_0(s) = e$. For the in-degree sequence, the expression holds by the change of notation $+\longrightarrow -$. 

For the reciprocal configuration model, three degree sequences are involved. Starting with the directed out-degree,
\begin{align}
    \kappa^+(s) & = \sum_{j\in \partial s}\max\left(\sum_{i\in \nu} (A_{ij}-A_{ji}),0\right) \\ 
    & = \frac{1}{2}\sum_{i\in \nu}\sum_{j\in \partial s} (A_{ij}-A_{ji}) + \frac{1}{2}\sum_{j\in\partial s }\left| \sum_{i\in \nu}(A_{ij} - A_{ji}) \right| \\ 
    & = \sum_{i\in \nu} \kappa^+_i - \kappa_0(s) ,
\end{align}
with
\begin{align}
    \kappa_0(s)& = e_d + \frac{1}{2}\sum_{j\in\partial s } \left(\sum_{i\in \nu} |A_{ij}-A_{ji}| -  \left| \sum_{i\in \nu}(A_{ij} - A_{ji}) \right|\right) .
\end{align}
The directed in-degree is identical  after the change of notation $+ \longrightarrow -$. 
The mutual degree of a supernode is simply given by its relationship to the configuration model's in- (or out-) degree and to the mutual degree,
\begin{align}
    \kappa^m(s) & = k^+(s) - \kappa^+(s) = k^-(s) - \kappa^-(s) \\
     & = \sum_{i\in s} (k_i^+ - \kappa_i^+) - [k_0(s) - \kappa_0(s)] \\ 
     & = \sum_{i\in s}\kappa^m_i - \kappa_0^m(s)
\end{align}
The supernode degree and node degrees of the subgraph's nodes are not the only sequence elements modified by the subgraph contraction. Properties of the subgraph's neighborhood are also likely to evolve if degree sequences are correlated, which is the case of the reciprocal configuration model. 
After the contraction of $s$, we will denote by $\Delta a_j(s)$, the variation of the element $a_j\longrightarrow a_j + \Delta a_j(s)$, associated to a network property of node $j\in\partial s$. Variations of the directed in- and out-degrees are equal: 
\begin{align}
    \Delta \kappa^+_j(s) &= \nonumber \max\left(\sum_{i\in \nu}(A_{ji} - A_{ij}),0 \right) - \sum_{i\in \nu}\max(A_{ji} - A_{ij},0)\\ \nonumber
     &= \frac{1}{2}\left|\sum_{i\in \nu} (A_{ji} - A_{ij})\right| - \frac{1}{2}\sum_{i\in \nu} |A_{ji} - A_{ij}| \\ 
     &= \Delta \kappa_j^-(s) \equiv \Delta \kappa_j(s)\label{eq:RCM-neighborhood}
\end{align}
The case of the mutual degree is deduced by the conservation of the in- and out-degree of the subgraph's neighbors:
\begin{align}
    \Delta k_j^+(s) &= \Delta k_j^-(s) = 0 \\
    \Leftrightarrow \Delta \kappa_j^m(s) &= - \Delta \kappa_j(s)
\end{align}
One can write how the distribution $\{r_\mu\}$ transforms into $\{r_\mu + \Delta r_\mu(s)\}$. 
\begin{equation}
    \Delta r_\mu(s) = \delta(a(s),\mu) - \sum_{i\in s}\delta(a_i,\mu) + \sum_{j\in \partial s}[\delta(a_j + \Delta a_j(s),\mu)-\delta(a_j,\mu)]
\end{equation}
Updates of $Q\longrightarrow Q + \Delta Q(s)$ and $q\longrightarrow q + \Delta q (s)$ are naturally determined by the evolution of the sequence $\mathbf{a}\longrightarrow \mathbf{a} + \Delta\mathbf{a}(s)$, and by how the distribution $\{r_\mu\}$ is shifted, due to the contraction of $s$, by the incremental distribution $\{\Delta r_\mu(s)\}$. 
Consider first the case of the update of the maximum. 
A first scenario could be an augmentation of the maximum, $\Delta Q(s) > 0$, by either the new supernode or an increase of a subgraph's neighbor degree, i.e., $\Delta a_j(s) > 0$. 
Let $Q'(s)$ be the maximum between a supernode degree and one of the updated subgraph's neighbors degrees: 
\begin{equation}
    Q'(s) = \max\left(a(s),\max_{j\in\partial s}\{a_j+\Delta a_j(s)\}\right)
\end{equation}
A second scenario, when $Q'(s) < Q$, is the possible extinction of the maximum, i.e., $\Delta r_Q(s) = -r_Q$. Introducing the quantity $\Delta Q'(s) = \max(Q'(s)-Q,0)$, the difference in maxima is expressed as
\begin{align}
    \Delta Q(s) &= \Delta Q'(s) + \delta(\Delta Q'(s),0)\delta (\Delta r_Q(s),-r_Q)  \sum_{\mu=0}^{Q-1}( \mu - Q)\mathfrak{g}_\mu^Q(s)\label{eq:delta_Q}
\end{align}
where $\mathfrak{g}_\mu^Q(s)$ in an indicator function that returns 1 when $\mu$ is the highest value below $Q$ in $\mathbf{a}$ after the update, and 0 otherwise. It can be formally written as:
\begin{equation}
    \mathfrak{g}_\mu^Q(s) = \left[1 - \delta(\Delta r_\mu(s),-r_\mu)\right]\prod_{\mu'=\mu+1}^{Q-1}\delta(\Delta r_{\mu '}(s),-r_{\mu'})
\end{equation}
The first term on the RHS of Eq.~\ref{eq:delta_Q}  corresponds to the case where the maximum of the sequence is increased by the insertion of a supernode or the restructuring of the subgraph's neighborhood. 
The second term is the alternative scenario where the maximum is decreased and must be searched within $\mathbf{a}$. 
The variation in minima $\Delta q(s)$ is naturally similar to Eq.~\ref{eq:delta_Q}. Let $q'(s)$ be the minimum between a supernode degree and one of the updated subgraph's neighbor degrees: 
\begin{equation}
    q'(s) = \min\left(a(s),\min_{j\in\partial s}\{a_j+\Delta a_j(s)\}\right)
\end{equation}
Introducing $\Delta q'(s) = -\min(q-q'(s),0)$, the difference in minima is expressed as 
\begin{equation}
\label{eq:delta_q}
    \Delta q(s) = \Delta q'(s) + \delta(\Delta q'(s),0)\delta(\Delta r_{q}(s),-r_q)\sum_{\mu=q+1}^{Q+\Delta Q(s)}(\mu - q)\mathfrak{g}_\mu^q(s)
\end{equation}
where $\mathfrak{g}_\mu^q(s)$ in an indicator function that returns 1 when $\mu$ is the lowest value greater than $q$ in $\mathbf{a}$ after the update, and 0 otherwise. It can be formally written as:
\begin{equation}
    \mathfrak{g}_\mu^q(s) = \left[1 - \delta(\Delta r_\mu(s),-r_\mu)\right]\prod_{\mu'=q+2}^{\mu}\delta(\Delta r_{\mu '}(s),-r_{\mu'})
\end{equation}
The first term on the RHS of Eq.~\ref{eq:delta_q}  corresponds to the case where the minimum of the sequence is decreased by the insertion of a supernode or the restructuring of the subgraph's neighborhood. 
The second term is the alternative scenario where the minimum is increased and must be searched within $\mathbf{a}$.
All necessary quantities involved in putative codelength differences of integer sequences have been determined.
Let us now give their exact expressions. 

\paragraph{Uniform code.} A uniform encoding of $\mathbf{a}$ corresponds to $N$ products of a uniform probability distribution over $q$ to $Q$, 
\begin{equation}
    L_U(\mathbf{a}) = N\log(Q-q+1) + L_\mathbb{N}(Q) + L_{\mathbb{N}}(q) .
\end{equation}
The codelength difference is 
\begin{align}\label{eq:subgraph-contraction-cost-sequence-uniform}
    \Delta L_U(\mathbf{a},s) &= -(N - n 
 + 1)\log\left(1+\frac{\Delta Q(s) - \Delta q(s)}{Q-q+1}\right) + (n-1)\log (Q - q  + 1) \\ 
    &+\Delta L_\mathbb{N}(Q,s) + \Delta L_\mathbb{N}(q,s)\nonumber .
\end{align}

\paragraph{Plug-in code.} The plug-in code is a function of $\{r_\mu\}$ and a hyperparameter $\lambda$, that constrains the shape of the prior.
Two different values for $\lambda$ were considered in the main text, $\lambda=1/2$ (Jeffreys prior) and $\lambda=1$ (uniform prior). 
The plug-in code of a sequence is characterized by three entities: $N$, $\{r_\mu\}_{q\leq\mu\leq Q}$, and $\Lambda\equiv\Lambda(Q,q) = (Q-q + 1)\lambda$:
\begin{equation}
     L_\lambda(\mathbf{a}) = \log\frac{\Gamma(N+\Lambda)}{\Gamma (\Lambda) } + (Q-q+1)\log\Gamma(\lambda) - \sum_{q\leq\mu\leq Q}\log\Gamma (r_\mu+\lambda) .
\end{equation}
Let $\Delta \Lambda(Q,q,s)$ the variation following the contraction of $s$, $\Lambda (Q,q)\longrightarrow \Lambda(Q,q) + \Delta \Lambda(Q,q,s)$. The latter is determined by how the maximum and minimum of $Q$ and $q$ are closer or more distant after the subgraph contraction. One can independently treat the case where $Q$ or $q$ changes. Thus, we adopt the following decomposition,
\begin{align}
    \Delta \Lambda(Q,q,s) & = \Delta \Lambda(Q,s) + \Delta\Lambda(q,s) \\ 
    & = [\Delta Q(s) - \Delta q(s)]\lambda 
\end{align}
The update of the plug-in code after a subgraph contraction is divided into multiple cases, depending on how the contraction of a subgraph $s$ affects $N$, $\{r_\mu\}_{q\leq\mu\leq Q}$, and $\Lambda(Q,q)$. All in all, the plug-in codelength difference is 

\begin{adjustwidth}{0in}{0in}
    \begin{align}\label{eq:subgraph-contraction-cost-sequence-DM}
    \Delta L_\lambda(\mathbf{a},s) &  = \log \frac{\Gamma(N + \Lambda)}{\Gamma(N - n + 1 + \Lambda + \Delta \Lambda(Q,q,s))}+ \log \frac{\Gamma(\Lambda + \Delta\Lambda(Q,q,s))}{\Gamma(\Lambda)}\\ & +\sum_{\mu=\mu_\textrm{min}}^{\mu_\textrm{max}} \log \frac{\Gamma(r_\mu + \Delta r_\mu(s) + \lambda)}{\Gamma(r_\mu + \lambda)} + \Delta L_\mathbb{N}(N,s)+ \Delta L_\mathbb{N}(Q,s) + \Delta L_\mathbb{N}(q,s) \nonumber ,
\end{align}   
\end{adjustwidth}
where $\mu_\textrm{min} = \min(q,q+\Delta q(s))$ and $\mu_\textrm{max} = \max(Q,Q+\Delta Q(s))$.

\subsection*{Erd\H{o}s-Rényi model}

For the ER model, $(\Vnumber,\Enumber)$ will change to $(\Vnumber-n+1,\Enumber-e)$ after the contraction of $s$. The putative codelength difference is then given by
\begin{align}\label{eq:subgraph-contraction-cost-ER}
    \compressibility{\thetaER}(H,\subgraph) & = e \log[(\Vnumber-n)(\Vnumber-n+1)] + \Enumber \log \frac{\Vnumber(\Vnumber-1)}{(\Vnumber-n)(\Vnumber-n+1)} \\ \nonumber
    &\quad - \log \frac{\Enumber !}{(\Enumber-\enumber)!} - \ell_\mathbf{A}(\subgraph) .
\end{align}

\subsection*{Reciprocal Erd\H{o}s-Rényi model}

For the reciprocal ER model, the variation of the number of mutual edges and directed edges do not only depend on $e_d$ and $e_m$ because the formation of multiedges (as stacked single edges) changes the number of mutual edges $E_m$ and the number of directed edges $E_d$ in $H$. 
The variations of the number of mutual edges $\Delta E_m(s)$ and of the number of directed edges $\Delta E_d(s)$ are given by
\begin{align}
    \Delta E_m(s) &= -e_m + \sum_{j\in\partial s}\left[\min\left(\sum_{i\in \nu}A_{ij},\sum_{i\in\nu} A_{ji}\right) - \sum_{i\in\nu}\min(A_{ij},A_{ji})\right] \\ 
    & = -e_m + \frac{1}{2}\sum_{j\in\partial s}\left(\sum_{i\in\nu}|A_{ij}-A_{ji}| - \left| \sum_{i\in\nu} (A_{ij} - A_{ji})\right| \right) ,
\end{align}
and
\begin{align}
    \Delta E_d(s) = & -e -2\Delta E_m(s) \\ 
                 = & -e_d  - \sum_{j\in\partial s}\left(\sum_{i\in\nu}|A_{ij}-A_{ji}| - \left| \sum_{i\in\nu} (A_{ij} - A_{ji})\right| \right) .
\end{align}
The codelength has two part, one for the directed edges and another for the mutual edges. For the directed edges, one can adapt Eq.~\ref{eq:subgraph-contraction-cost-ER}, and replace $e$ by $\Delta E_d(s)$: 
\begin{align}\label{eq:directed-ER-contraction-cost}
    \compressibility{(\Vnumber,\Ednumber)}(H^{\mathrm{asym}},\subgraph) & = -\Delta \Ednumber(s)\log\left[(\Vnumber-n)(\Vnumber-n+1)\right] + \Ednumber \log \frac{\Vnumber(\Vnumber-1)}{(\Vnumber-n)(\Vnumber-n+1)} \nonumber \\ 
    &\quad - \log \frac{\Ednumber !}{(\Ednumber+\Delta \Ednumber(s))!} - \ell_{\mathbf{A}^\mathrm{asym}}(\subgraph) .
\end{align}
For the mutual part, the putative codelength difference is: 
\begin{align}\label{eq:mutual-ER-contraction-cost}
    \compressibility{(\Vnumber,\Emnumber)}(H^{\mathrm{sym}},\subgraph) & = -\Delta E_m(s)\log\left[\frac{(\Vnumber-n)(\Vnumber-n+1)}{2}\right] + \Emnumber \log \frac{\Vnumber(\Vnumber-1)}{(\Vnumber-n)(\Vnumber-n+1)} \nonumber \\ 
    &\quad - \log \frac{\Emnumber !}{(\Emnumber+\Delta E_m(s))!} - \ell_{\mathbf{A}^\mathrm{sym}}(\subgraph) .
\end{align}
where $\ell_{\mathbf{A}^\mathrm{sym}}(s)$ is the undirected version of Eq.~\ref{eq:dL-common}, where only one of two terms inside the $\log$ needs to be kept. 
Finally, the codelength difference can be written as
\begin{equation}\label{eq:subgraph-contraction-cost-RER}
    \compressibility{\thetaRER}(H,\subgraph) = \compressibility{(\Vnumber,\Ednumber)}(H^{\mathrm{asym}},\subgraph)+\compressibility{(\Vnumber,\Emnumber)}(H^{\mathrm{sym}},\subgraph)
\end{equation}

\subsection*{Configuration model}
The codelength difference when contracting a subgraph $\subgraph$ for the configuration model is
\begin{align}\label{eq:subgraph-contraction-cost-CM}
    \compressibility{\thetaConf}(H,\subgraph)
    & = \log\frac{\Enumber!}{(\Enumber -\enumber)!} + \log\left[\frac{k^+(\subgraph)!}{\prod_{i\in\vset}k^+_i!}\times\frac{k^-(\subgraph)!}{\prod_{i\in\vset}k^-_i!}\right] - \ell_\mathbf{A}(\subgraph),
\end{align}
where $\outdegree_i = \outdegree_i(H)$, and $\indegree_i = \indegree_i(H)$ as above, and 
\begin{align}
    k^\pm(\subgraph) &\ =  \sum_{i\in\vset} k^\pm_i(s) = \sum_{i\in\vset} k^\pm_i - e .
\end{align}

\subsection*{Reciprocal configuration model}

Finally, the codelength difference for the reciprocal configuration model is equal to
\begin{align}\label{eq:subgraph-contraction-cost-RCM}
    \nonumber \compressibility{\thetaRConf}(H,\subgraph) 
    & = \log \frac{\Ednumber!}{[\Ednumber+\Delta\Ednumber]!}  + \log \frac{(2\Emnumber-1)!!}{[2(\Emnumber+\Delta\Emnumber)-1]!!} \nonumber \\ 
    &\quad + \log\left[ \frac{\kappa^+(\subgraph)!}{\prod_{i\in \vset}\kappa_i^+!} \times \frac{\kappa^-(\subgraph)!}{\prod_{i\in\vset}\kappa_i^-!} \times\ \frac{\kappa^m(\subgraph)!}{\prod_{i\in\vset}\kappa^m_i!}\right]  \nonumber \\ \nonumber  
    &\quad + \sum_{j\in\partial s}\log\left[ \frac{(\kappa^+_j+\Delta\kappa_j(\subgraph))!}{\kappa_j^+!} \times \frac{(\kappa^-_j+\Delta\kappa_j(\subgraph))!}{\kappa_j^-!} \times\ \frac{(\kappa^m_j-\Delta\kappa_j(\subgraph))!}{\kappa^m_j!}\right]
    \\
    &\quad -\ell_{\mathbf{A}^\text{asym}}(\subgraph) -\ell_{\mathbf{A}^\text{sym}}(\subgraph) .
\end{align}
where 
\begin{align}
    \kappa^\pm(s) & = \sum_{i\in\nu}\kappa^\pm_i +\frac{1}{2}\Delta E_d(s) - \frac{e_d}{2} \\ 
    \kappa^m(s) &= \sum_{i\in\nu}\kappa_i^m + \Delta E_m(s) 
\end{align}
are respectively the directed and mutual degrees of the future supernode.  The $\{\Delta\kappa_j(s)\}_{j\in\partial s}$ are respectively variations of the directed degrees of the subgraph neighborhood due to its contraction (see Eq.~\ref{eq:RCM-neighborhood} for their expressions). 

\subsection*{Motif-based code}

Based on the previous subsections, we can give the complete putative codelength difference when contracting a subgraph. 
As a reminder, the codelength of our model is
\begin{equation}
    \codelength{}(G, \params) 
    = \graphletCost
    + \baseCost 
    + \supernodeCost
    + \reconstructionCost ,
\end{equation}
where $\theta=\{ H,\baseParams,\subgraphSet, \supernodes, \graphletSet\}$. $H$ is the reduced multigraph, $\graphletSet$ is the set of all discovered graphlets, $\subgraphSet$ is the graphlet multiset (a proxy for a set of subgraphs), $\supernodes$ are $H$'s node labels identifying supernodes, and $\baseParams$ are the parameters of the dyadic base model. 
Let us give the subgraph-contraction-induced cost for all terms of the above equation.

The update of the encoding cost of the graphlet set and multiset, $\graphletCost$ (see Eq.~(6) in the ``Methods'' section), is seen as an extension of $\subgraphSet$ by $\alpha$, the label of the graphlet to which $s$ is isomorphic. We choose to encode $\subgraphSet$ as an ordered multiset of elements, that are independently sampled from $\graphletSet$ and their respective frequency in $\subgraphSet$ is encoded by a uniform distribution over the range one to $m_\textrm{max}$. The minimum value of $m_\textrm{max}$ is one. Two exclusive scenarios may occur for a non-zero update cost. Either an occurrence of the most represented graphlet in $\subgraphSet$ is again selected and leads to an incremental increase of $m_\textrm{max}$, or $s$ is isomorphic to a different $\alpha\notin \subgraphSet$. Denoting by $\graphletSubset$ the unique set of elements of $\subgraphSet$, 
    \begin{align}
        \Delta \codelength{}(\graphletSet,\subgraphSet,s)& =    
       -\sum_{\graphlet\in\graphletSubset}\mathbb{I}(s \cong g_\graphlet) \biggl\{\delta(m_\alpha,m_\text{max}) \left[|\graphletSubset|\log\left(1+\frac{1}{m_\textrm{max}}\right) + \log\left(1+\frac{2}{m_\textrm{max}}\right)\right]\biggr.\\  & + \biggl. \delta(m_\alpha,0)\left(\log |\graphletSet| + \log m_{\text{max}}\right)\biggr\}\nonumber
    \end{align}
where $\mathbb{I}(s \cong g_\graphlet)$ is an indicator function that is one if $s$ is isomorphic to the graphlet of canonical label $\graphlet$.

The update of the encoding of the supernode labels, $\supernodeCost$ (see Eq.~(7) in the ``Methods'' section), is, again, affected by the graph size, the growth of the supernode number and the incremental increase of a graphlet occurrence. Denoting by $M \equiv \sum_{\alpha'}m_{\alpha'} = |\subgraphSet|$ the number of supernodes, 
\begin{align}
    \Delta \codelength{}(\supernodes,s|H,\subgraphSet) =  \log \binom{N}{M} - \log\binom{N-n+1}{M+1} + \sum_{\graphlet\in\graphletSubset}\mathbb{I}(s \cong g_\graphlet)\log\frac{m_\alpha+1}{M+1}
\end{align}

The update of reconstruction cost from $H$ to $G$ depends on the reduced graph size, the associated graphlet orientation number and the subgraph's neighborhood:
\begin{align}
    \Delta\codelength{}(G,s|H,\supernodes,\subgraphSet,\graphletSet)  & =  \log \frac{(N-n+1)!}{N!} + \sum_{\graphlet\in\graphletSubset}\mathbb{I}(s \cong g_\graphlet)\frac{n_\alpha!}{|\mathrm{Aut}(\alpha)|} \\ \vspace{0.1cm}
    & + \sum_{j\in\mathcal{N}(H)\backslash \supernodes} \log\binom{n}{\sum_{i\in\nu}A_{ij}}\binom{n}{\sum_{i\in\nu}A_{ji}} + \sum_{j_{s'}\in\supernodes}\log \binom{nn_{j_s'}}{\sum_{i\in\nu}A_{ij_{s'}}}\binom{nn_{j_s'}}{\sum_{i\in\nu}A_{j_{s'}i}} \nonumber,
\end{align}
where $n_{j_s'}$ is the subgraph size relative to the supernode $j_{s'}\in\supernodes$, replacing the subgraph $s'$. The two sums represent the encoding of the nodes' neighbors within $s$, i.e., how to distribute the multiedge among the nodes that would be deleted. The first sum corresponds to regular node neighbors, while the second sum corresponds to supernode neighbors. 

Finally, $\Delta\codelength{\baseParams}(H,s)$ being given by Eqs.~\eqref{eq:subgraph-contraction-cost-ER},\eqref{eq:subgraph-contraction-cost-RER},\eqref{eq:subgraph-contraction-cost-CM},\eqref{eq:subgraph-contraction-cost-RCM} and $\Delta\codelength{}(\baseParams,s)$ by Eqs.~\eqref{eq:subgraph-contraction-cost-integer},\eqref{eq:subgraph-contraction-cost-sequence-uniform},\eqref{eq:subgraph-contraction-cost-sequence-DM} the complete putative codelength difference is: 
\begin{equation}
    \Delta\codelength{}(G,\theta,s) = \Delta\codelength{\baseParams}(H,s) + \Delta\codelength{}(\baseParams,s) + \Delta\codelength{}(G,s|H,\supernodes,\subgraphSet,\graphletSet) + \Delta \codelength{}(\supernodes,s|H,\subgraphSet)  + \Delta \codelength{}(\graphletSet,\subgraphSet,s)
\end{equation}

\clearpage
\section*{S5 Text: Generating random graphs from the null models}
\labelname{S5 Text}
\label{SIsec:null-model-sampling}

{We generate random graphs from the four different null models by employing Markov Chain Monte Carlo (MCMC) edge switching procedures that constrain the corresponding graph features while maximally randomizing the graph structure under this constraint.}

{\paragraph*{Erd\H{o}s-Rényi model.} 
To sample graphs from the Erd\H{o}s-Rényi (ER) model starting from a given network $G$, we switch in each iteration of the edge swapping a random edge $(i,j)\in\mathcal{E}(G)$ with a random non-edge $(k,l)\in \mathcal{E}(\Bar{G})$, where $\Bar{G}$ is the complement graph of $G$, i.e., $A_{ij}(\Bar{G}) = 1 - A_{ij}(G)$~\cite{orsini_quantifying_2015}. 
The procedure conserves $N$ and $E$, but otherwise generates maximally random networks.}

{\paragraph*{Reciprocal Erd\H{o}s-Rényi model.}
The procedure for generating random graphs from the reciprocal ER (RER) model is very similar to the one for the ER model, except that we additionally enforce the conservation of the numbers of mutual and single edges. This is done by explicitly distinguishing two types of edge switching, selected randomly at every step, one that switches single edges, the other that switches mutual edges.  
Unconnected node pairs are sampled rather than non-edges because we must ensure that a directed edge switch will not lead to the creation of new mutual edge.
The procedure thus conserves $\Vnumber$, $\Ednumber$, and $\Emnumber$.}

{\paragraph*{Configuration model.}
To sample the configuration model (CM), we employ the ``Maslov-Sneppen'' edge-swapping algorithm algorithm to generate random graphs that share a fixed degree sequence. 
Let $(i,j)$ and $(k,l)$ be two edges of $G$, then the edge-swap is defined by the transformations $(i,j)\longrightarrow (i,l)$ and $(k,l) \longrightarrow (k,j)$. If the edge swap leads to a loop, i.e., $i = l$ or $k = j$, then the swap is rejected \cite{orsini_quantifying_2015}.}

{\paragraph*{Reciprocal configuration model.}
The generative procedure for sampling the reciprocal CM (RCM) combines those of the CM and the RER model. 
As for the RER model, each step of the algorithm is either a mutual or single (directed) edge swap selected at random. 
The edge swap is then performed following the Maslov-Sneppen procedure as described above, either between a pair of directed or a pair of reciprocal edges.
If the edge swap is directed, the reciprocal connection of the newly formed edge must be empty, otherwise, the swap is rejected~\cite{milo_superfamilies_2004}.
}

\clearpage
\section*{S6 Text: Measures of graphlet topology}
\labelname{S6 Text}
\label{SIsec:gpr}
\setcounter{figure}{0}

\paragraph{Density.}
The \textit{density} $\rho$ measures the fraction of node pairs in a simple graph $G = (\Vnumber,\Enumber)$ that are connected by an edge~\cite{hagberg2020networkx},
\begin{equation}
    \rho = \frac{\Enumber}{\Vnumber  (\Vnumber - 1)} .
\end{equation}
A simple graph is said to be \textit{sparse} if its density is close to zero, and \textit{dense} if its density is close to one.

\paragraph{Reciprocity.}
The \textit{reciprocity} $r$ measures the fraction of edges in a graph $G$ that are reciprocated~\cite{hagberg2020networkx},  
\begin{equation}
    r = \frac{1}{E}\sum_{ij} A_{ij}A_{ji} ,
\end{equation}
where $A_{ij} \in \{0,1\}$ are the entries of the adjacency matrix for $G$.

\paragraph{Number of cycles.}
The \textit{number of cycles} of a simple graph $G$ is the number of \textit{distinct} closed paths in $G$ where no node appears twice, and where two cycles are distinct if one is not a cyclic permutation of the other. We calculate the number of cycles using Johnson's algorithm~\cite{johnson1975finding} implemented in the Python module NetworkX \cite{hagberg2020networkx}. 

\begin{figure}[h!]
    \centering
         \includegraphics[height=0.41\textheight]{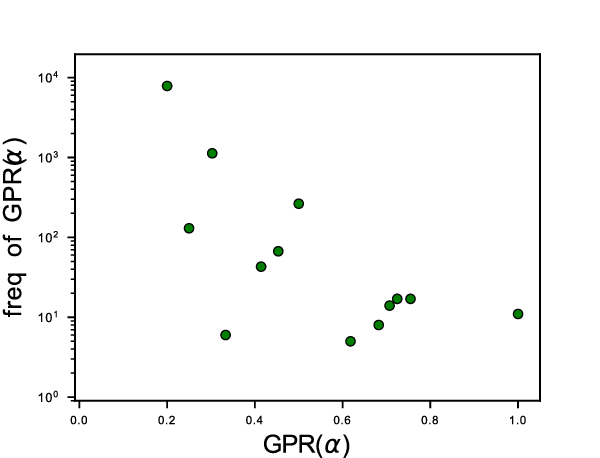} 
    \caption{\small\textbf{Distribution of graph polynomial root (GPR) values of all 3- to 5-node graphlets.} The minimum value of the GPR is 1/5 for five-node graphlets. It would be 0 in an infinite, maximally asymmetric graph, e.g., one where the automorphism group is a singleton. A GPR of 1, i.e., its maximum value for any graph size, represent maximally symmetric graphs, i.e., cliques or empty graphs. The symmetry of inferred motif sets in Fig 5 in the ``Results'' section should be interpreted knowing that the GPR is bounded between 0.2 and 1.
    }
    \label{SIfig:hist_gpr}
\end{figure}

\paragraph{Graph polynomial root.}
The \textit{graph polynomial root} (GPR) is a measure of the symmetry of a graph. 
It is related to the so-called orbit-polynomial \cite{dehmer2020orbit} $\Pi_G(z)$ and allows a ranking of graphs based on the distribution of their orbit sizes. 
Let $c_{o_l}$ be the number of orbits of size $o_l$, where $l\in\{1,2,\ldots,L\}$ and  $L$ is the number of different orbit sizes. 
The graph polynomial is then defined as
\begin{equation}
    \Pi_G(z) = \sum_{l=1}^L c_{o_l} z^{o_l} .
\end{equation}
The GPR, denoted $z^*$, is the unique solution of the following equation
\begin{equation}
    \Pi_G(z^*) = 1 ,
\end{equation}
which can be solved numerically. Orbit sizes are determined using McKay's \texttt{nauty} algorithm \cite{mckay2014practical}. A strong degree of symmetry is affiliated with a high GPR, while an asymmetric structure corresponds to a low GPR. 
Fig~\ref{SIfig:hist_gpr}. shows the distribution of values of the GPR of all 9\,364 3- to 5-node graphlets. 

\clearpage
\section*{Supplementary figures.}
\renewcommand{\thefigure}{S\arabic{figure}}
\setcounter{figure}{0}

\vspace{5cm}
\begin{figure}[ht]
    \centering
    \begin{adjustwidth}{-.7in}{0.in}
        \includegraphics[height=0.3\textheight]{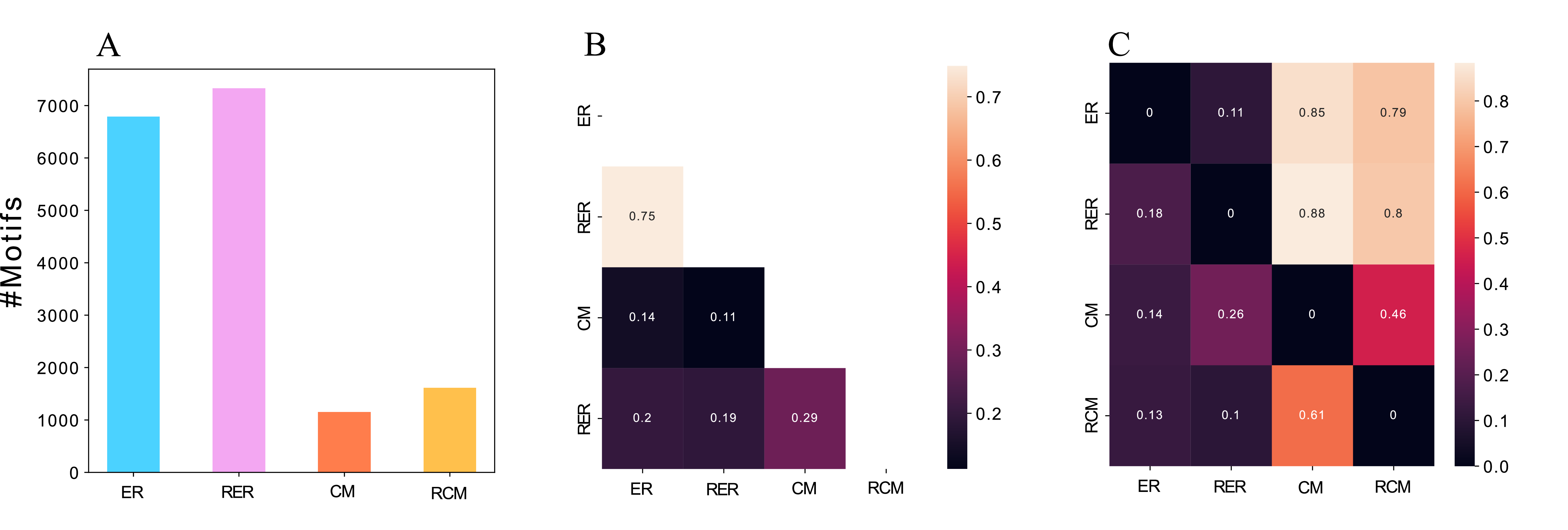}
    \end{adjustwidth}
    \caption{{\bf Differences in the motifs inferred using hypothesis testing when using different null models.}
    (A) Number of apparent motifs inferred in the \textit{Drosophila} larva right mushroom body connectome when using each of the four null models.
    Note in particular that even though the reciprocal models are strictly more constrained than their directed counterparts, more motifs are found with these null models than with the less constrained ones.
    (B) Overlap (Jaccard index) between the inferred graphlets using the different null models. 
    (C) Per null model, fraction of uniquely found motifs compared to another null model. Formally, denoting by $\mathcal{M}_i$, the motif set WRT the null model in the $i$-th row, and $\mathcal{M}_j$, the motif set WRT the null model in the $j$-th column, matrix entries are computed as  $|\mathcal{M}_i\backslash \mathcal{M}_j | / |\mathcal{M}_i|$. A low ratio indicates that $\mathcal{M}_j$ contains most of $\mathcal{M}_i$, while a high ratio expresses strong dissimilarities between the two emerged motif sets.}
    \labelname{S1}
    \label{SIfig:analysis-classic-testing}
\end{figure}

\begin{figure}[ht!]
    \centering
    \begin{adjustwidth}{-.75in}{0in}
        \includegraphics[height=0.6\textheight]{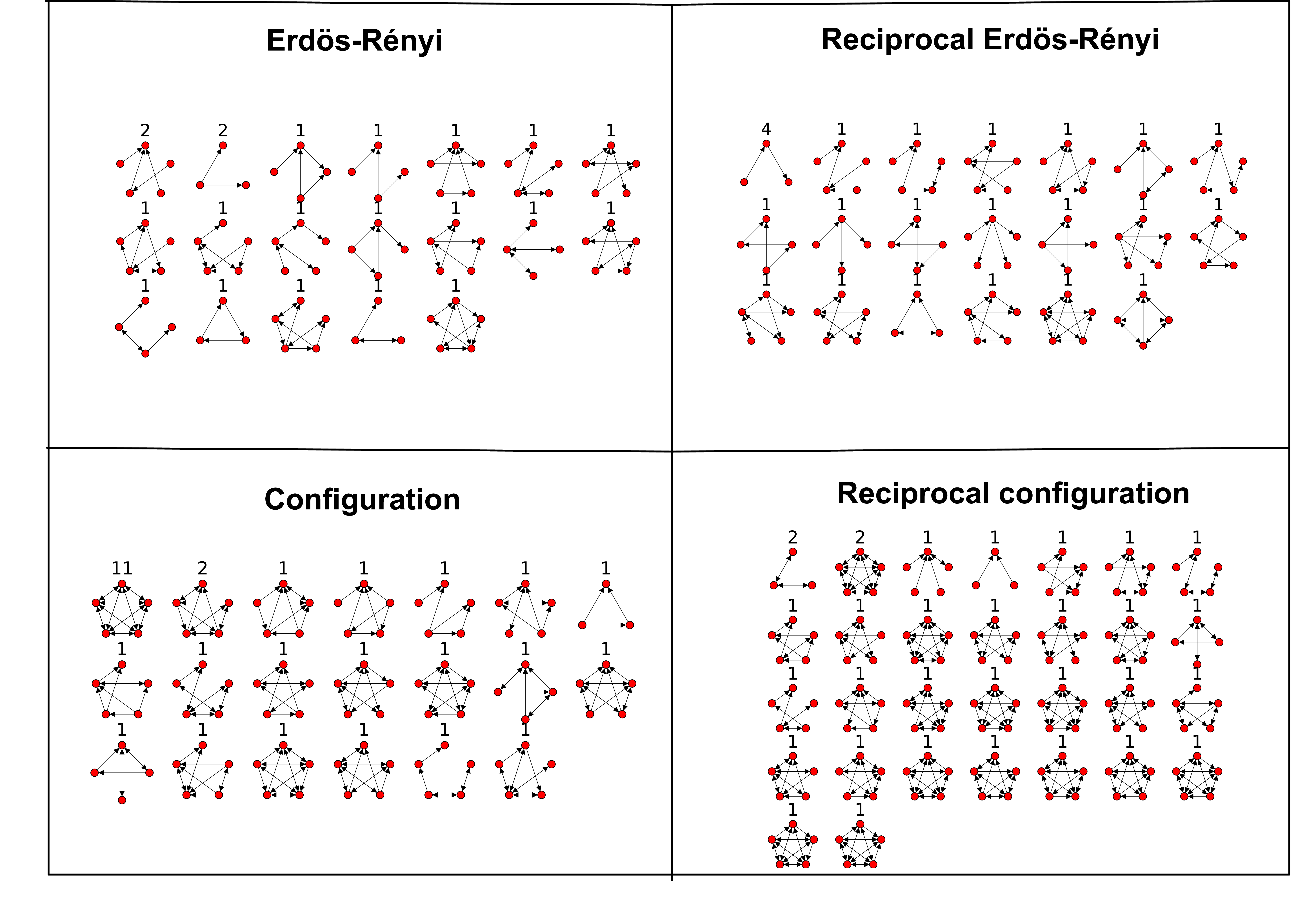}
    \end{adjustwidth}
    \caption{\textbf{Different motif sets obtained with the four base models.} 
    Inferred motif sets of the best model for the right hemisphere of the \textit{Drosophila} larva MB connectome. In this specific application, over all inferences across base models, the configuration model has the lowest codelength.  
    We observe a particularly clear distinction in the main types of motifs between Erd\H{o}s-Rényi-like and configuration like models.}
    \labelname{S2}
    \label{SIfig:base-model-diff}
\end{figure}

\begin{figure}
    \centering
    \begin{adjustwidth}{-.75in}{}
        \includegraphics[height=0.55\textheight]{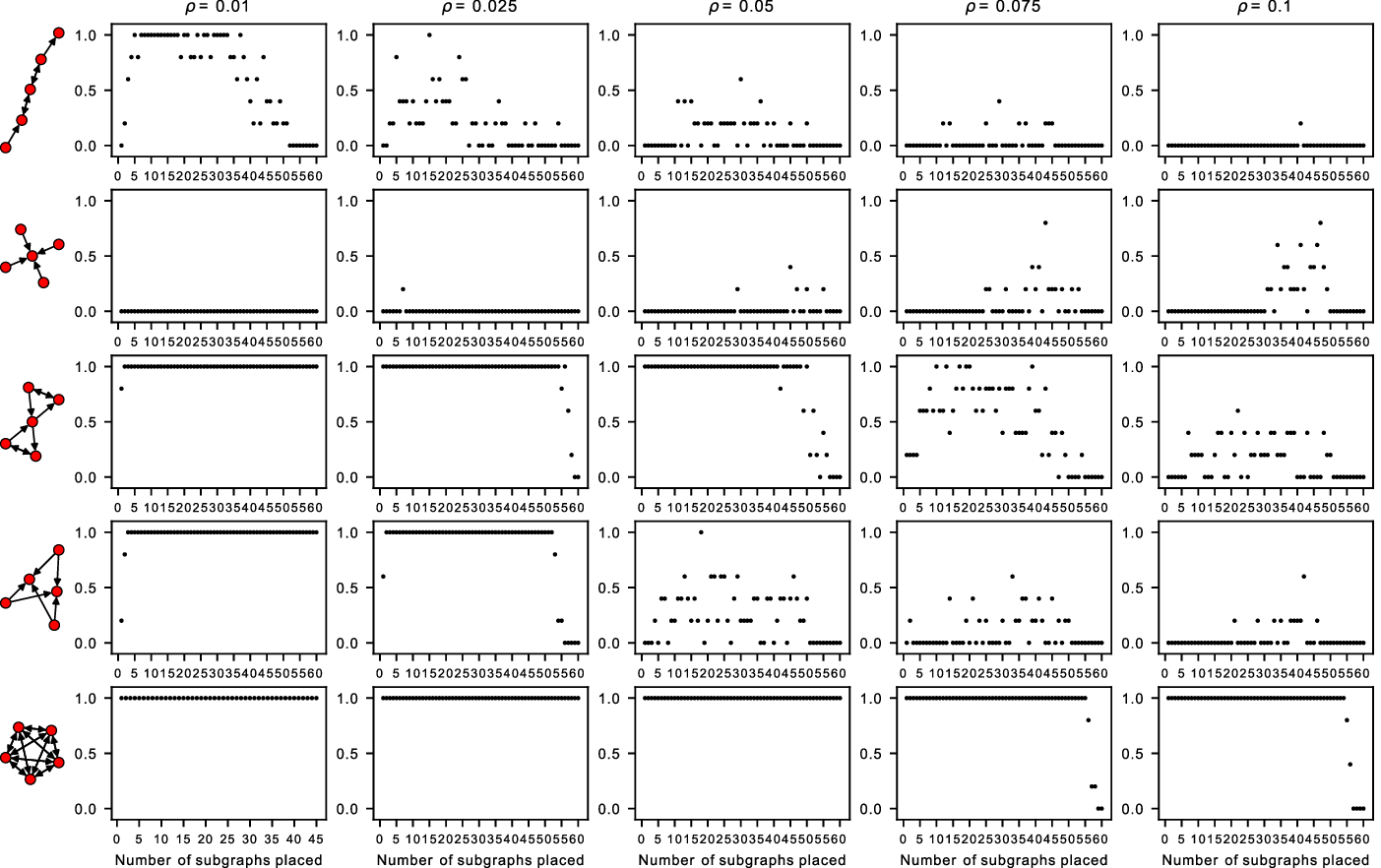}
    \end{adjustwidth}
    \caption{\textbf{Probability of correctly identifying the embedded motif in the planted motif model ($N = 300$).}
    Probability of the inferred  motif set containing at least one repetition of the true planted motif as a function of the number of times the motif is planted for five different planted motifs and for different network densities.  
    The generated networks contain $N = 300$ nodes and the edge density ranges from {$\rho = 0.01$} (leftmost) to {$\rho=0.1$} (rightmost). 
    Each point is an average over five independently generated graphs.
    Note that the maximum value of motifs that can be inserted depend both on the number of nodes in the network and on the networks density, as well as that of the motif; hence the range of the x-axis can vary.}
    \labelname{S3}
    \label{SIfig:motif_detection_rate_300}
\end{figure}

\begin{figure}
    \centering
    \begin{adjustwidth}{-.75in}{}
        \includegraphics[height=0.55\textheight]{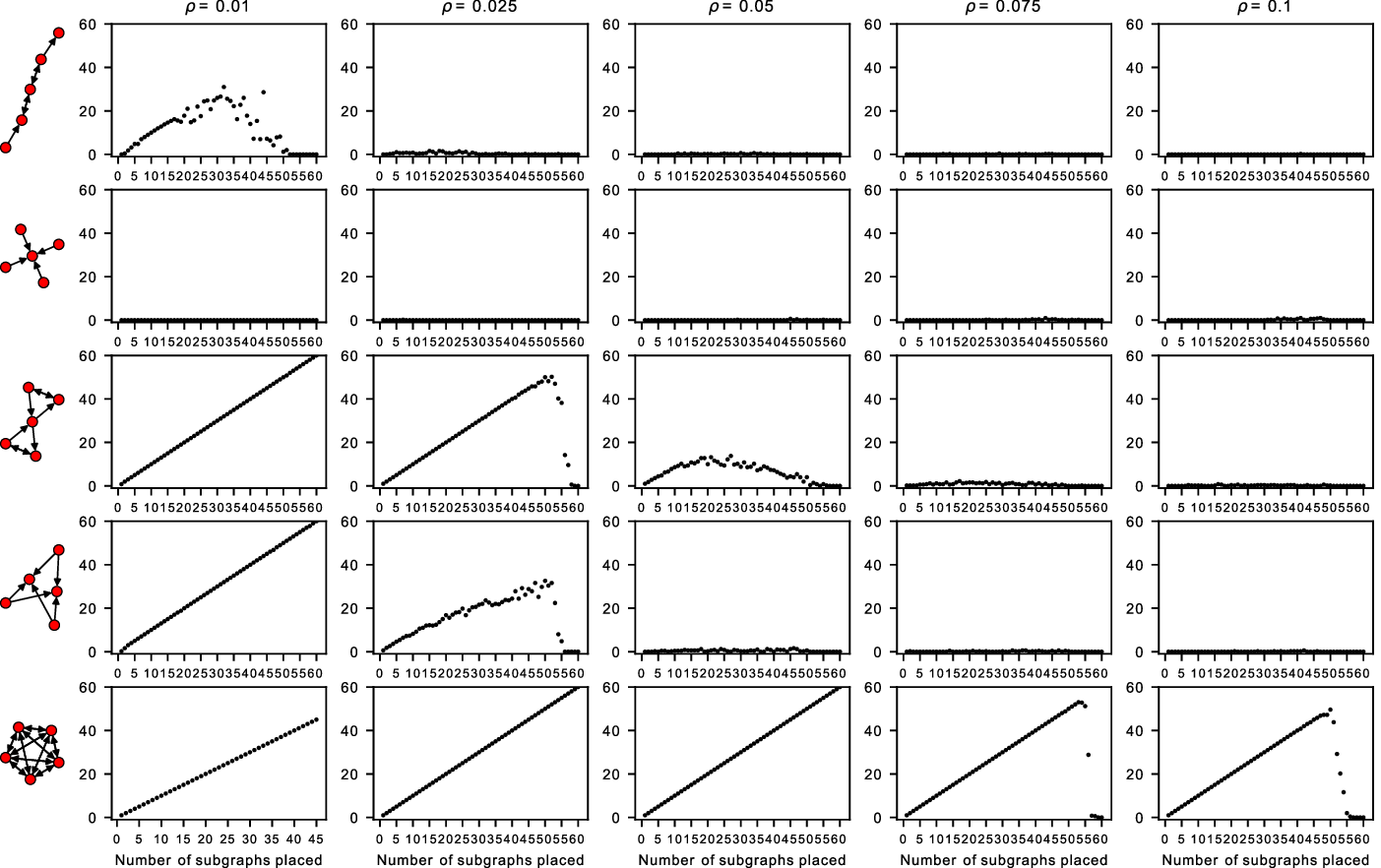}
    \end{adjustwidth}
    \caption{\textbf{Number of occurrences of the planted motif inferred ($N = 300$).} The number of insertions in the generated graphs is plotted on the x-axis, and this inferred number, averaged over five independent graphs, on the y-axis. 
    The generated networks contain {$N = 300$} nodes and the edge density ranges from {$\rho = 0.01$} (leftmost) to {$\rho=0.1$} (rightmost). 
    Each point is an average over five independently generated graphs.     
    Note that the maximum value of motifs that can be inserted depend both on the number of nodes in the network and on the networks density as well as that of the motif; hence the range of the x-axis can vary.}
    \labelname{S4}
    \label{SIfig:motifs_detected_300}
\end{figure}

\begin{figure}
    \centering
    \begin{adjustwidth}{-.75in}{}
        \includegraphics[height=0.55\textheight]{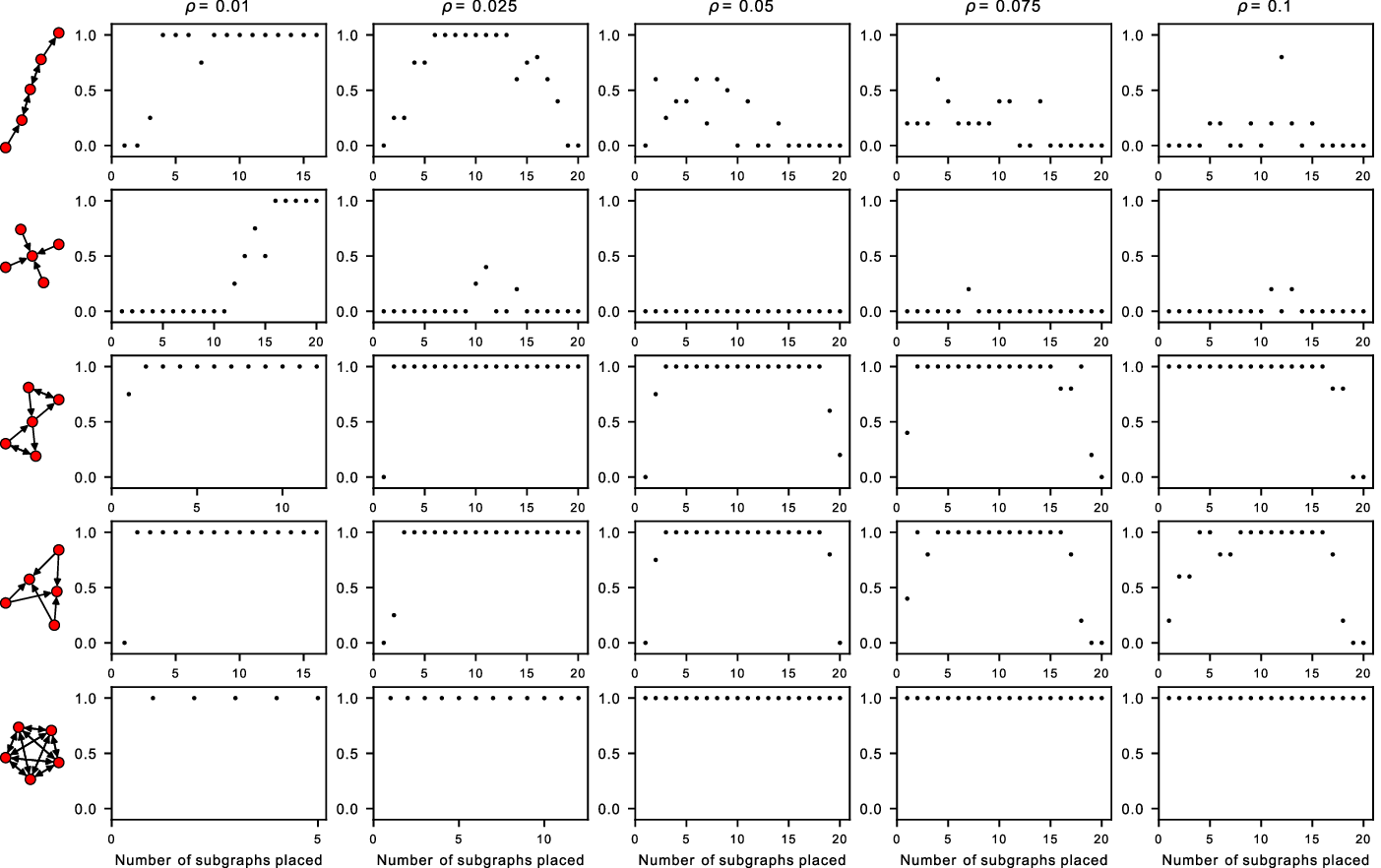}
    \end{adjustwidth}
    \caption{\textbf{Probability of correctly identifying the embedded motif in the planted motif model ($N = 100$).}
    Probability of the inferred  motif set containing at least one repetition of the true planted motif as a function of the number of times the motif is planted for five different planted motifs and for different network densities.  
    The generated networks contain $N = 100$ nodes and the edge density ranges from {$\rho = 0.01$} (leftmost) to {$\rho=0.1$} (rightmost). 
    Each point is an average over five independently generated graphs.     
    Note that the maximum value of motifs that can be inserted depend both on the number of nodes in the network and on the networks density as well as that of the motif; hence the range of the x-axis can vary.}
    \labelname{S5}
    \label{SIfig:motif_detection_rate_100}
\end{figure}

\begin{figure}
    \centering
    \begin{adjustwidth}{-.75in}{}
        \includegraphics[height=0.55\textheight]{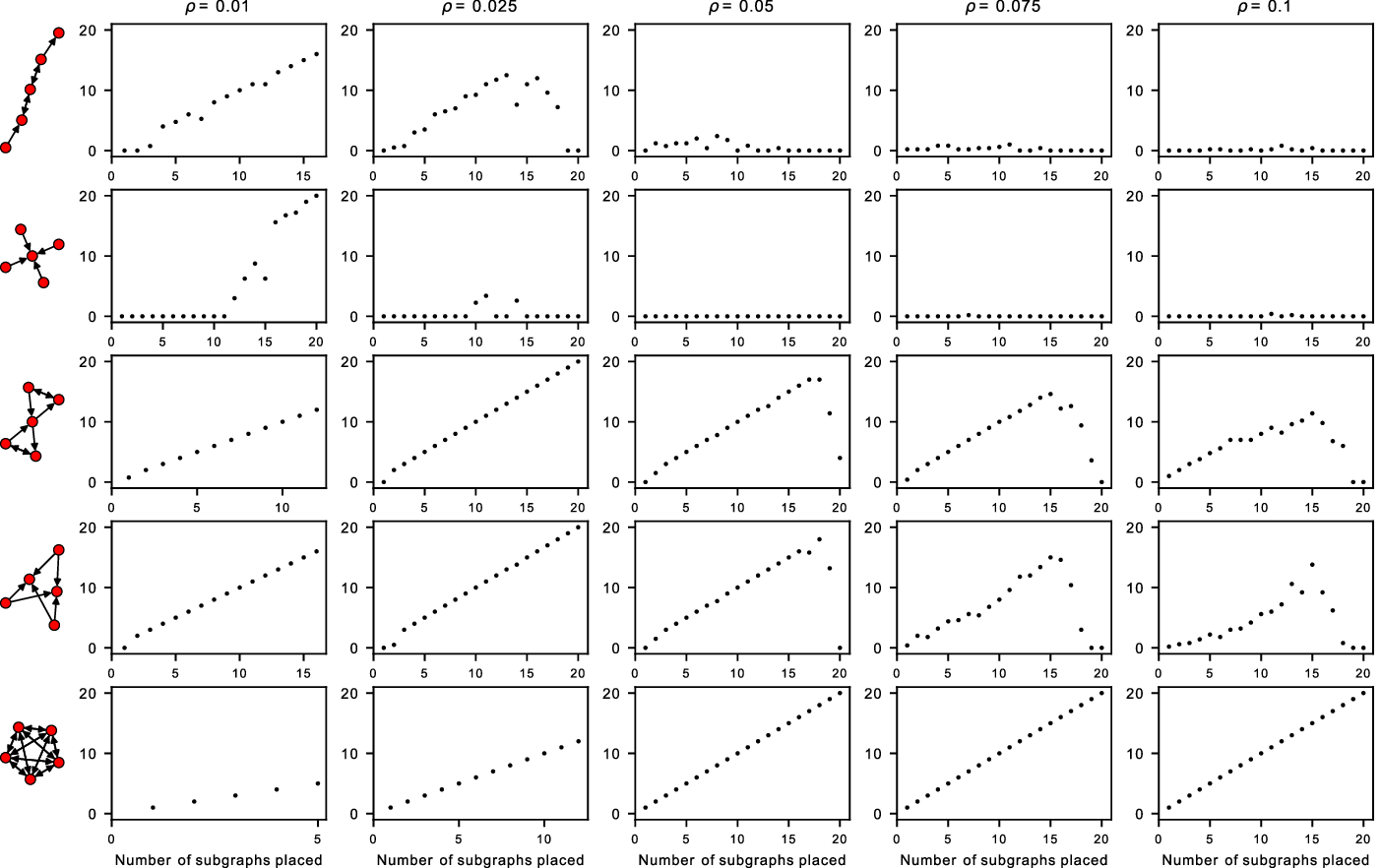}
    \end{adjustwidth}
    \caption{\textbf{Number of occurrences of the planted motif inferred ($N = 100$).} The number of insertions in the generated graphs is plotted on the x-axis, and this inferred number, averaged over five independent graphs, on the y-axis. 
    The generated networks contain {$N = 100$} nodes and the edge density ranges from {$\rho = 0.01$} (leftmost) to {$\rho=0.1$} (rightmost).
    Each point is an average over five independently generated graphs.     
    Note that the maximum value of motifs that can be inserted depend both on the number of nodes in the network and on the networks density as well as that of the motif; hence the range of the x-axis can vary.}
    \labelname{S6}
    \label{SIfig:motifs_detected_100}
\end{figure}

\begin{figure}
    \centering
    \begin{adjustwidth}{-.75in}{}
        \includegraphics[height=0.42\textheight]{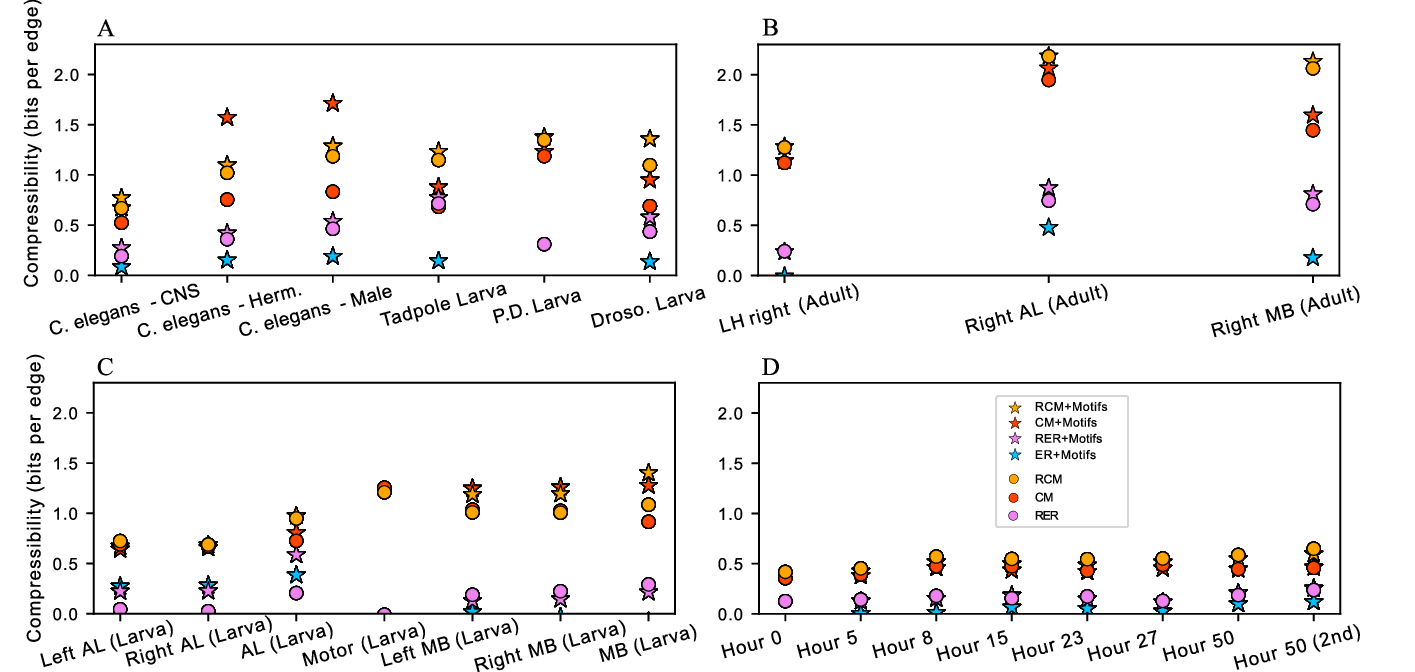}
    \end{adjustwidth}
    \caption{\textbf{Compressibility per edge of the connectomes obtained with the different base models, with and without motifs.}
    Difference in codelength between the simple Erd\H{o}s-R\'enyi (ER) model and each of the other seven models (RER: reciprocal ER model, CM: configuration model, RCM: reciprocal configuration model, ER+Motifs: ER base model with motifs, RER+Motifs: reciprocal ER base with motifs, CM+Motifs: configuration model with motifs, RCM+Motifs: reciprocal configuration model with motifs).}
    \labelname{S7}
    \label{SIfig:compressibility_B}
\end{figure}

\begin{figure}
    \centering
    \begin{adjustwidth}{.6in}{}
        \includegraphics[height=0.3\textheight]{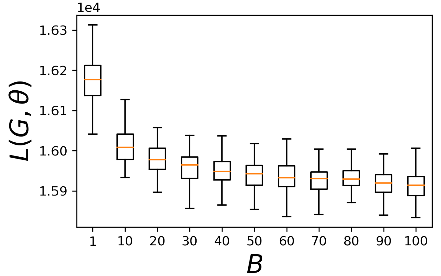}
    \end{adjustwidth}
    \caption{\textbf{Dependence of the optimum model on the batch size.}
    Mean codelength of the inferred model ($\pm$ SD) for different minibatch sizes $B$, where $B$ is the number of occurrences of each graphlet sampled. The inference is performed on the \textit{Drosophila} larva right MB and run 100 times independently for each $B$ value.}
    \labelname{S8}
    \label{SIfig:minibatch_size_effect}
\end{figure}

\end{document}